\documentclass{article}%
\usepackage{graphicx}
\usepackage{amsmath}
\usepackage{amsfonts}
\usepackage{amssymb}%
\providecommand{\U}[1]{\protect\rule{.1in}{.1in}}

\begin{document}

\author{Antony Valentini\\Augustus College}

\begin{center}
{\LARGE Quantum gravity and quantum probability}

\bigskip

\bigskip

\bigskip

Antony Valentini\footnote{email: antonyv@clemson.edu}

\textit{Augustus College,}

\textit{14 Augustus Road, London SW19 6LN, UK.}

\textit{Department of Physics and Astronomy,}

\textit{Clemson University, Kinard Laboratory,}

\textit{Clemson, SC 29634-0978, USA.}

\bigskip

\bigskip
\end{center}

We argue that in quantum gravity there is no Born rule. The quantum-gravity
regime, described by a non-normalisable Wheeler-DeWitt wave functional $\Psi$,
must be in quantum nonequilibrium with a probability distribution
$P\neq\left\vert \Psi\right\vert ^{2}$ (initially and always). A Born rule can
emerge only in the semiclassical regime of quantum systems on a classical
spacetime background, with normalisable Schr\"{o}dinger wave functions $\psi$.
Conditioning on the underlying quantum-gravitational ensemble yields a
nonequilibrium distribution $\rho\neq\left\vert \psi\right\vert ^{2}$ at the
beginning of the semiclassical regime, with quantum relaxation $\rho
\rightarrow\left\vert \psi\right\vert ^{2}$ taking place only afterwards.
Quantum gravity naturally creates an early nonequilibrium universe. We also
show how small corrections to the Schr\"{o}dinger equation yield an
intermediate regime in which the Born rule is unstable: an initial
distribution $\rho=\left\vert \psi\right\vert ^{2}$ can evolve to a final
distribution $\rho\neq\left\vert \psi\right\vert ^{2}$. These results arise
naturally in the de Broglie-Bohm pilot-wave formulation of quantum gravity. We
show that quantum instability during inflation generates a large-scale deficit
$\sim1/k^{3}$ in the primordial power spectrum at wavenumber $k$, though the
effect is too small to observe. Similarly we find an unobservably large
timescale for quantum instability in a radiation-dominated universe. Quantum
instability may be important in black-hole evaporation, with a final burst of
Hawking radiation that violates the Born rule. Deviations from the Born rule
can also be generated for atomic systems in the gravitational field of the
earth, though the effects are unlikely to be observable. The most promising
scenario for the detection of Born-rule violations appears to be in radiation
from exploding primordial black holes.

\bigskip

\bigskip

\bigskip

\bigskip

\bigskip

\section{Introduction}

An unfinished task of theoretical physics concerns how to combine the two
great theories of the twentieth century -- general relativity and quantum
mechanics -- into a coherent theory of quantum gravity. One approach has been
to apply the methods of canonical quantisation to the gravitational field.
Despite some initial confusion, by the late 1960s a formal theory was arrived
at in which the quantum state $\Psi\lbrack g_{ij},\phi]$ is a functional of
the 3-metric $g_{ij}$ (together with other quantum fields $\phi$) and obeys a
time-independent Wheeler-DeWitt equation of the form $\mathcal{\hat{H}}\Psi=0$
(for an appropriate operator $\mathcal{\hat{H}}$) \cite{DeW67,Kief12}. A
consensus was reached on the formal equations defining the theory, but to this
day there remains widespread controversy over how to interpret them -- a
controversy which is often summarised under the general heading of the
`problem of time' [2--8]. The underlying theory appears to be timeless, and
yet in some semiclassical approximation it must yield the standard
time-dependent quantum mechanics of systems on a classical spacetime
background. There are many approaches to solving this problem, and authors
differ as to whether or not the problem has in fact been solved. Controversy
over the physical interpretation of the theory lingers despite the fundamental
equations having been written down more than half a century ago. Much progress
has been made in recent decades rewriting the equations in terms of
alternative sets of variables, resulting in a version of the theory widely
known as loop quantum gravity, which has certain technical advantages while
retaining similar interpretational problems [9--12]. The problems of
interpretation seem unrelated to questions concerning ultraviolet divergences
and perturbative non-renormalisability, arising as they do even in
finite-dimensional minisuperspace models. Instead the problems appear to
signal a basic conceptual difficulty that afflicts the formalism. It has long
been suspected that progress in understanding quantum gravity might require
some change in our understanding of quantum mechanics. Some authors have
proposed, for example, that we must reformulate quantum mechanics itself as a
timeless theory [13--15].

In this paper we suggest that some of the conceptual difficulties in
understanding quantum gravity arise from the unwarranted assumption that the
Born rule is still relevant in the deep quantum-gravity regime. Our everyday
laboratory experience tells us that quantum systems have time-dependent wave
functions $\psi$ and the Born rule states that $\left\vert \psi\right\vert
^{2}$ is a probability density. It has long been assumed that some analogue of
the Born rule must hold even at the Planck scale. Some early workers assumed
that the Wheeler-DeWitt wave functional $\Psi$ yields a probability density
$\left\vert \Psi\right\vert ^{2}$, but this `naive Schr\"{o}dinger
interpretation' encountered difficulties and has been widely abandoned. Among
other problems the density $\left\vert \Psi\right\vert ^{2}$ is
non-normalisable -- and not for merely technical reasons but for the deep
physical reason that the Wheeler-DeWitt equation has the structure of a
Klein-Gordon-like wave equation on configuration space. Even so the many
approaches to solving the problem of time generally assume that some form of
Born rule must be applied. We argue here that this is a mistake, that in the
deep quantum-gravity regime there is no such thing as the Born rule, and that
the Born rule emerges \textit{only} in the semiclassical regime for quantum
systems on a classical spacetime background.

To make sense of this proposal requires a formulation of quantum mechanics in
which the Born rule is not an axiom or law but is merely emergent. Such a
formulation is provided by the pilot-wave theory of de Broglie and Bohm
[16--21] -- at least when correctly interpreted [22--30]. In pilot-wave theory
a quantum system has an evolving configuration $q(t)$ whose motion is
determined by the configuration-space wave function $\psi(q,t)$. For standard
systems the components of the velocity $dq/dt$ are proportional to the
gradient of the phase $S=\operatorname{Im}\ln\psi$. This nonclassical theory
of motion -- originally due to de Broglie\footnote{Pilot-wave theory was first
presented by de Broglie, for a many-body system, at the 1927 Solvay conference
\cite{deB28}. For a complete translation of the conference proceedings, and a
detailed historical analysis, see ref. \cite{BV09}.} -- provides a
deterministic theory of trajectories for quantum systems. But the equations
determine a trajectory $q(t)$ only given the initial configuration $q(t_{i})$
(and the initial wave function $\psi(q,t_{i})$) at some initial time $t_{i}$.
Over an ensemble of systems with the same $\psi(q,t_{i})$ we will have some
initial distribution $\rho(q,t_{i})$ of configurations. If we \textit{assume}
that $\rho(q,t_{i})=\left\vert \psi(q,t_{i})\right\vert ^{2}$ it follows from
the equations that this initial Born-rule density is preserved in time in the
sense that we obtain $\rho(q,t)=\left\vert \psi(q,t)\right\vert ^{2}$ at later
times $t$. The Born rule describes a state of `quantum equilibrium'. With this
assumption about the initial conditions the empirical predictions of textbook
quantum mechanics are recovered (as first shown in full detail by Bohm)
[18--21]. However, the status of the Born rule as an initial condition in
pilot-wave theory has been disputed. Most authors take it as an extra
postulate or law of the theory (along with the Schr\"{o}dinger equation for
$\psi$ and the de Broglie `guidance equation' for the velocity $dq/dt$) [19,
31--33]. This author has long argued that this is a mistake [20--30, 34].
There is a basic conceptual difference between initial conditions on the one
hand and laws of motion on the other. For an ensemble of systems with the same
$\psi(q,t_{i})$, the actual initial distribution $\rho(q,t_{i})$ of
configurations is in principle arbitrary (just as, in classical mechanics, for
an ensemble of systems with the same Hamiltonian the actual initial
distribution $\rho(q,p,t_{i})$ on phase space is in principle arbitrary). In
particular $\rho(q,t_{i})$ may or may not be equal to $\left\vert \psi
(q,t_{i})\right\vert ^{2}$. If we take pilot-wave theory seriously as a
physical theory, it tells us that more general non-Born-rule distributions
$\rho(q,t_{i})\neq\left\vert \psi(q,t_{i})\right\vert ^{2}$ (corresponding to
`quantum nonequilibrium') are possible at least in principle, resulting in a
wider physics beyond that allowed by the usual quantum formalism [20--30,
34--36]. In other words, quantum theory is merely a special case of a much
wider physics \cite{AV09}.

It has been argued that in pilot-wave theory the Born rule emerges by a
dynamical process of `quantum relaxation', broadly analogous to thermal
relaxation in classical physics, a view which has been supported by extensive
numerical simulations [22, 24, 26, 30, 38--44]. On this view relaxation
$\rho(q,t)\rightarrow\left\vert \psi(q,t)\right\vert ^{2}$ to the Born rule
took place (on a coarse-grained level) some time in the remote past, probably
in the very early universe [23--26, 30, 34, 37], and may have left traces in
the cosmic microwave background [34, 36, 45--50] or in relic cosmological
particles \cite{AV01,AV07,AV08a,UV15,UV20}. Simulations of quantum relaxation
have been carried out for a variety of systems on a classical spacetime
background, including quantum scalar fields in a cosmological setting. It has
been shown that quantum relaxation tends to be suppressed for long-wavelength
(super-Hubble)\ field modes on expanding space [36, 45--47, 49], but in most
respects numerical studies have confirmed the general picture of the Born rule
emerging by dynamical relaxation.

In non-gravitational physics quantum equilibrium is stable in the sense that
it is preserved in time. Under standard operations and interactions an initial
equilibrium state $\rho=\left\vert \psi\right\vert ^{2}$ will evolve into a
final equilibrium state $\rho=\left\vert \psi\right\vert ^{2}$. Thus, for
example, the Born rule continues to hold in high-energy collisions (as probed
by scattering cross-sections). In pilot-wave theory this stability is a simple
consequence of the dynamics, which evolves an initial Born-rule state to a
final Born-rule state. We might then ask what happens to quantum equilibrium
in the presence of gravitational processes. It has been suggested that the
Born rule could become unstable during the formation and complete evaporation
of a black hole, so as to compensate for the apparent information loss that
would otherwise occur \cite{AV07,AV04b,KV20}. On this scenario an evaporating
black hole would emit Hawking radiation in a state of quantum nonequilibrium
$\rho\neq\left\vert \psi\right\vert ^{2}$. However, the arguments given are
semiclassical, for quantum fields on a classical spacetime background. The
suggested mechanism, which involves entanglement between ingoing and outgoing
field modes, depends on the assumption that quantum nonequilibrium is somehow
generated behind the horizon (close to the singularity). This lacks a sound
theoretical basis. To make further progress requires the application of
pilot-wave theory to the gravitational field itself.

A pilot-wave theory of quantum gravity, in which the Wheeler-DeWitt wave
functional $\Psi\lbrack g_{ij},\phi]$ is supplemented by a de Broglie-Bohm
trajectory $(g_{ij}(t),\phi(t))$, has been considered by a number of authors.
Beginning with the early papers of Vink \cite{Vink92}, Horiguchi \cite{Hor94}
and Shtanov \cite{Sht96}, the theory has been extensively applied to quantum
cosmology in particular by Pinto-Neto and collaborators [58--61]. An
alternative approach to pilot-wave quantum gravity considers a
Schr\"{o}dinger-like equation for a time-dependent wave functional
$\Psi\lbrack g_{ij},\phi,t]$ with a preferred time parameter $t$
\cite{AV92,AV96,RV14}. In such a theory the Born-rule distribution
$P[g_{ij},\phi,t]=\left\vert \Psi\lbrack g_{ij},\phi,t]\right\vert ^{2}$
should be stable by construction. However the consistency and completeness of
the latter approach remains in doubt \cite{AVBook}.

Here we restrict ourselves to pilot-wave theory as applied to the timeless
Wheeler-DeWitt equation with a time-independent wave functional $\Psi\lbrack
g_{ij},\phi]$. In previous work the focus has been on properties of the
trajectories for the evolving 3-metric (for example whether or not the
trajectories are singularity-avoiding in cosmological scenarios \cite{NonSing}%
), while little has been done in considering probabilities and the Born rule.
As we shall see, the usual problems emerge when trying to interpret the
non-normalisable (and static) Wheeler-DeWitt density $\left\vert
\Psi\right\vert ^{2}$ as a probability density. Perhaps for this reason most
authors in the field have confined themselves to considering properties of the
trajectories alone, without attempting to define a theory of ensembles or of probabilities.

In this paper we propose a new approach to understanding the Born rule in
quantum gravity.\footnote{Related and preliminary versions of these proposals
were given in refs. \cite{AV14,AV18}. For a concise overview of the present
work see ref. \cite{AV21}.} Adopting a pilot-wave theory of gravity with the
Wheeler-DeWitt equation, we argue that at the fundamental level a probability
density $P$ must be, and always remains, \textit{unequal} to the apparent
`Born-rule' density $\left\vert \Psi\right\vert ^{2}$, since $P$ is
normalisable (by construction)\ while $\left\vert \Psi\right\vert ^{2}$ is
not. Quantum relaxation cannot take place in the usual way because there is no
well-defined equilibrium state. In effect the deep quantum-gravity regime is
in a perpetual state of quantum nonequilibrium $P\neq\left\vert \Psi
\right\vert ^{2}$. Even so, the Born rule can be recovered in the
semiclassical regime, with the Schr\"{o}dinger approximation for an effective
time-dependent (and normalisable) wave function $\psi$. In that regime quantum
relaxation $\rho\rightarrow\left\vert \psi\right\vert ^{2}$ can proceed in the
usual way and over time we recover the Born rule $\rho=\left\vert
\psi\right\vert ^{2}$ as an equilibrium state. At the beginning of the
semiclassical regime, however, the system is expected to be in a state of
nonequilibrium $\rho\neq\left\vert \psi\right\vert ^{2}$, with $\rho$ arising
as a conditional probability from the underlying quantum-gravitational
nonequilibrium state $P\neq\left\vert \Psi\right\vert ^{2}$. Relaxation
$\rho\rightarrow\left\vert \psi\right\vert ^{2}$ takes place only afterwards.
In this way the hypothesis of primordial quantum nonequilibrium [22--27, 30,
34, 36, 37] is derived as a consequence of quantum gravity.

We also study an intermediate regime, with quantum-gravitational corrections
to the Schr\"{o}dinger approximation, following methods developed by Kiefer
and collaborators [68--71]. A long-standing difficulty in the field has been
the appearance of small, quantum-gravitationally induced, non-Hermitian
corrections in the effective Hamiltonian (for a field system on a classical
background). In standard quantum theory such terms violate the conservation of
probability. For this reason, in such calculations usually only the Hermitian
corrections are kept and the non-Hermitian terms are discarded. We show that,
when reformulated in terms of pilot-wave theory, probabilities are fully
conserved even in the presence of non-Hermitian terms, whose effect is instead
to generate a small instability of the Born rule: an initial equilibrium
distribution $\rho=\left\vert \psi\right\vert ^{2}$ can in principle evolve to
a nonequilibrium distribution $\rho\neq\left\vert \psi\right\vert ^{2}$ (at
least in circumstances where relaxation is relatively negligible). In this
paper we study various systems where such `quantum instability' can occur: a
scalar field on de Sitter space, a field on a radiation-dominated background,
a field in the spacetime of an evaporating black hole, and an atomic system
with a rapidly-changing Hamiltonian in a curved background. In general the
effects are found to be extremely small, with the possible exception of the
final stages of black-hole evaporation where the effects may well be significant.

An outline of this paper now follows. We employ units with $\hbar=c=16\pi
G=1$. In Section 2 we review the status of the Born rule in pilot-wave theory
on a classical spacetime background. In Section 3 we review the difficulties
with understanding and applying the Born rule in canonical quantum gravity in
both its conventional and pilot-wave versions. Our new approach to the Born
rule is explained and outlined in general terms in Section 4. In Section 5 we
consider quantum-gravitational corrections to the Schr\"{o}dinger
approximation and in Section 6 we show how the non-Hermitian part of such
corrections can be understood in pilot-wave theory as generating a small
instability of the Born rule. In Section 7 we apply these proposals to quantum
cosmology and implement a simplified model to enable tractable calculations.
This model is applied in Section 8 to study the gravitational creation of
quantum nonequilibrium for a scalar field on de Sitter space. In Section 9
these results are employed to derive an approximate correction to the
cosmological primordial power spectrum, in the form of a small power deficit
scaling with wave number $k$ as $1/k^{3}$, whose magnitude is however
estimated to be too small to be observable in the cosmic microwave background.
In Section 10 we study similar effects for a scalar field in a
radiation-dominated expanding universe. In Section 11 we show how quantum
nonequilibrium is expected to be created by quantum-gravitational effects in
the spacetime of an evaporating black hole, where the effects are estimated to
be significant only in the final stages of evaporation when the mass $M$ of
the hole approaches the Planck mass $m_{\mathrm{P}}$. Finally, in Section 12
we consider how comparable effects can occur for atomic systems with
rapidly-changing Hamiltonians in the gravitational field of the earth, however
the effects are so tiny as to be seemingly of theoretical interest only. Our
conclusions are drawn in Section 13.

\section{Pilot-wave theory and the Born rule}

In this section we outline the status of the Born rule in pilot-wave theory.

\subsection{Pilot-wave theory and quantum equilibrium}

In pilot-wave theory a general system has an evolving configuration $q(t)$
with a velocity law [16--21]%
\begin{equation}
\frac{dq}{dt}=v(q,t)\label{deB1}%
\end{equation}
where $v(q,t)$ is determined by the wave function $\psi(q,t)$. The time
parameter $t$ is associated with a foliation of spacetime by spacelike
hypersurfaces. For a system evolving on a classical spacetime background, $q$
is the configuration of fields and particles on a curved 3-space (as defined
by the hypersurfaces). The de Broglie velocity field $v(q,t)$ is defined as
follows. The usual Schr\"{o}dinger equation on configuration
space,\footnote{Systems with spin have multi-component wave functions. It will
not be necessary to consider such systems here.}%
\begin{equation}
i\frac{\partial\psi}{\partial t}=\hat{H}\psi\ ,\label{Sch1}%
\end{equation}
implies a continuity equation%
\begin{equation}
\frac{\partial\left\vert \psi\right\vert ^{2}}{\partial t}+\partial_{q}\cdot
j=0\label{Contj}%
\end{equation}
for a density $\left\vert \psi\right\vert ^{2}$ and a current $j=j\left[
\psi\right]  =j(q,t)$, where $\partial_{q}$ is a generalised gradient and $j$
satisfies%
\begin{equation}
\partial_{q}\cdot j=2\operatorname{Re}\left(  i\psi^{\ast}\hat{H}\psi\right)
\ .\label{div_j}%
\end{equation}
The explicit expression for $j$ in terms of $\psi$ is determined by the form
of the Hamiltonian $\hat{H}$, and such a current exists whenever $\hat{H}$ is
given by a differential operator \cite{SV09}. Given an expression for $j$, the
de Broglie velocity field is defined by%
\begin{equation}
v(q,t)=\frac{j(q,t)}{|\psi(q,t)|^{2}}\ .\label{v}%
\end{equation}
The equation of motion (\ref{deB1}) then determines a trajectory $q(t)$ in
configuration space, given an initial configuration $q(0)$ and an initial wave
function $\psi(q,0)$. Physical Hamiltonians are often quadratic in the
canonical momenta, in which case the components $v_{a}$ of $v$ are
proportional to the components of the phase gradient:%
\begin{equation}
v_{a}\propto\partial_{q_{a}}S=\operatorname{Im}\left(  \frac{\partial_{q_{a}%
}\psi}{\psi}\right)  \label{vst}%
\end{equation}
(where $\psi=\left\vert \psi\right\vert e^{iS}$). It is important to note that
the `pilot wave' $\psi$ is a field on configuration space that guides the
deterministic motion of an individual system. Fundamentally $\psi$ has nothing
to do with probability.

We can consider an ensemble of systems with the same wave function $\psi
(q,t)$. The ensemble will have an evolving distribution $\rho(q,t)$ of
configurations $q(t)$. The initial distribution $\rho(q,t_{i})$ at time
$t_{i}$ need not be equal to the squared-amplitude $\left\vert \psi
(q,t_{i})\right\vert ^{2}$ of the initial pilot wave $\psi(q,t_{i})$. Because
the individual configurations evolve by the equation of motion (\ref{deB1}),
the ensemble distribution $\rho(q,t)$ will necessarily evolve by the
continuity equation%
\begin{equation}
\frac{\partial\rho}{\partial t}+\partial_{q}\cdot\left(  \rho v\right)  =0\ .
\label{cont1}%
\end{equation}
This is the same as the continuity equation (\ref{Contj}) for the time
evolution of $\left\vert \psi\right\vert ^{2}$. We then have a simple `quantum
equilibrium theorem': if $\rho$ and $\left\vert \psi\right\vert ^{2}$ happen
to be equal at an initial time $t_{i}$ then they will remain equal at later
times $t$. Thus an initial distribution $\rho(q,t_{i})=\left\vert \psi
(q,t_{i})\right\vert ^{2}$ evolves into a final distribution%
\begin{equation}
\rho(q,t)=\left\vert \psi(q,t)\right\vert ^{2}\ . \label{qe1}%
\end{equation}
This is the state of quantum equilibrium, in which the ensemble obeys the Born rule.

It is sometimes useful to consider how the ratio%
\begin{equation}
f=\frac{\rho}{|\psi|^{2}} \label{f}%
\end{equation}
evolves along a trajectory. From (\ref{cont1}) and (\ref{Contj}) it is easy to
show that%
\begin{equation}
\frac{df}{dt}=0\ , \label{fdot}%
\end{equation}
where $d/dt=\partial/\partial t+v\cdot\partial_{q}$ is the time derivative
along a trajectory with local velocity $v$.

As was first shown in detail by Bohm \cite{B52}, in the state (\ref{qe1}) of
quantum equilibrium the statistical predictions for the outcomes of general
quantum measurements agree with the usual predictions of textbook quantum
theory.\footnote{See refs. \cite{Holl93,AVOUP} for detailed accounts of
textbook quantum mechanics in terms of pilot-wave theory.} In contrast, for a
`quantum nonequilibrium' ensemble with a non-Born-rule distribution
$\rho(q,t)\neq\left\vert \psi(q,t)\right\vert ^{2}$, the statistical
predictions generally differ from those of quantum theory [20--26, 28--30, 34,
36, 37]. Such ensembles entail new physics outside the domain of conventional
quantum physics. If pilot-wave theory is taken seriously, we must conclude
that quantum physics is merely an effective theory of an equilibrium state --
and that, at least in principle, there is a much wider nonequilibrium physics
beyond the physics that is currently known.

For all systems that are currently accessible to us, experiments have
confirmed the Born rule $\rho=\left\vert \psi\right\vert ^{2}$. This state can
be understood as having arisen from a dynamical process of quantum relaxation
(analogous to thermal relaxation). Because $\rho$ and $\left\vert
\psi\right\vert ^{2}$ obey the same continuity equation, the fine-grained
$H$-function%
\begin{equation}
H(t)=\int dq\ \rho\ln(\rho/\left\vert \psi\right\vert ^{2}) \label{Hfn}%
\end{equation}
(minus the relative entropy of $\rho$ with respect to $\left\vert
\psi\right\vert ^{2}$) is constant in time, $dH/dt=0$. But if we assume that
$\rho$ and $\left\vert \psi\right\vert ^{2}$ have no fine-grained structure at
some initial time $t_{i}$, the coarse-grained $H$-function%
\begin{equation}
\bar{H}(t)=\int dq\ \bar{\rho}\ln(\bar{\rho}/\overline{\left\vert
\psi\right\vert ^{2}}) \label{Hfnbar}%
\end{equation}
obeys an $H$-theorem%
\begin{equation}
\bar{H}(t)\leq\bar{H}(t_{i})\ , \label{Hthm}%
\end{equation}
where $\bar{H}(t)$ is bounded below by zero and is equal to zero if and only
if $\bar{\rho}=\overline{\left\vert \psi\right\vert ^{2}}$ everywhere
\cite{AV91a,AV92,AV01}. Coarse-graining is needed because of the fine-grained
conservation (\ref{fdot}) of $f$, which is analogous to the classical
Liouville theorem on phase space. As in classical thermal relaxation for an
isolated system, we must assume that the initial state has no fine-grained
structure.\footnote{For a detailed discussion see ref. \cite{Allori20}.}

Wide-ranging numerical simulations show that, when $\psi$ is a superposition
of energy eigenstates, there is rapid coarse-grained relaxation $\bar{\rho
}\rightarrow\overline{\left\vert \psi\right\vert ^{2}}$ [22, 24, 26, 30,
38--44], with $\bar{H}(t)$ decaying approximately exponentially towards zero
\cite{VW05,TRV12,ACV14},%
\begin{equation}
\bar{H}(t)\approx\bar{H}(0)\exp(-t/\tau_{\mathrm{relax}})\ ,
\end{equation}
where the quantum relaxation timescale $\tau_{\mathrm{relax}}$ depends on
$\psi$ (as well as on the coarse-graining length $\varepsilon$) \cite{TRV12}.
Numerical results for two-dimensional systems have yielded values
$\tau_{\mathrm{relax}}$ roughly comparable to the quantum timescale over which
$\psi$ evolves, however there is no general relation between the two
timescales. As expected relaxation tends to be faster for larger numbers $M$
of superposed modes (as well as for larger $\varepsilon$). For particles in a
two-dimensional box, for example, there is strong numerical evidence for an
approximate inverse scaling $\tau_{\mathrm{relax}}\propto1/M$ (at fixed
$\varepsilon$) \cite{TRV12}. Similar results are found in pilot-wave field
theory, where a single scalar field mode is mathematically equivalent to a
two-dimensional oscillator. For fields on expanding space the results are
somewhat modified: quantum relaxation takes place efficiently at short
(sub-Hubble)\ wavelengths but is suppressed at long (super-Hubble) wavelengths
\cite{AV07,AV08a,CV13}.

It has been suggested that quantum relaxation took place in the very early
universe [23--26, 30, 34, 37]. Ordinary laboratory systems have a long and
violent astrophysical history, and are expected to have reached quantum
equilibrium a long time ago. For such systems we can expect to see the Born
rule today to exceedingly high accuracy. However, quantum nonequilibrium in
the early universe can leave an observable imprint in the cosmic microwave
background (CMB) \cite{AV10,AV07,AV08a,CV13,AV15}, with a specific signature
\cite{CV15,CV16} (caused by super-Hubble suppression of relaxation) which has
been searched for in recent CMB data \cite{VPV19}. Furthermore, early
nonequilibrium might survive to the present day in relic cosmological
particles if they decoupled at sufficiently early times
\cite{AV01,AV07,AV08a,UV15}. This could potentially be observed in the form of
anomalous spectra for decaying or annihilating dark matter \cite{UV20}. Thus
the wider physics of quantum nonequilibrium may have existed in the very early
universe, before quantum relaxation took place, possibly leaving traces in the
CMB and in relic particles today.

Finally, we mention an alternative approach to understanding the Born rule in
pilot-wave theory, which will be relevant later. Beginning with ref.
\cite{DGZ92} there has developed a distinctive school of de Broglie-Bohm
theory called `Bohmian mechanics' \cite{DT09}.\footnote{This terminology
should properly be applied to Bohm's 1952 reformulation \cite{B52} of de
Broglie's original 1927 dynamics \cite{deB28}. Bohm's version of the theory
(based on an equation for acceleration with a pseudo-Newtonian `quantum
potential') has been shown to be unstable \cite{CV14}.} This school has been
particularly influential among philosophers \cite{Tum18,Gold17}. The Bohmian
mechanics school claims that a fundamental role is played by the initial
Born-rule measure $\left\vert \Psi(q,0)\right\vert ^{2}$ for the universe,
where $\Psi(q,0)$ is the initial universal wave function at $t=0$. It is
claimed that $\left\vert \Psi(q,0)\right\vert ^{2}$ provides a fundamental
measure of `typicality' (or equivalently, of probability) for the initial
universal configuration $q(0)$, implying Born-rule probabilities for
subsystems. In effect, the Born rule is asserted to have a law-like status at
the beginning of the universe, and the Born rule we see at later times is
simply a consequence of the Born rule at $t=0$. However, a different choice of
initial measure, such as $\left\vert \Psi(q,0)\right\vert ^{4}$, yields
non-Born rule distributions for subsystems at $t=0$ \cite{AV96,AV01,Allori20}.
To obtain the Born rule at later times we would then have to appeal to some
form of dynamical relaxation. But the Bohmian mechanics school claims that
$\left\vert \Psi(q,0)\right\vert ^{2}$ is the natural choice at $t=0$, and
that this suffices as an explanation, rendering dynamical relaxation
superfluous. The argument is, however, circular and unjustified.\footnote{See
ref. \cite{Allori20} for a detailed critique of this approach -- as well as of
the Bohmian mechanics school generally.} The Born rule is assumed to apply to
the whole universe, in order to obtain the Born rule for subsystems. As we
will see in Section 4, even leaving the circularity aside, a careful analysis
of the role of probability in pilot-wave quantum gravity definitively resolves
the dispute in favour of dynamical relaxation.

\subsection{Quantum equilibrium on a globally-hyperbolic spacetime}

We have seen that the existence of a quantum equilibrium state is a trivial
consequence of the structure of pilot-wave dynamics for any system that obeys
a Schr\"{o}dinger equation with an associated conserved current $j$. Given
such a current, $\rho$ and $\left\vert \psi\right\vert ^{2}$ will by
construction obey the same continuity equation (with the same velocity field
$v=j/\left\vert \psi\right\vert ^{2}$) and the quantum equilibrium theorem
immediately follows. It is instructive to illustrate this for field theory on
a classical curved spacetime background -- assuming that the spacetime is
globally hyperbolic \cite{AV04b}.

A globally hyperbolic spacetime can always be foliated (usually nonuniquely)
by spacelike hypersurfaces $\Sigma(t)$ labelled by a global time function $t$.
The line element $d\tau^{2}=\,^{(4)}g_{\mu\nu}dx^{\mu}dx^{\nu}$ (with 4-metric
$^{(4)}g_{\mu\nu}$) can then be written in the form \cite{ADM62}%
\begin{equation}
d\tau^{2}=(N^{2}-N_{i}N^{i})dt^{2}-2N_{i}dx^{i}dt-g_{ij}dx^{i}dx^{j}\ ,
\label{ADM}%
\end{equation}
where $N$ is the lapse function, $N^{i}$ is the shift vector, and $g_{ij}$ is
the 3-metric on $\Sigma(t)$. We have a proper time element $d\tau=Ndt$ normal
to the slices $\Sigma(t)$, where the normal deviates from lines of constant
$x^{i}$ by $dx^{i}=-N^{i}(x^{j},t)dt$. For simplicity we can take $N^{i}=0$ so
that lines $x^{i}=\mathrm{const}.$ are normal to the slices (this can always
be done provided such lines do not meet singularities).

Consider for example a massless, minimally-coupled real scalar field $\phi$
with Lagrangian density%
\begin{equation}
\mathcal{L}=\frac{1}{2}\sqrt{-\,^{(4)}g}\,^{(4)}g^{\mu\nu}\partial_{\mu}%
\phi\partial_{\nu}\phi
\end{equation}
(with $^{(4)}g=\det g_{\mu\nu}$). This implies a canonical momentum density%
\begin{equation}
\pi=\frac{\partial\mathcal{L}}{\partial\dot{\phi}}=\frac{\sqrt{g}}{N}\dot
{\phi}%
\end{equation}
(with $g=\det g_{ij}$ and taking $N^{i}=0$) and a Hamiltonian%
\begin{equation}
H=\int d^{3}x\;\frac{1}{2}N\sqrt{g}\left(  \frac{1}{g}\pi^{2}+g^{ij}%
\partial_{i}\phi\partial_{j}\phi\right)  \ .
\end{equation}
The system may then be quantised in the usual way, with field operators
$\hat{\phi}(x)$ and $\hat{\pi}(x)$ on $\Sigma(t)$. In the functional
Schr\"{o}dinger picture, with the realisations $\hat{\phi}(x)\rightarrow
\phi(x)$ and $\hat{\pi}(x)\rightarrow-i\delta/\delta\phi(x)$, the wave
functional $\Psi\lbrack\phi,t]$ obeys the Schr\"{o}dinger equation\footnote{In
this context it is usual to implicitly assume some form of regularisation --
for example, dimensional regularisation \cite{Guv89}.}%
\begin{equation}
i\frac{\partial\Psi}{\partial t}=\int d^{3}x\;\frac{1}{2}N\sqrt{g}\left(
-\frac{1}{g}\frac{\delta^{2}}{\delta\phi^{2}}+g^{ij}\partial_{i}\phi
\partial_{j}\phi\right)  \Psi\ . \label{Sch2}%
\end{equation}

So far we have simply written down a standard quantum field theory on a curved
background. To convert this into a pilot-wave theory we note that (\ref{Sch2})
implies a continuity equation%
\begin{equation}
\frac{\partial\left\vert \Psi\right\vert ^{2}}{\partial t}+\int d^{3}%
x\;\frac{\delta}{\delta\phi}\left(  \left\vert \Psi\right\vert ^{2}\frac
{N}{\sqrt{g}}\frac{\delta S}{\delta\phi}\right)  =0 \label{cont2}%
\end{equation}
(where $\Psi=\left\vert \Psi\right\vert e^{iS}$) with a current%
\begin{equation}
j=\left\vert \Psi\right\vert ^{2}\frac{N}{\sqrt{g}}\frac{\delta S}{\delta\phi
}\ . \label{j2}%
\end{equation}
From (\ref{v}) we then have a de Broglie velocity%
\begin{equation}
\frac{\partial\phi}{\partial t}=\frac{N}{\sqrt{g}}\frac{\delta S}{\delta\phi
}\ . \label{deB2}%
\end{equation}
In addition to the evolving wave $\Psi\lbrack\phi,t]$ on configuration space
we also have an evolving field $\phi(x,t)$ on 3-space.

Equations (\ref{Sch2}) and (\ref{deB2}) define the dynamics for an individual
field on a curved background. We can also consider a theoretical ensemble of
fields guided by the same wave functional $\Psi$ (on the same curved
background). The ensemble will have some arbitrary initial distribution
$P[\phi,t_{i}]$, which need not be equal to $\left\vert \Psi\lbrack\phi
,t_{i}]\right\vert ^{2}$. Because each field has the velocity (\ref{deB2}),
the distribution $P[\phi,t]$ will necessarily evolve by the continuity
equation%
\begin{equation}
\frac{\partial P}{\partial t}+\int d^{3}x\;\frac{\delta}{\delta\phi}\left(
P\frac{N}{\sqrt{g}}\frac{\delta S}{\delta\phi}\right)  =0\ . \label{cont2'}%
\end{equation}
Again, by construction, this is the same continuity equation (\ref{cont2})
that is satisfied by $\left\vert \Psi\right\vert ^{2}$ and so the quantum
equilibrium theorem immediately follows: if $P=\left\vert \Psi\right\vert
^{2}$ holds at some initial time then $P=\left\vert \Psi\right\vert ^{2}$ will
hold at future times. In this way we can easily establish the existence of a
quantum equilibrium state for a field on a classical globally-hyperbolic spacetime.

Pilot-wave theory with a stable Born rule covers a wide range of physics,
including high-energy field theory on a curved spacetime background --
provided the background is globally hyperbolic.\footnote{For completeness we
note that, in pilot-wave theory, fermions can be described in terms of a Dirac
sea with particle trajectories generated by a many-body Dirac wave function
\cite{BH93,CCS}. A less well-developed approach describes fermions in terms of
anti-commuting Grassmann fields \cite{AV92,AV96,AVBook}.} It is, however,
unclear if a similar construction can be given when the background spacetime
is not globally hyperbolic. It has been argued that such a spacetime is
generated by the formation and complete evaporation of a black hole
\cite{Hawk76}. Such arguments remain controversial, but if they are correct we
may be forced to rethink the idea of a stable Born rule in a gravitational
context. It has been suggested that quantum equilibrium could be unstable for
fields and particles propagating on a background non-globally-hyperbolic
spacetime \cite{AV07,AV04b,KV20}. Hawking radiation from an evaporating black
hole could then be in a state of quantum nonequilibrium, even if the original
collapsing matter was initially in quantum equilibrium. Because nonequilibrium
radiation can contain more information than is possible for conventional
radiation, this opens up a new approach to the (still controversial) puzzle of
information loss in black holes. The results reported in this paper support
the suggestion that evaporating black holes can create quantum nonequilibrium
(see Section 11).

\section{Quantum gravity and the Born rule}

In this section we review canonical quantum gravity, in both its conventional
and pilot-wave versions, focussing on the difficulties with understanding and
applying the Born rule.

\subsection{Canonical quantum gravity}

We begin by outlining the standard canonical quantisation of general
relativity \cite{DeW67,Kief12}, whose starting point is the classical
Hamiltonian formulation of Einstein's field equations \cite{MTW73,Wald84}. We
first review the formalism for pure gravitation, with the classical field
equations $^{(4)}G_{\mu\nu}=0$ in free space, before generalising to
gravitation with a matter field $\phi$.

Classically, second time-derivatives of the metric appear only in the
space-space components $^{(4)}G_{ij}=0$, which are the dynamical part of the
field equations. The other components, $^{(4)}G_{0\mu}=0$, contain only
first-order time derivatives and are constraints on the initial data (defined
on an initial spacelike slice $\Sigma(0)$). To rewrite this system in
Hamiltonian form, the 4-metric is first split into the 3+1 form (\ref{ADM}).
After dropping surface terms the Einstein-Hilbert action%
\begin{equation}
I=-\int d^{4}x\ (-^{(4)}g)^{1/2}\ ^{(4)}R
\end{equation}
becomes%
\begin{equation}
I=\int dt\int d^{3}x\ Ng^{1/2}(K_{ij}K^{ij}-K^{2}+R)\ ,
\end{equation}
where%
\begin{equation}
K_{ij}=\frac{1}{2N}\left(  -\frac{\partial g_{ij}}{\partial t}+D_{i}%
N_{j}+D_{j}N_{i}\right)  \label{Kij}%
\end{equation}
(with $D_{i}$ the 3-covariant derivative with respect to $x^{i}$) is the
extrinsic curvature tensor, $K=K_{i}^{i}$ and $R$ is the 3-scalar curvature
(the intrinsic curvature of $\Sigma$).

From the Lagrangian density%
\begin{equation}
\mathcal{L}=Ng^{1/2}(K_{ij}K^{ij}-K^{2}+R) \label{Lag}%
\end{equation}
we obtain the canonical momentum density%
\begin{equation}
p^{ij}=\frac{\partial\mathcal{L}}{\partial\dot{g}_{ij}}=-g^{1/2}(K^{ij}%
-g^{ij}K)\ . \label{pij}%
\end{equation}
This relation can be inverted:%
\begin{equation}
K^{ij}=-g^{-1/2}(p^{ij}-{\frac{1}{2}}g^{ij}p)\ . \label{Kij2}%
\end{equation}
From (\ref{Kij}) and (\ref{Kij2}) we then have%
\begin{equation}
\frac{\partial g_{ij}}{\partial t}=2NG_{ijkl}p^{kl}+D_{i}N_{j}+D_{j}N_{i}\ ,
\label{gij_dot}%
\end{equation}
where%
\begin{equation}
G_{ijkl}={\frac{1}{2}}g^{-1/2}(g_{ik}g_{jl}+g_{il}g_{jk}-g_{ij}g_{kl})\ .
\end{equation}

The functions $N$, $N^{i}$ are not dynamical variables and their canonical
momenta vanish. The gravitational Hamiltonian is then%
\begin{equation}
H=\int d^{3}x\ (p^{ij}\dot{g}_{ij}-\mathcal{L})=\int d^{3}x\ (N\mathcal{H}%
+N_{i}\mathcal{H}^{i}) \label{Ham}%
\end{equation}
where%
\begin{equation}
\mathcal{H}=G_{ijkl}p^{ij}p^{kl}-g^{1/2}R \label{H}%
\end{equation}
and%
\begin{equation}
\mathcal{H}^{i}=-2D_{j}p^{ij}\ . \label{Hi}%
\end{equation}
Treating $N$, $N^{i}$ as Lagrange multipliers, the conditions $\delta H/\delta
N=0$, $\delta H/\delta N_{i}=0$ imply the respective Hamiltonian and momentum
constraints%
\begin{equation}
\mathcal{H}=0\ ,\ \ \ \ \ \mathcal{H}^{i}=0\ . \label{cons's}%
\end{equation}
These are respectively equivalent to the initial-value constraints
$^{(4)}G_{00}=0$ and $^{(4)}G_{0i}=0$.

We can now write down Hamilton's first-order dynamical equations%
\begin{equation}
\dot{g}_{ij}=\frac{\delta H}{\delta p^{ij}}\ ,\ \ \ \ \ \dot{p}^{ij}%
=-\frac{\delta H}{\delta g_{ij}}\ . \label{Ham's eqns}%
\end{equation}
The first is equivalent to (\ref{gij_dot}). This can be used to eliminate
$p^{ij}$ from the second, yielding the second-order result $^{(4)}G_{ij}=0$.

To quantise this Hamiltonian system, the canonical variables $g_{ij}$,
$p^{ij}$ are promoted to operators $\hat{g}_{ij}$, $\hat{p}^{ij}$ satisfying
the usual commutation relations on the hypersurface $\Sigma$. We employ the
functional Schr\"{o}dinger picture with the configuration-space operator
realisations%
\begin{equation}
\hat{g}_{ij}(x)\rightarrow g_{ij}(x),\ \ \ \ \ \hat{p}^{ij}(x)\rightarrow
-i\frac{\delta}{\delta g_{ij}(x)} \label{realns}%
\end{equation}
(where $x$ labels a spatial point on $\Sigma$). We might have expected the
formal wave functional $\Psi\lbrack g_{ij},t]=\langle g_{ij}|\Psi(t)\rangle$
to satisfy a Schr\"{o}dinger equation $i\partial\Psi/\partial t=\hat{H}\Psi$,
where $t$ is the time function labelling the hypersurfaces $\Sigma$. But,
following the method of Dirac, the classical constraints (\ref{cons's}) are
promoted to operator constraints on $\Psi$:%
\begin{equation}
\mathcal{\hat{H}}\Psi=0\ ,\ \ \ \ \ \mathcal{\hat{H}}^{i}\Psi=0\ .
\label{cons's2}%
\end{equation}
A formal Schr\"{o}dinger equation then reads $i\partial\Psi/\partial t=\hat
{H}\Psi=0$. The functional $\Psi=\Psi\lbrack g_{ij}]$ depends on the 3-metric
only and not on $t$ -- a first hint that the quantum-gravitational state is a
different kind of thing from a conventional quantum state. Note further that,
according to (\ref{cons's2}), $\Psi$ is restricted not merely to eigenstates
of zero energy but to eigenstates of zero energy density.

In configuration space the constraints (\ref{cons's2}) take the form%
\begin{equation}
\left(  -G_{ijkl}\frac{\delta^{2}}{\delta g_{ij}\delta g_{kl}}-g^{1/2}%
R\right)  \Psi=0\ , \label{WD}%
\end{equation}%
\begin{equation}
D_{j}\left(  \frac{\delta\Psi}{\delta g_{ij}}\right)  =0\ . \label{momcon}%
\end{equation}
The first is the Wheeler-DeWitt equation (or Hamiltonian constraint). We have
written it with a specific operator ordering in the kinetic term, but in fact
the ordering is ambiguous and this should be understood. Different orderings
will be considered later. The second is the momentum constraint, which ensures
that $\Psi$ is really a function on the space of coordinate-independent
3-geometries (that is, on superspace).

It is straightforward to write down the generalisation to quantum gravity in
the presence of a scalar matter field $\phi$ with potential $\mathcal{V}%
(\phi)$ \cite{Hor94,PN05}. The wave functional $\Psi\lbrack g_{ij},\phi]$
obeys an extended Wheeler-DeWitt equation%
\begin{equation}
(\mathcal{\hat{H}}_{g}+\mathcal{\hat{H}}_{\phi})\Psi=0\ , \label{W-D_ext}%
\end{equation}
where%
\begin{equation}
\mathcal{\hat{H}}_{g}=-G_{ijkl}\frac{\delta^{2}}{\delta g_{ij}\delta g_{kl}%
}-\sqrt{g}R
\end{equation}
and%

\begin{equation}
\mathcal{\hat{H}}_{\phi}=\frac{1}{2}\sqrt{g}\left(  -\frac{1}{g}\frac
{\delta^{2}}{\delta\phi^{2}}+g^{ij}\partial_{i}\phi\partial_{j}\phi\right)
+\sqrt{g}\mathcal{V}(\phi)
\end{equation}
are respectively the gravitational and matter-field Hamiltonian constraint
operators. The momentum constraint now reads%
\begin{equation}
-2D_{j}\frac{\delta\Psi}{\delta g_{ij}}+\partial^{i}\phi\frac{\delta\Psi
}{\delta\phi}=0
\end{equation}
(where classically $\mathcal{H}^{i}=-2D_{j}p^{ij}+\pi_{\phi}\partial^{i}\phi$).

In our general considerations below we often work for simplicity with the
purely gravitational wave functional $\Psi\lbrack g_{ij}]$, where it is
understood that the extension to a system including a matter field is straightforward.

\subsection{The problem of time and of probability}

The time independence of the Wheeler-DeWitt wave functional $\Psi\lbrack
g_{ij}]$ sets quantum gravity apart from other quantum theories. For a
non-gravitational system with configuration $q$ we usually have a
time-dependent Schr\"{o}dinger equation (\ref{Sch1}) for a wave function(al)
$\psi=\psi(q,t)$. For a general quantum observable $\hat{\omega}=f(\hat
{q},\hat{p})$ (where symbolically $\hat{p}=-i\partial_{q}$ is the canonical
momentum) we have a time-dependent expectation value%
\begin{equation}
\left\langle \hat{\omega}\right\rangle =\int dq\ \psi^{\ast}%
(q,t)f(q,-i\partial_{q})\psi(q,t)\ .
\end{equation}
In quantum gravity, in contrast, we have a seemingly static theory. Formally,
an arbitrary observable $\hat{\omega}=f[\hat{g}_{ij},\hat{p}^{ij}]$ appears to
have a time-independent expectation value%
\begin{equation}
\left\langle \hat{\omega}\right\rangle =\int Dg\ \Psi^{\ast}[g_{ij}%
]f[g_{ij},-i\delta/\delta g_{ij}]\Psi\lbrack g_{ij}]
\end{equation}
(where $\int Dg$ is an appropriate functional integral over the 3-metric).

Despite the apparent lack of time evolution, it is usually assumed that
conventional quantum mechanics with the Born rule still applies in some
(perhaps modified) form. However, even taking into account the impressive
technical advances of recent decades, it is fair to say that the physical
interpretation of canonical quantum gravity remains controversial. The
difficulties are usually discussed under the general heading of the `problem
of time'. This is essentially the problem of recovering an approximate
time-dependent Schr\"{o}dinger-like evolution and time-dependent probabilities
-- for example in the limit of quantum fields, perhaps including metric
perturbations, on a classical spacetime background -- from an underlying
theory with no time. There have been many attempts to solve this problem since
the pioneering work of DeWitt and Wheeler in the late 1960s
\cite{DeW67,Wheel68}. Numerous approaches have been tried, with varying
degrees of success. Particularly insightful reviews of the problem, and of
various potential solutions, were given by Unruh and Wald \cite{UW89}, Isham
\cite{Ish91,Ish93} and Kucha\v{r} \cite{Kuch92,Kuch99}. An exhaustive and
up-to-date review has recently been given by Anderson \cite{Anders17}.

Some authors seek to avoid the problem from the outset by identifying a
preferred time parameter prior to quantisation. This approach has often been
called the `internal Schr\"{o}dinger interpretation'. The aim is to obtain,
after quantisation, a time-dependent Schr\"{o}dinger-like equation with an
appropriately-defined Hamiltonian. This approach has a long history with its
own problems. Two pilot-wave theory models along these lines have been
proposed. The first adopts a slicing with a uniform lapse function $N=1$
\cite{AV92}, while the second assumes that the preferred slices are foliations
of constant York time \cite{AV96,RV14}. The resulting Schr\"{o}dinger-like
equations should ensure that the Born rule has the usual status as an
equilibrium state and there is no instability. In fact these models resemble
some formulations of Ho\v{r}ava-Lifshitz gravity \cite{Hor09}, for which
Lorentz invariance is broken at very high energies and there is a preferred
foliation of spacetime. But the consistency and completeness of such models
remains a matter for future research \cite{AVBook}.

Many authors in the field take the view that time is fundamentally undefined
in the deep quantum-gravity regime and that an effective time evolution
emerges only in certain conditions and in certain approximations [1, 13--15,
84, 85]. In these approaches it is common to suppose that an effective
physical `time' is hidden in the 3-metric $g_{ij}$ \cite{DeW67,Wheel68}. On
this view some function $\mathcal{T}$ of $g_{ij}$ must be extracted to play
the role of time (where now the extraction of a time function is attempted
\textit{after} quantisation). The functional $\Psi\lbrack g_{ij}]$ then takes
the schematic form $\Psi\lbrack\sigma,\mathcal{T}]$, where $\sigma$ represents
the remaining metric variables. In appropriate conditions we might recover an
effective and well-defined time evolution.

Such approaches, with no fundamental notion of time, often run into what some
authors regard as natural limitations and other authors regard as conceptual
difficulties. For example, in quantum cosmology a common choice for
$\mathcal{T}$ is the cosmological scale factor $a$. In a minisuperspace model
we might have a wave function of the form $\psi(\phi,\sigma,a)$, where $\phi$
is a homogeneous matter field, $\sigma$ are reduced metric degrees of freedom
(perhaps representing perturbations around a homogeneous and isotropic
3-metric), and $a$ is regarded as an effective time. In a closed universe that
expands and recontracts, the `time' $a$ can have pathological properties:
distinct states can be associated with the same value of `time' and certain
`times' might never be reached at all. It is not obvious that a parameter
measuring spatial volume can be consistently reinterpreted as a parameter that
measures time.\footnote{It has been argued that such problems will be generic
for any realistic degree of freedom $\mathcal{T}$ hidden in $g_{ij}$
\cite{UW89}.} We might be able to derive a conventional time evolution in some
local region of configuration space (corresponding for example to an expanding
cosmological phase), but globally we are likely to find pathologies. Some
authors are unconcerned by the pathologies, concluding that our conventional
ideas about time have limited validity and emerge only in certain restricted
circumstances. Other authors are troubled by the pathologies and argue that
the formalism suffers from a deep conceptual problem.

Issues also arise concerning the application of basic rules of quantum
mechanics in a fundamentally timeless theory \cite{Rov04,Rov91,Hall,Mondra}.
Again, authors differ on the significance of these questions.

It is not our purpose here to review or critique the numerous approaches to
the problem of time developed over more than half a century \cite{Anders17}.
Proposals that are still being actively pursued include, for example, evolving
constants of motion and conditional probability interpretations (to name just
two among many). It is noteworthy that this area remains active and
controversial with as yet no definitive resolution.

There are, however, three well-known approaches that are particularly relevant
for our purposes. The first, known as the `Klein-Gordon interpretation',
exploits the mathematical parallel between the Wheeler-DeWitt equation (on
superspace) and the Klein-Gordon equation for a single particle (on a curved
space with an arbitrary potential), with the hope of obtaining well-defined
time-dependent probabilities at least in some regime. This approach was
studied in particular by Kucha\v{r} and was shown to have seemingly
unresolvable difficulties. The second approach, which came to be known as the
`naive Schr\"{o}dinger interpretation', was championed in particular by
Hawking and collaborators in the 1980s but was also shown to have serious
difficulties and has been widely abandoned. The third or `WKB' approach dates
from the 1960s and is still widely used (in particular in quantum cosmology).
It will be helpful to review these three approaches before we proceed.

\subsubsection{Klein-Gordon interpretation}

As is well known the DeWitt metric $G_{ijkl}$ defines a manifold with
hyperbolic signature $-+++++$ (for a 6-dimensional space of `points' $g_{ij}$
at each spatial point $x$) \cite{DeW67}. For this reason the Wheeler-DeWitt
equation is formally analogous to an infinite-dimensional Klein-Gordon
equation (with a `mass-squared' term $g^{1/2}R$). By extracting an appropriate
time functional $\mathcal{T}[g_{ij},x)$ (a functional of $g_{ij}$ at each
$x$), the Wheeler-DeWitt equation (\ref{WD}) for $\Psi\lbrack g_{ij}]$ can be
written as a Klein-Gordon-like equation \cite{Ish93}%
\begin{equation}
\left(  -\frac{\delta^{2}}{\delta\mathcal{T}^{2}}+\mathcal{F}^{ab}\frac
{\delta^{2}}{\delta\sigma^{a}\delta\sigma^{b}}-g^{1/2}R\right)  \Psi=0
\label{WD_KG}%
\end{equation}
for $\Psi\lbrack\sigma,\mathcal{T}]$, where $\sigma^{a}[g,x)$ ($a=1,..,5$) are
the remaining metric variables and $\mathcal{F}^{ab}=\mathcal{F}%
^{ab}[\mathcal{T},\sigma;x)$.

The Klein-Gordon interpretation attempts to understand the physics of the
Wheeler-DeWitt equation by exploiting the formal analogy with the
single-particle Klein-Gordon equation%
\begin{equation}
\left(  -\frac{\partial^{2}}{\partial t^{2}}+\delta^{ij}\frac{\partial^{2}%
}{\partial x^{i}\partial x^{j}}-m^{2}\right)  \psi=0 \label{K-G}%
\end{equation}
(written for simplicity on Minkowski spacetime). This implies a continuity
equation $\partial_{t}\rho_{\mathrm{KG}}+\partial_{i}j_{\mathrm{KG}}^{i}=0$
with a density%
\begin{equation}
\rho_{\mathrm{KG}}=i(\psi^{\ast}\dot{\psi}-\psi\dot{\psi}^{\ast})=-2\left\vert
\psi\right\vert ^{2}\dot{S}\
\end{equation}
and a spatial current%
\begin{equation}
j_{\mathrm{KG}}^{i}=-i(\psi^{\ast}\partial_{i}\psi-\psi\partial_{i}\psi^{\ast
})=2\left\vert \psi\right\vert ^{2}\partial_{i}S
\end{equation}
(with $\psi=\left\vert \psi\right\vert e^{iS}$). The global quantity $\int
d^{3}x\ \rho_{\mathrm{KG}}$ is preserved in time. However $\rho_{\mathrm{KG}}$
is not positive definite and so cannot be interpreted as a probability
density. If we multiply $\rho_{\mathrm{KG}}$ and $j_{\mathrm{KG}}^{i}$ by
$1/2m$, the current appears to take a standard form $\left\vert \psi
\right\vert ^{2}(\partial_{i}S/m)$ and we might attempt to take $\left\vert
\psi\right\vert ^{2}$ as the true probability density. But in that case we
find that $\int d^{3}x\ \left\vert \psi\right\vert ^{2}$ is not preserved in
time. For these reasons it is unclear how to associate probabilities with the
single-particle Klein-Gordon equation.

Similar problems arise for the Wheeler-DeWitt equation written in the
Klein-Gordon form (\ref{WD_KG}). The infinite-dimensional analogue of the
Klein-Gordon density is again not positive definite \cite{DeW67}. In
developing the Klein-Gordon interpretation it was hoped that the density would
turn out to be positive on some appropriately-defined subspace of solutions
$\Psi$, but unfortunately this program was beset with difficulties and has
been widely abandoned \cite{Kuch92,Ish93}. Even so, as we shall see, important
mathematical and physical insights can be gained by considering the
Wheeler-DeWitt equation in the Klein-Gordon form (\ref{WD_KG}).

\subsubsection{Naive Schr\"{o}dinger interpretation}

According to the naive Schr\"{o}dinger interpretation we can treat $\left\vert
\Psi\lbrack g_{ij}]\right\vert ^{2}$ directly as the probability density for
the 3-metric $g_{ij}$.\footnote{Unruh and Wald \cite{UW89} were the first to
call this `the "naive interpretation" of canonical quantum gravity'.} More
precisely, $\left\vert \Psi\lbrack g_{ij}]\right\vert ^{2}$ is taken to be the
probability density (on superspace) for finding a spacelike hypersurface with
3-geometry represented by $g_{ij}$. In a minisuperspace model with a wave
function $\psi(\phi,\sigma,a)$, the quantity $\left\vert \psi(\phi
,\sigma,a)\right\vert ^{2}$ is taken to be the probability density for finding
a matter field $\phi$, metric perturbations $\sigma$, and a scale factor $a$.
This is analogous to taking $\left\vert \psi(x,t)\right\vert ^{2}$ for a
single particle to be the probability density for finding a time $t$ with the
particle at $x$. This kind of reasoning was applied by Hawking and his school
to argue that certain universes are more likely than others (concluding, for
example, that the universe is likely to be flat and large) \cite{Hawketal}.

One difficulty with this approach is that it seems unable to answer what we
might call dynamical questions, specifically, how to calculate the probability
of finding a measured value given the outcome of a preceding measurement.
Attempts have been made to refine the interpretation using conditional
probabilities, but this `conditional probability interpretation' raises
difficult questions about which variables can be appropriately chosen as
conditional statements for the remaining variables \cite{Kuch92,Ish93}.

The naive Schr\"{o}dinger interpretation, and attempts to refine it, in any
case founder on one insuperable problem: solutions $\Psi\lbrack g_{ij}]$ to
the Wheeler-DeWitt equation (\ref{WD}) are generally not normalisable (that
is, not square-integrable). To see why, consider the Wheeler-DeWitt equation
written in the Klein-Gordon form (\ref{K-G}). A solution $\Psi\lbrack g_{ij}]$
of (\ref{WD}) corresponds to a solution $\Psi\lbrack\sigma,\mathcal{T}]$ of
(\ref{K-G}), with a change of variables from $g_{ij}$ to $\sigma,\mathcal{T}$.
Therefore we can write%
\begin{equation}
\int Dg\ \left\vert \Psi\lbrack g_{ij}]\right\vert ^{2}\sim\int D\sigma\int
D\mathcal{T}\ \left\vert \Psi\lbrack\sigma,\mathcal{T}]\right\vert ^{2}%
=\infty\ . \label{sq-norm}%
\end{equation}
The right-hand side of (\ref{sq-norm}) diverges because it is just a
higher-dimensional analogue of the integral%
\begin{equation}
\int d^{3}x\int_{-\infty}^{+\infty}dt\ \left\vert \psi(x,t)\right\vert
^{2}=\infty\label{K-G-int}%
\end{equation}
for a solution $\psi(x,t)$ of the single-particle Klein-Gordon equation
(\ref{K-G}), in which we integrate not only over $x$ but also over $t$. Thus
the Klein-Gordon-like character of the Wheeler-DeWitt equation ensures that
solutions $\Psi\lbrack g_{ij}]$ are not square-integrable.

It might be thought that the diverging integral (\ref{sq-norm}) could be
rendered finite by some appropriate regularisation (for example replacing
continuous 3-space with a discrete lattice). But for as long as we integrate
over the whole `time axis' the integral (\ref{sq-norm}) will remain divergent
just like its lower-dimensional counterpart (\ref{K-G-int}). The divergence
reflects a basic fact about wave propagation: a solution of the wave equation
can be confined (or decay rapidly) with respect to $x$ but not with respect to
$t$.

Because $\left\vert \Psi\lbrack g_{ij}]\right\vert ^{2}$ is not
square-integrable it cannot be employed as a probability density. Similar
problems afflict the conditional probability interpretation. For this reason
the naive Schr\"{o}dinger interpretation fails. Some authors have tried to
view the divergence of (\ref{sq-norm}) as a technical issue to be resolved by
more sophisticated mathematics. In this paper we argue instead that the
divergence of (\ref{sq-norm}) points to an important physical fact: that there
is no fundamental Born rule in quantum gravity.

To be clear, we should remark that there are technical issues with
normalisation and the definition of integration measures that afflict any
continuous field theory. Even in classical physics it would be mathematically
delicate to define a probability density on the space of continuous
electromagnetic fields. For a classical scalar field, for example, we might
consider a formal probability density $P[\phi(x)]$ on the space of field
configurations $\phi(x)$ and try to define the functional integral $\int
D\phi$ (with $\int D\phi\ P[\phi]=1$) as a limiting integral $\int\int...\int
d\phi_{1}d\phi_{2}...d\phi_{n}...$ over field values $\phi_{i}$ on a discrete
lattice of spatial points $x_{i}$. But, as is well known, the Lebesgue measure
$d\phi_{1}d\phi_{2}...d\phi_{n}...$ is not well-defined in the continuum
limit. But in practice such problems are routinely evaded, for example, by
introducing periodic boundary conditions and moving to Fourier space, and
perhaps adding a high-frequency cutoff, so as to make the system effectively
discrete and finite. Similar techniques are routinely applied in quantum field
theory. Despite such technical issues with measures we can, for example,
calculate the probability distribution for vacuum field fluctuations in
inflationary cosmology and compare successfully with observation. But the
normalisability problem for the Wheeler-DeWitt equation is much deeper: the
very structure of the equation ensures what is in effect a wavelike
propagation in configuration space, so that attempting to normalise a
Wheeler-DeWitt functional $\Psi\lbrack g_{ij}]$ in superspace is as misguided
as attempting to normalise a Klein-Gordon function $\psi(x,t)$ in spacetime.
We might say that $\Psi\lbrack g_{ij}]$ is not merely `technically'
non-normalisable but `intrinsically' so.

To conclude, the mathematical structure of the Wheeler-DeWitt equation ensures
a wave-like propagation in configuration space, just as if some of the degrees
of freedom played the role of time in analogy with the Klein-Gordon equation
on spacetime. For this reason solutions of the Wheeler-DeWitt equation are
intrinsically non-normalisable and the naive Schr\"{o}dinger interpretation
fails. While there is a general consensus in the literature that the naive
Schr\"{o}dinger interpretation is unworkable, in this paper we suggest that
the physical reason for its failure has not been properly identified. It is
often thought, for example, that it fails because some of the metric degrees
of freedom really play the role of time. As we shall see, from the point of
view of pilot-wave theory, the naive Schr\"{o}dinger interpretation fails
because there is no such thing as quantum equilibrium or the Born rule in the
deep quantum-gravity regime. This observation will form the starting point for
our new approach to the Born rule in quantum gravity (Section 4).

\subsubsection{WKB approach and Schr\"{o}dinger approximation}

One of the most successful and still widely-used approaches to the problem of
time employs the semiclassical WKB method, in which trajectories for an
evolving 3-geometry are associated with a WKB wave functional $\Psi$.

If we write $\Psi=|\Psi|e^{iS}$ and assume that $|\Psi|$ varies slowly
compared with $S$, the Wheeler-DeWitt equation (\ref{WD}) implies an
approximate time-independent Hamilton-Jacobi equation%
\begin{equation}
G_{ijkl}\frac{\delta S}{\delta g_{ij}}\frac{\delta S}{\delta g_{kl}}%
-g^{1/2}R=0 \label{HJe}%
\end{equation}
for $S$. In the WKB approach it is assumed that a solution $S$ generates
classical trajectories with a canonical momentum%
\begin{equation}
p^{ij}=\frac{\delta S}{\delta g_{ij}}\ . \label{deB}%
\end{equation}
From (\ref{gij_dot}) we then have a first-order equation of motion%
\begin{equation}
\frac{\partial g_{ij}}{\partial t}=2NG_{ijkl}\frac{\delta S}{\delta g_{kl}%
}+D_{i}N_{j}+D_{j}N_{i} \label{gdot2}%
\end{equation}
for the trajectories. If we differentiate (\ref{gdot2}) with respect to time,
then from (\ref{HJe}) (together with the momentum constraint (\ref{momcon}),
which implies two similar constraints on $|\Psi|$ and $S$) it is possible to
recover the classical Einstein equations \cite{DeW67}.

Thus, given a WKB wave functional $\Psi\lbrack g_{ij}]$, we can define
approximately classical trajectories for the 3-metric and recover an
approximately classical spacetime background. If we include a scalar matter
field $\phi$, we can also recover an approximate time-dependent
Schr\"{o}dinger equation%
\begin{equation}
i\frac{\partial\psi\lbrack\phi,t]}{\partial t}=\hat{H}_{\mathrm{eff}}%
\psi\lbrack\phi,t] \label{Sch_approxn_1}%
\end{equation}
for the propagation of $\phi$ on the classical spacetime background (where
$\hat{H}_{\mathrm{eff}}$ is an effective Hamiltonian). Details are given in
Section 3.4. To summarise the method, we begin with the extended
Wheeler-DeWitt equation (\ref{W-D_ext}) for $\Psi\lbrack g_{ij},\phi]$. We
take an approximate solution of the form%
\begin{equation}
\Psi\lbrack g_{ij},\phi]\approx\Psi_{\mathrm{WKB}}[g_{ij}]\psi\lbrack
\phi,g_{ij}]\ , \label{WKB}%
\end{equation}
where $\Psi_{\mathrm{WKB}}$ is a WKB solution for the 3-metric alone whose
phase $S_{\mathrm{WKB}}=\operatorname{Im}\ln\Psi_{\mathrm{WKB}}$ satisfies the
classical Hamilton-Jacobi equation (\ref{HJe}). We can then use
$S_{\mathrm{WKB}}$ to generate trajectories for the classical background via
(\ref{gdot2}). The wave function $\psi\lbrack\phi,g_{ij}]$ of the matter field
is evaluated along a trajectory $g_{ij}=g_{ij}(t)$ for the background and so
we can define an effective time-dependent function%
\begin{equation}
\psi_{\mathrm{eff}}[\phi,t]=\psi\lbrack\phi,g_{ij}(t)]\ .
\end{equation}
In a small time $\delta t$ the quantity $\psi_{\mathrm{eff}}$ will change by%
\[
\delta\psi_{\mathrm{eff}}=\int d^{3}x\ \frac{\delta\psi_{\mathrm{eff}}}{\delta
g_{ij}}\delta g_{ij}=\int d^{3}x\ \frac{\delta\psi_{\mathrm{eff}}}{\delta
g_{ij}}\dot{g}_{ij}\delta t\ .
\]
We then have a time derivative%
\begin{equation}
\frac{\partial}{\partial t}=\int d^{3}x\ \dot{g}_{ij}\frac{\delta}{\delta
g_{ij}}\ , \label{WKB_t}%
\end{equation}
where $\dot{g}_{ij}$ is given by (\ref{gdot2}).

Similarly, in a minisuperspace model with wave function $\psi(\phi,\sigma,a)$,
we have a scale factor $a=a(t)$ for the classical background and an effective
time-dependent wave function%
\[
\psi_{\mathrm{eff}}(\phi,\sigma,t)=\psi(\phi,\sigma,a(t))
\]
with time derivative%
\[
\frac{\partial}{\partial t}\equiv\dot{a}\frac{\partial}{\partial a}\ ,
\]
where $\dot{a}$ is found from the minisuperspace analogue of the relation
(\ref{gdot2}).

Note the crucial role played by the WKB trajectories for the classical
background. They allow us to define an effective time parameter $t$ (often
called `WKB time' \cite{Zeh88}) and an effective time-dependent wave
functional for a matter field on the background. The WKB trajectories are in
fact de Broglie-Bohm trajectories evaluated in the WKB approximation, and so
the above construction is entirely natural in pilot-wave theory.

\subsection{Pilot-wave theory and quantum gravity}

We now consider canonical quantum gravity with the Wheeler-DeWitt equation
supplemented by a de Broglie-Bohm trajectory for the 3-geometry [55--58].

\subsubsection{Pilot-wave geometrodynamics}

Collecting together the basic general equations, the pilot wave $\Psi\lbrack
g_{ij}]$ obeys the Wheeler-DeWitt equation $\mathcal{\hat{H}}\Psi=0$ or%
\begin{equation}
-G_{ijkl}\frac{\delta^{2}\Psi}{\delta g_{ij}\delta g_{kl}}-g^{1/2}R\Psi=0
\label{WD2}%
\end{equation}
(subject to the constraint (\ref{momcon}) or $D_{j}\left(  \delta\Psi/\delta
g_{ij}\right)  =0$), while the trajectories $g_{ij}(t)$ for the evolving
3-geometry are determined by the de Broglie guidance equation%
\begin{equation}
\frac{\partial g_{ij}}{\partial t}=2NG_{ijkl}\frac{\delta S}{\delta g_{kl}%
}+D_{i}N_{j}+D_{j}N_{i} \label{deB3}%
\end{equation}
(where $S=\operatorname{Im}\ln\Psi$). As in the WKB approach (\ref{deB3}) can
be motivated from the classical canonical relations (\ref{deB}) and
(\ref{gij_dot}). However note that here (\ref{deB3}) is assumed to be valid
for any Wheeler-DeWitt wave functional $\Psi$ -- even outside the WKB limit.
Note also the (implicitly understood) ambiguity in the ordering of the kinetic
term in (\ref{WD2}).

Equations (\ref{WD2}) and (\ref{deB3}) (with (\ref{momcon})) are the
fundamental laws of pilot-wave geometrodynamics for pure gravitation. Together
they define a dynamics for an individual system, with no mention of ensembles
or probabilities. As in non-gravitational pilot-wave theory, $\Psi$ is a
physical object in configuration space that guides the motion of an individual
system -- it has no intrinsic connection with probability.

If we substitute $\Psi=\left\vert \Psi\right\vert e^{iS}$ into (\ref{WD2}) the
real part yields a modified Hamilton-Jacobi equation%
\begin{equation}
G_{ijkl}\frac{\delta S}{\delta g_{ij}}\frac{\delta S}{\delta g_{kl}}%
-g^{1/2}R+q=0\ , \label{H-J-q}%
\end{equation}
where%
\begin{equation}
q=-\frac{1}{\left\vert \Psi\right\vert }G_{ijkl}\frac{\delta^{2}\left\vert
\Psi\right\vert }{\delta g_{ij}\delta g_{kl}} \label{qpd}%
\end{equation}
is the quantum potential density, while the imaginary part yields the equation%
\begin{equation}
G_{ijkl}\frac{\delta}{\delta g_{ij}}\left(  |\Psi|^{2}\frac{\delta S}{\delta
g_{kl}}\right)  =0\ . \label{W-D_cont}%
\end{equation}

The results (\ref{qpd}) and (\ref{W-D_cont}) follow if we adopt the explicit
ordering $G_{ijkl}(\delta/\delta g_{ij})(\delta/\delta g_{kl})$ in (\ref{WD2})
but are modified for other orderings. For example with the alternative
ordering $(\delta/\delta g_{ij})G_{ijkl}(\delta/\delta g_{kl})$ the result
(\ref{W-D_cont}) instead reads%
\begin{equation}
\frac{\delta}{\delta g_{ij}}\left(  |\Psi|^{2}G_{ijkl}\frac{\delta S}{\delta
g_{kl}}\right)  =0\ . \label{W-D_cont2}%
\end{equation}
DeWitt \cite{DeW67} advocated a rule whereby $\delta/\delta g_{ij}$ is
understood to give zero when acting on $g_{kl}$ at the same spatial point, in
which case the two forms (\ref{W-D_cont}) and (\ref{W-D_cont2}) are the same.
Horiguchi \cite{Hor94} employs the general form $g^{-p}(\delta/\delta
g_{ij})g^{p}G_{ijkl}(\delta/\delta g_{kl})$ for a fixed parameter $p$ (where
$p=-1/2$ corresponds to a Laplace-Beltrami operator ordering).

Instead of motivating the guidance equation (\ref{deB3}) from the classical
relations (\ref{deB}) and (\ref{gij_dot}), alternatively it can be motivated
as the natural velocity field appearing in (\ref{W-D_cont2}), which can be
rewritten as%
\begin{equation}
\frac{\delta}{\delta g_{ij}}\left(  |\Psi|^{2}\frac{\partial g_{ij}}{\partial
t}\right)  =0\ \label{W-D_cont3}%
\end{equation}
(for simplicity setting $N_{i}=0$). However we must be careful not to
over-interpret (\ref{W-D_cont3}). As we shall discuss in Section 3.3.2, some
workers integrate (\ref{W-D_cont3}) over $x$ and (mistakenly) regard the
result as a physical continuity equation for a probability density $|\Psi
|^{2}$ \cite{DS20}. But the static functional $|\Psi|^{2}$ has pathological
properties: as we saw in Section 3.2.2 it is not integrable and cannot be a
physical probability density.

An important issue concerns the status of the spacetime foliation in the above
dynamics. It seems fair to say that at present this aspect of the theory is
still not fully understood. The arbitrary choice of lapse and shift functions
$N$ and $N^{i}$, which appear in the guidance equation (\ref{deB3}), should
not affect the overall 4-geometry that is traced out by the time evolution of
the 3-metric (for given initial conditions on a spacelike slice). Otherwise
the initial-value problem would not be well-posed. Shtanov \cite{Sht96}
presented an example where the 4-geometry seemed to depend on $N$ and
suggested on this basis that the dynamics breaks foliation invariance. One
possible way to proceed would then be to impose a specific choice of $N$ as
part of the theory.\footnote{Alternatively, as we have noted, we might abandon
foliation invariance already at the classical level and obtain a
time-dependent Schr\"{o}dinger equation for $\Psi\lbrack g_{ij},t]$ with a
preferred time parameter $t$ \cite{AV92,AV96,RV14}.} But later work by
Pinto-Neto and Santini \cite{PNS02} suggests that the above pilot-wave
dynamics of geometry is well-posed for arbitrary $N$. The question is
addressed by writing the dynamics as a classical Hamiltonian system with
(\ref{Ham}) replaced by%
\begin{equation}
H_{q}=\int d^{3}x\ (N\mathcal{H}_{q}+N_{i}\mathcal{H}^{i})\ , \label{Ham_q}%
\end{equation}
where $\mathcal{H}_{q}=\mathcal{H}+q$ and $q$ is the quantum potential density
(\ref{qpd}). If we include $p^{ij}=\delta S/\delta g_{ij}$ as an initial
condition on the momenta, Hamilton's equations then generate the same
trajectories as pilot-wave dynamics with the guidance equation $p^{ij}=\delta
S/\delta g_{ij}$ applied at all times. The question is whether the time
evolution of an initial 3-geometry will yield the same 4-geometry for any
choice of $N$, $N^{i}$. Classically this question is answered by the following
theorem: Hamilton's equations, with a Hamiltonian $\bar{H}=\int d^{3}%
x\ (N\mathcal{\bar{H}}+N_{i}\mathcal{\bar{H}}^{i})$, generate a unique
locally-Lorentzian 4-geometry if and only if $\mathcal{\bar{H}}$ and
$\mathcal{\bar{H}}^{i}$ satisfy the Dirac-Teitelboim algebra \cite{DiracTeit}.
The classical quantities $\mathcal{H}$ and $\mathcal{H}^{i}$ (appearing in
(\ref{Ham})) satisfy this algebra and so the classical Hamiltonian dynamics
indeed generates a 4-geometry that is independent of $N$ and $N^{i}$. In
contrast, for a system with Hamiltonian (\ref{Ham_q}), the algebra satisfied
by $\mathcal{H}_{q}$ and $\mathcal{H}^{i}$ is found to be closed for all
$\Psi$ (when evaluated on de Broglie-Bohm trajectories) but modified when
$q\neq0$ \cite{PNS02}. Pinto-Neto and Santini conclude that the 4-geometry is
again independent of $N$ and $N^{i}$ but now forms a non-Lorentzian spacetime.
On this interpretation the nonlocality associated with $q\neq0$ breaks local
Lorentz invariance for individual trajectories. A locally-Lorentzian spacetime
is obtained only in the classical limit $q\rightarrow0$ where the nonlocality
vanishes. The physical implication appears to be that, for a given solution
$\Psi$ of the Wheeler-DeWitt equation and for a given initial 3-geometry,
there is in effect a preferred foliation of the resulting spacetime.

Intuitively this is consistent with the `Aristotelian' structure of pilot-wave
dynamics, where de Broglie's law of motion determines velocity rather than
acceleration \cite{AV97}. For standard systems of fields and particles the
pilot wave $\Psi$ determines the canonical momentum for a given configuration.
In the case of gravity the canonical momentum density $p^{ij}$, expressed by
(\ref{pij}), is essentially equal to the extrinsic curvature tensor $K^{ij}$.
Thus for a given 3-metric $\Psi$ determines $K^{ij}$, which describes how the
3-geometry is embedded in spacetime. In effect the slicing is determined by
the 3-geometry and wave functional -- just as the momentum of a particle is
determined by its position and wave function. A preferred foliation for a
quantum-gravitational system is also consistent with a preferred foliation for
the limiting case of a field on a classical spacetime background
\cite{Holl93,AV92,AV96,BHK87}. On this issue we note, finally, that for
quantum nonequilibrium ensembles of entangled systems, nonlocality can
manifest as statistical nonlocal signalling \cite{AV91b,AV02a,AV02b}, which
arguably requires a physical preferred foliation of spacetime \cite{AV08b}.

In this paper we focus on the question of probability in pilot-wave quantum
gravity. Equations (\ref{WD2}) and (\ref{deB3}) are assumed to define a
pilot-wave dynamics of 3-geometry for an individual system with wave
functional $\Psi$. We can then consider a theoretical ensemble of such
systems, with the same wave functional $\Psi$, and with an arbitrary initial
probability distribution $P[g_{ij},t_{i}]$ at some initial time $t_{i}$.
Because the 3-metric evolves in time, with a velocity $\partial g_{ij}%
/\partial t$ given by (\ref{deB3}), by construction the distribution
$P[g_{ij},t]$ will evolve according to the continuity equation%
\begin{equation}
\frac{\partial P}{\partial t}+\int d^{3}x\ \frac{\delta}{\delta g_{ij}}\left(
P\frac{\partial g_{ij}}{\partial t}\right)  =0\ .
\end{equation}

The above equations are readily generalised to include matter fields. We
simply take the extended Wheeler-DeWitt equation (\ref{W-D_ext}) for
$\Psi\lbrack g_{ij},\phi]$, the same guidance equation (\ref{deB3}) for
$g_{ij}(t)$, and add the guidance equation (\ref{deB2}) for the matter field
trajectory $\phi(x,t)$. For a theoretical ensemble with the same wave
functional $\Psi$, a general distribution $P[g_{ij},\phi,t]$ will then evolve
by the extended continuity equation
\begin{equation}
\frac{\partial P}{\partial t}+\int d^{3}x\ \frac{\delta}{\delta g_{ij}}\left(
P\frac{\partial g_{ij}}{\partial t}\right)  +\int d^{3}x\ \frac{\delta}%
{\delta\phi}\left(  P\frac{\partial\phi}{\partial t}\right)  =0\ .
\label{cont_P}%
\end{equation}

A crucial question remains: how do we construct the theory of a quantum
equilibrium ensemble? As we saw in Section 2, for a non-gravitational system
this is usually straightforward: the continuity equation (\ref{cont1}) for a
general distribution $\rho$ coincides with the continuity equation
(\ref{Contj}) for $\left\vert \psi\right\vert ^{2}$ (derived from the
time-dependent Schr\"{o}dinger equation), enabling us to deduce that
$\rho=\left\vert \psi\right\vert ^{2}$ is an equilibrium state. As we shall
see, for a gravitational system obeying the Wheeler-DeWitt equation, this
reasoning breaks down.

\subsubsection{Naive Schr\"{o}dinger interpretation in pilot-wave theory}

Some workers try to interpret $\left\vert \Psi\right\vert ^{2}$ as the
equilibrium probability density for pilot-wave theory with a Wheeler-DeWitt
wave functional $\Psi$. Even though $\left\vert \Psi\right\vert ^{2}$ is
static, individual trajectories depend on time and we might hope to recover
time-dependent probabilities for subsystems. This approach was originally
suggested by Vink \cite{Vink92} in the context of a minisuperspace model, and
it has recently been advocated in general terms by D\"{u}rr and Struyve
\cite{DS20}. However, this amounts to applying the naive Schr\"{o}dinger
interpretation to pilot-wave quantum gravity. While having time-dependent
trajectories adds a novel element, the essential difficulty of the naive
Schr\"{o}dinger interpretation remains.\footnote{For completeness we note that
Horiguchi \cite{Hor94} tries to implement a form of the Klein-Gordon
interpretation.}

To see the difficulty let us include a matter field $\phi$, so that
$\Psi\lbrack g_{ij},\phi]$ obeys the extended Wheeler-DeWitt equation
(\ref{W-D_ext}), and consider how D\"{u}rr and Struyve attempt to motivate
$\left\vert \Psi\lbrack g_{ij},\phi]\right\vert ^{2}$ as the equilibrium
density \cite{DS20}. Writing $\Psi=\left\vert \Psi\right\vert e^{iS}$ we find
from (\ref{W-D_ext}) (with an appropriate operator ordering) that $\left\vert
\Psi\lbrack g_{ij},\phi]\right\vert ^{2}$ satisfies%
\begin{equation}
\frac{\delta}{\delta g_{ij}}\left(  |\Psi|^{2}\frac{\partial g_{ij}}{\partial
t}\right)  +\frac{\delta}{\delta\phi}\left(  |\Psi|^{2}\frac{\partial\phi
}{\partial t}\right)  =0 \label{W-Dcont}%
\end{equation}
at each spatial point $x$. This is not yet a continuity equation. Comparing
with the standard continuity equation for a single particle, which for a
static density ($\partial\rho/\partial t=0$) takes the form $\mathbf{\nabla
}\cdot\mathbf{j}=0$, equation (\ref{W-Dcont}) is analogous to the set of
equations $\partial_{x}j_{x}=\partial_{y}j_{y}=\partial_{z}j_{z}=0$ (one for
each degree of freedom). A proper continuity equation instead takes the form
(\ref{cont_P}) satisfied by a general (time-dependent) distribution
$P[g_{ij},\phi,t]$. However, following D\"{u}rr and Struyve, such an equation
for $\left\vert \Psi\right\vert ^{2}$ can be obtained by integrating
(\ref{W-Dcont}) over $x$. Noting that $\left\vert \Psi\right\vert ^{2}$ has no
explicit time dependence, $\partial\left\vert \Psi\right\vert ^{2}/\partial
t=0$, we can then write what we shall refer to as a `pseudo-continuity
equation'%
\begin{equation}
\frac{\partial\left\vert \Psi\right\vert ^{2}}{\partial t}+\int d^{3}%
x\ \frac{\delta}{\delta g_{ij}}\left(  \left\vert \Psi\right\vert ^{2}%
\frac{\partial g_{ij}}{\partial t}\right)  +\int d^{3}x\ \frac{\delta}%
{\delta\phi}\left(  \left\vert \Psi\right\vert ^{2}\frac{\partial\phi
}{\partial t}\right)  =0\ . \label{false_cont}%
\end{equation}
This is now formally the same as the physical continuity equation
(\ref{cont_P}) satisfied by $P$. It might then be thought that we can deduce a
quantum equilibrium state $P=\left\vert \Psi\right\vert ^{2}$ in the usual
way: because $P$ and $\left\vert \Psi\right\vert ^{2}$ obey the same evolution
equation, if they are equal initially they will be equal later, and so
$P=\left\vert \Psi\right\vert ^{2}$ is an equilibrium state -- just as usual
for non-gravitational systems. Indeed D\"{u}rr and Struyve write down the
equation (\ref{false_cont}) (without the vanishing first term) and assert on
this basis that $\left\vert \Psi\right\vert ^{2}$ is a quantum equilibrium
measure for the system \cite{DS20}. It is claimed further that $\left\vert
\Psi\right\vert ^{2}$ can be employed as a measure of `typicality' for initial
configurations of the universe, from which the Born rule for subsystems can be
derived.\footnote{This mimics the general approach to the Born rule advocated
by the Bohmian mechanics school of de Broglie-Bohm theory (noted at the end of
Section 2.1).} But this is another version of the naive Schr\"{o}dinger
interpretation, albeit in the context of pilot-wave theory. It fails for the
same reason as before: $\left\vert \Psi\right\vert ^{2}$ is not normalisable
and so cannot be used to define an equilibrium measure. As we saw in Section
3.2.1, and as has been known since the 1960s, the Wheeler-DeWitt equation has
the character of a Klein-Gordon equation. For this reason trying to normalise
a solution $\Psi\lbrack g_{ij},\phi]$ with respect to the gravitational
degrees of freedom $g_{ij}$ is mathematically analogous to trying to normalise
a single-particle Klein-Gordon wave function $\psi(x,t)$ in spacetime.

D\"{u}rr and Struyve suggest that the non-normalisability of $\left\vert
\Psi\right\vert ^{2}$ for the Wheeler-DeWitt equation is analogous to the
non-normalisability of $\left\vert \Psi\right\vert ^{2}$ encountered in
relational theories of dynamics \cite{DGZ20} and that it can be handled in the
same way. But the two cases are not analogous. In relational theories $\Psi$
is non-normalisable because of symmetries associated with unobservable and
unphysical (absolute) structure. As discussed in ref. \cite{DGZ20} this can be
handled essentially by factoring out the unobservable structure. A correct
analogy might be drawn between relational theories and quantum gauge theories,
where for the latter we need to factor out an infinite unobservable gauge
volume when normalising the relevant wave functional (though there the gauge
volume can be rendered finite by the simple device of putting the field theory
on a lattice \cite{WeinVolII}). But the case of the Wheeler-DeWitt wave
functional is fundamentally different: the non-normalisability is not caused
by unobservable structure but by the wave-like propagation of $\Psi$ in
configuration space (Section 3.2.2).

While the pseudo-continuity equation (\ref{false_cont}) is strictly speaking
correct, it disguises the fact that the Wheeler-DeWitt equation implies not
just the single equation (\ref{false_cont}) but the infinity of equations
(\ref{W-Dcont}) (one per space point $x$). Summing the equations
(\ref{W-Dcont})\ to produce the one equation (\ref{false_cont}), in order to
give the appearance that $\left\vert \Psi\right\vert ^{2}$ obeys a standard
continuity equation, is analogous to summing the equations $\partial_{x}%
j_{x}=\partial_{y}j_{y}=\partial_{z}j_{z}=0$ for a single particle to produce
$\mathbf{\nabla}\cdot\mathbf{j}=0$. The result is mathematically correct but
physically misleading. If $\left\vert \Psi\right\vert ^{2}$ really did behave
like a conventional density, as suggested by equation (\ref{false_cont}), we
would expect it to be normalisable. But in fact the Klein-Gordon-like
structure of the Wheeler-DeWitt equation shows that $\left\vert \Psi
\right\vert ^{2}$ evidently cannot be normalised and is not a physical
density. Thus we should beware of artificial attempts to make $\left\vert
\Psi\right\vert ^{2}$ take on the appearance of a conventional density, when
in fact it is a different kind of object. As numerous authors have pointed
out, the naive Schr\"{o}dinger interpretation is physically misguided, and
this remains true in pilot-wave theory.

It will be argued below that there is no concept of quantum equilibrium in the
deep quantum-gravity regime, and that this is the physical significance of the
non-normalisability of the Wheeler-DeWitt wave functional $\Psi$. To interpret
$\left\vert \Psi\right\vert ^{2}$ as a Born-rule probability measure (or
typicality measure) is a category mistake. At the fundamental level there is
no Born rule. The Born rule can emerge only in the Schr\"{o}dinger
approximation (Section 4.3).

\subsection{Schr\"{o}dinger approximation for a matter field}

It will be useful to consider carefully how the Schr\"{o}dinger approximation,
with a time-dependent effective wave function, arises from the underlying
quantum-gravitational formalism.

\subsubsection{Semiclassical expansion of the Wheeler-DeWitt equation}

To convey the general method we first outline the original treatment given by
Kiefer and Singh \cite{KS91}. We consider quantum gravity in the presence of a
scalar field $\phi$. The extended Wheeler-DeWitt equation (\ref{W-D_ext}) for
$\Psi\lbrack g_{ij},\phi]$ is solved by successive approximation, by means of
a semiclassical expansion in powers of a parameter $\mu=c^{2}/32\pi G$
(dimensionally a mass per length),%
\begin{equation}
\Psi=\exp i\left(  \mu S_{0}+S_{1}+\mu^{-1}S_{2}+...\right)  \ , \label{B-O}%
\end{equation}
where the terms in the round brackets are generally complex. The expansion
(\ref{B-O}) is inserted into the left-hand side of (\ref{W-D_ext}), terms of
the same order in $\mu$ are collected and the sum is set equal to zero.

The highest order that appears is $\mu^{2}$, for which it is found that%
\begin{equation}
\left(  \frac{\delta S_{0}}{\delta\phi}\right)  ^{2}=0\ . \label{OM2}%
\end{equation}
Thus $S_{0}=S_{0}[g_{ij}]$ depends only on $g_{ij}$. The function $S_{0}$ is
associated with a classical spacetime background which does not depend on the
matter field (or matter-field perturbations) propagating on it.

At order $\mu$ we obtain a classical Hamilton-Jacobi equation%
\begin{equation}
G_{ijkl}\frac{\delta S_{0}}{\delta g_{ij}}\frac{\delta S_{0}}{\delta g_{kl}%
}-g^{1/2}R=0\ . \label{H-J}%
\end{equation}
As we have noted this is equivalent to the vacuum Einstein equations. A
realistic model would include a matter term in (\ref{H-J}), but for present
purposes (\ref{H-J}) describes a classical spacetime background on which
quantum perturbations propagate.

At order $\mu^{0}$ we obtain an equation for $S_{1}$:%
\begin{equation}
G_{ijkl}\frac{\delta S_{0}}{\delta g_{ij}}\frac{\delta S_{1}}{\delta g_{kl}%
}-\frac{i}{2}G_{ijkl}\frac{\delta^{2}S_{0}}{\delta g_{ij}\delta g_{kl}}%
+\frac{1}{2\sqrt{g}}\left(  \frac{\delta S_{1}}{\delta\phi}\right)  ^{2}%
-\frac{i}{2\sqrt{g}}\frac{\delta^{2}S_{1}}{\delta\phi^{2}}+u=0\ , \label{OM0}%
\end{equation}
where%
\begin{equation}
u=\frac{1}{2}\sqrt{g}g^{ij}\partial_{i}\phi\partial_{j}\phi+\sqrt
{g}\mathcal{V}(\phi)\ .
\end{equation}
This can be written as an effective time-dependent Schr\"{o}dinger equation
for a matter wave functional $\psi^{(0)}[\phi,g_{ij},t]$, which we may write
more simply as $\psi^{(0)}[\phi,t]$ and which describes the quantum evolution
of $\phi$ on a classical background with metric $g_{ij}$ (where the background
is determined by (\ref{H-J})).

This is accomplished by using the trajectories of the classical background to
define an effective time parameter $t$ by means of the relation%
\begin{equation}
\frac{\partial}{\partial t}=\int d^{3}x\ 2NG_{ijkl}\frac{\delta S_{0}}{\delta
g_{kl}}\frac{\delta}{\delta g_{ij}}\ .
\end{equation}
The factor $2NG_{ijkl}\delta S_{0}/\delta g_{kl}$ coincides with the WKB (or
de Broglie) velocity $\dot{g}_{ij}$ generated by $S_{0}$ (with $N_{i}=0$).
Thus this definition is the same as the WKB time defined above in equation
(\ref{WKB_t}). Here $t$ parameterises integral curves of the de Broglie
velocity field associated with the classical background. If we then define%
\begin{equation}
\psi^{(0)}=D[g_{ij}]\exp(iS_{1})\ , \label{psi_(0)}%
\end{equation}
where $D$ is chosen so as to satisfy the condition%
\begin{equation}
G_{ijkl}\frac{\delta S_{0}}{\delta g_{ij}}\frac{\delta D}{\delta g_{kl}}%
-\frac{1}{2}G_{ijkl}\frac{\delta^{2}S_{0}}{\delta g_{ij}\delta g_{kl}}D=0\ ,
\label{D_condn}%
\end{equation}
it is found that%
\begin{equation}
i\frac{\partial\psi^{(0)}}{\partial t}=\int d^{3}x\ \mathcal{\hat{H}}_{\phi
}\psi^{(0)} \label{Sch_0}%
\end{equation}
(of the general form advertised in (\ref{Sch_approxn_1})). This equation
describes a quantum field $\phi$ evolving on a classical spacetime background,
where $\psi^{(0)}[\phi,t]$ is the zeroth-order or uncorrected wave functional.

Note that, to this order, $\Psi$ does indeed take the WKB form (\ref{WKB})
with%
\begin{equation}
\Psi_{\mathrm{WKB}}[g_{ij}]=\frac{1}{D[g_{ij}]}\exp\left(  iMS_{0}\right)
\end{equation}
and $\psi\lbrack\phi,g_{ij}]=\psi^{(0)}[\phi,t]$.

\subsubsection{Semiclassical expansion of the de Broglie velocity}

We have just reviewed the semiclassical reduction of the extended
Wheeler-DeWitt equation (\ref{W-D_ext}) for $\Psi\lbrack g_{ij},\phi]$ to an
effective time-dependent Schr\"{o}dinger equation (\ref{Sch_0}) for
$\psi^{(0)}[\phi,t]$. In pilot-wave theory we must also consider the de
Broglie guidance equations. If the metric $g_{ij}$ is coupled to a scalar
matter field $\phi$, then as well as the guidance equation (\ref{deB3}) for
$g_{ij}$ we also have the guidance equation (\ref{deB2}) for $\phi$. We must
consider what happens to (\ref{deB2}) under the semiclassical reduction, and
in particular how the field velocity $\dot{\phi}$ is related to the effective
wave functional $\psi^{(0)}[\phi,t]$.

Note that the presence of a de Broglie trajectory $q(t)=(g_{ij}(t),\phi(t))$
for the combined metric-plus-field system does not affect the equations for
the guiding wave functional $\Psi\lbrack g_{ij},\phi]$. Thus the semiclassical
reduction of the Wheeler-DeWitt equation proceeds exactly as before (though
arguably our understanding of WKB time is improved by the explicit presence of
a de Broglie-Bohm trajectory). The question is only what form the guidance
equation (\ref{deB2}) for $\phi$ takes after performing the semiclassical
expansion (\ref{B-O}) of $\Psi\lbrack g_{ij},\phi]$.

Inserting the expansion (\ref{B-O}) into the guidance equation (\ref{deB2}) we
find%
\begin{equation}
\frac{\partial\phi}{\partial t}=\frac{N}{\sqrt{g}}\frac{\delta}{\delta\phi
}\left(  \mu\operatorname{Re}S_{0}+\operatorname{Re}S_{1}+\mu^{-1}%
\operatorname{Re}S_{2}+...\right)  \label{phidot1}%
\end{equation}
(where $S=\operatorname{Im}\ln\Psi$ and $\operatorname{Im}%
(iz)=\operatorname{Re}z$). From (\ref{OM2}) we know that $S_{0}$ does not
depend on $\phi$ and so the first term in (\ref{phidot1}) vanishes. Thus we
have%
\begin{equation}
\frac{\partial\phi}{\partial t}=\frac{N}{\sqrt{g}}\frac{\delta}{\delta\phi
}\left(  \operatorname{Re}S_{1}+\mu^{-1}\operatorname{Re}S_{2}+...\right)  \ .
\label{phidot2}%
\end{equation}
From equation (\ref{OM0}) it is clear that $S_{1}$ is generally complex.
Furthermore, in (\ref{psi_(0)}) we have defined a zeroth-order wave function
$\psi^{(0)}=D\exp(iS_{1})$, where $D$ satisfies the condition (\ref{D_condn})
and can be chosen to be real. Thus $\operatorname{Re}S_{1}$ is equal to the
phase of $\psi^{(0)}$ -- that is, $\operatorname{Re}S_{1}=\operatorname{Im}%
\ln\psi^{(0)}$. If we keep only the first term in (\ref{phidot2}), to lowest
order we then have the usual de Broglie velocity,%
\begin{equation}
\left(  \frac{\partial\phi}{\partial t}\right)  ^{(0)}=\frac{N}{\sqrt{g}}%
\frac{\delta}{\delta\phi}\operatorname{Re}S_{1}=\frac{N}{\sqrt{g}}\frac
{\delta}{\delta\phi}\left(  \operatorname{Im}\ln\psi^{(0)}\right)  \ ,
\label{deB_0}%
\end{equation}
generated in the usual way by $\psi^{(0)}$. To lowest order the de Broglie
velocity for the matter field is unchanged.

\subsubsection{The problem of the Born rule}

Given the effective Schr\"{o}dinger equation (\ref{Sch_0}) for the wave
function $\psi^{(0)}$ of the matter field $\phi$, writing $\psi^{(0)}%
=\left\vert \psi^{(0)}\right\vert e^{iS^{(0)}}$ we can readily show that
$\left\vert \psi^{(0)}\right\vert ^{2}$ satisfies a continuity equation%
\begin{equation}
\frac{\partial\left\vert \psi^{(0)}\right\vert ^{2}}{\partial t}+\int
d^{3}x\;\frac{\delta}{\delta\phi}\left(  \left\vert \psi^{(0)}\right\vert
^{2}\frac{N}{\sqrt{g}}\frac{\delta S^{(0)}}{\delta\phi}\right)  =0\ .
\label{contA}%
\end{equation}
This implies that the squared-norm $\int D\phi\ \left\vert \psi^{(0)}%
\right\vert ^{2}$ is conserved in time. In standard quantum mechanics we might
then simply assume that $\left\vert \psi^{(0)}\right\vert ^{2}$ represents a
physical probability density for the field $\phi$ propagating on the classical
background. But it would be more satisfactory if we could derive this
interpretation from first principles starting from the underlying
quantum-gravitational description.

If we begin with the Wheeler-DeWitt wave functional $\Psi\lbrack g_{ij},\phi
]$, it may seem natural to interpret $\left\vert \psi^{(0)}\right\vert ^{2}$
as a conditional probability density derived from an underlying joint density
$\left\vert \Psi\lbrack g_{ij},\phi]\right\vert ^{2}$ -- where the probability
for $\phi$ is conditional on the metric coinciding with the classical
background. Recalling the standard conditional probability formula
$p(A\ |\ B)=p(A\cap B)/p(B)$ for events $A$ and $B$, we might consider the
conditional probability (at a given time $t$)%
\begin{equation}
p(\phi\in D\phi\ |\ g_{ij}\in Dg_{ij})=\frac{p(\phi\in D\phi\cap g_{ij}\in
Dg_{ij})}{p(g_{ij}\in Dg_{ij})}%
\end{equation}
to find $\phi$ in the interval $D\phi$ given that $g_{ij}$ lies in the
interval $Dg_{ij}$, where for given $g_{ij}$ we can define a density
$\rho\lbrack\phi,t]$ by%
\begin{equation}
\rho\lbrack\phi,t]D\phi=p(\phi\in D\phi\ |\ g_{ij}\in Dg_{ij})\ .
\end{equation}
If we write down the formal expressions%
\begin{equation}
p(\phi\in D\phi\cap g_{ij}\in Dg_{ij})=\left\vert \Psi\lbrack g_{ij}%
,\phi]\right\vert ^{2}D\phi Dg_{ij}%
\end{equation}
and%
\begin{equation}
p(g_{ij}\in Dg_{ij})=\left(  \int\left\vert \Psi\lbrack g_{ij},\phi
]\right\vert ^{2}D\phi\right)  Dg_{ij}%
\end{equation}
we find%
\begin{equation}
\rho\lbrack\phi,t]=\frac{\left\vert \Psi\lbrack g_{ij},\phi]\right\vert ^{2}%
}{\left(  \int\left\vert \Psi\lbrack g_{ij},\phi]\right\vert ^{2}D\phi\right)
}\ .
\end{equation}
If we take the WKB approximation $\Psi\lbrack g_{ij},\phi]\approx
\Psi_{\mathrm{WKB}}[g_{ij}]\psi^{(0)}[\phi,t]$ we then have%
\begin{equation}
\rho^{(0)}[\phi,t]\approx\left\vert \psi^{(0)}[\phi,t]\right\vert ^{2}%
\end{equation}
(assuming that $\psi^{(0)}$ is normalised).

It may then appear that we have succeeded in deriving the Born rule
$\rho^{(0)}=\left\vert \psi^{(0)}\right\vert ^{2}$ for the effective
Schr\"{o}dinger regime from an underlying Born-rule density $\left\vert
\Psi\lbrack g_{ij},\phi]\right\vert ^{2}$ for the joint Wheeler-DeWitt system.
Unfortunately, however, the derivation rests on treating $\left\vert
\Psi\lbrack g_{ij},\phi]\right\vert ^{2}$ as if it were a well-defined
physical probability density, which as we saw in Section 3.2.2 is impossible,
reflecting the well-known and long-standing difficulties with the naive
Schr\"{o}dinger interpretation. The Wheeler-DeWitt wave functional
$\Psi\lbrack g_{ij},\phi]$ is intrinsically non-normalisable and the quantity
$\left\vert \Psi\lbrack g_{ij},\phi]\right\vert ^{2}$ cannot be interpreted as
a physical probability density. Therefore the above derivation of the
effective Born rule $\rho^{(0)}=\left\vert \psi^{(0)}\right\vert ^{2}$ is in
fact without foundation. Even if the resulting conditional probability $\rho$
is normalisable, it makes no physical sense to derive it from a parent density
$\left\vert \Psi\right\vert ^{2}$ which is not a physical probability.

In pilot-wave theory we can go a little further. The velocity field appearing
in the continuity equation (\ref{contA}) for $\left\vert \psi^{(0)}\right\vert
^{2}$ coincides with the de Broglie velocity (\ref{deB_0}) for $\phi$ obtained
from the Wheeler-DeWitt equation. We can then readily show that $\left\vert
\psi^{(0)}\right\vert ^{2}$ is a state of quantum equilibrium. A general
distribution $\rho^{(0)}$ will by construction satisfy%
\begin{equation}
\frac{\partial\rho^{(0)}}{\partial t}+\int d^{3}x\;\frac{\delta}{\delta\phi
}\left(  \rho^{(0)}\frac{N}{\sqrt{g}}\frac{\delta S^{(0)}}{\delta\phi}\right)
=0\ . \label{contB}%
\end{equation}
This takes the same form as (\ref{contA}). If $\rho^{(0)}$ and $\left\vert
\psi^{(0)}\right\vert ^{2}$ happen to be equal at some initial time then they
will necessarily remain equal at later times. Thus we have an equilibrium
state $\rho^{(0)}=\left\vert \psi^{(0)}\right\vert ^{2}$. But a basic question
still remains. We have a well-defined Born rule and equilibrium state in the
effective Schr\"{o}dinger regime, but this can only be an emergent
approximation. What happens to the Born rule and to the concept of quantum
equilibrium at the underlying level of the Wheeler-DeWitt equation? The
quantity $\left\vert \Psi\lbrack g_{ij},\phi]\right\vert ^{2}$ cannot be a
physical density and there are no other obvious candidates for a general Born
rule in the quantum-gravity regime. There is then a logical gap: we appear to
have a well-defined Born rule at the emergent level but not at the fundamental level.

To make sense of this, we must reconsider the nature of probability in quantum
gravity from first principles. We shall see that, if we are careful to follow
the internal logic of pilot-wave theory, the problem resolves itself naturally
and simply.

\section{New approach to the Born rule in quantum gravity}

In this paper we propose a new approach to the Born rule in quantum gravity.
We begin by accepting that the time-independent and non-normalisable
Wheeler-DeWitt wave functional $\Psi$ is a different kind of thing from the
time-dependent and normalisable Schr\"{o}dinger wave functions $\psi$ that we
are used to from non-gravitational physics. In particular we take the
non-normalisability of $\Psi$ as an indication that $\left\vert \Psi
\right\vert ^{2}$ can never be equal to a physical probability density $P$. To
make sense of this, we will consider the pilot-wave theory of the
Wheeler-DeWitt equation on its own terms, without making any undue assumptions
taken from standard quantum theory. We find that there is no physical
equilibrium or Born-rule state in the deep quantum-gravity regime. Furthermore
this implies the presence of quantum nonequilibrium at the beginning of the
emergent semiclassical regime. Quantum gravity naturally creates an early
nonequilibrium universe. Once the Schr\"{o}dinger approximation is
established, quantum relaxation to the Born rule can take place in the usual
way. Fundamentally, however, the physics of the deep quantum-gravity regime
undermines the Born rule as we know it, rendering it unstable.

\subsection{Absence of an equilibrium state in the deep quantum-gravity
regime}

The starting point for our new approach is to recognise that the
non-integrable quantity $\left\vert \Psi\right\vert ^{2}$ is not and cannot be
a physical Born-rule probability density $P$. As we saw in Section 3.2.2,
$\Psi$ is intrinsically non-normalisable and for a deep reason: the
Wheeler-DeWitt equation has the character of a Klein-Gordon equation. This
problem cannot be eliminated by a standard device such as discretising the
system. Instead the problem suggests that something is deeply wrong with our
interpretation of the formalism. In our view the failure of the naive
Schr\"{o}dinger interpretation shows that $\Psi$ is not an ordinary wave
function. If we accept this starting point we nevertheless need to discuss
probabilities for ensembles -- and somehow recover an effective Born rule for
systems propagating on a classical spacetime background. By carefully
following the internal logical of pilot-wave theory, this turns out to be straightforward.

We begin with the deterministic pilot-wave dynamics of an individual
gravitational system with wave functional $\Psi\lbrack g_{ij},\phi]$, where
$\Psi\lbrack g_{ij},\phi]$ satisfies the extended Wheeler-DeWitt equation
(\ref{W-D_ext}). Each gravitational system has a trajectory determined by the
de Broglie guidance equations (\ref{deB3}) and (\ref{deB2}) for $g_{ij}$ and
$\phi$ respectively. The wave functional $\Psi\lbrack g_{ij},\phi]$ acts as a
`pilot wave' on configuration space guiding the motion of the system. The
question is how to relate this dynamics to ensembles and how to recover an
effective Born rule in some limit.

Let us consider a theoretical ensemble of similar systems each guided by the
same wave functional $\Psi\lbrack g_{ij},\phi]$. At time $t$ each element of
the ensemble has an evolving configuration $q(t)=(g_{ij}(x,t),\phi(x,t))$.
Over the ensemble we will then have an evolving distribution $P[g_{ij}%
,\phi,t]$ of configurations. As usual in pilot-wave theory, at an initial time
$t_{i}$, the initial distribution $P[g_{ij},\phi,t_{i}]$ is in principle
arbitrary. By definition, of course, $P$ is a physical probability density and
must be normalisable,%
\begin{equation}
\int\int DgD\phi\ P[g_{ij},\phi,t]=1\ ,
\end{equation}
at all times $t$. There are the usual technical issues concerning the rigorous
definition of the measure $DgD\phi$. These might be avoided by some form of
discretisation; in any case we leave them aside and focus on the conceptual
questions. Each configuration $q(t)=(g_{ij}(x,t),\phi(x,t))$ evolves with the
velocities (\ref{deB3}) and (\ref{deB2}). It follows that $P[g_{ij},\phi,t]$
evolves in time according to the continuity equation (\ref{cont_P}). In this
way pilot-wave dynamics defines the time evolution of an arbitrary ensemble of
gravitational systems.

As we saw in Section 3.3.2, $\left\vert \Psi\lbrack g_{ij},\phi]\right\vert
^{2}$ formally satisfies the same continuity equation (\ref{false_cont}).
However $\left\vert \Psi\lbrack g_{ij},\phi]\right\vert ^{2}$ is
non-integrable and cannot count as a physical density.

Now we have seen that, in non-gravitational pilot-wave theory, the initial
density $\rho(q,t_{i})$ need not be equal to the initial equilibrium density
$\left\vert \psi(q,t_{i})\right\vert ^{2}$. The Born rule $\rho
(q,t)=\left\vert \psi(q,t)\right\vert ^{2}$ can nevertheless emerge at later
times $t$ by a process of quantum relaxation (Section 2.1). Here we argue
that, for gravitational systems in the deep quantum-gravity regime, we must go
a step further: the density $P[g_{ij},\phi,t]$ can \textit{never} be equal to
$\left\vert \Psi\lbrack g_{ij},\phi]\right\vert ^{2}$, neither at the initial
time nor subsequently. In other words, in quantum gravity there is no such
thing as `quantum equilibrium'. By definition $P[g_{ij},\phi,t]$ is a physical
probability density which (subject to some mathematical caveats) can be
normalised. In contrast, the failure of the naive Schr\"{o}dinger
interpretation shows that $\left\vert \Psi\lbrack g_{ij},\phi]\right\vert
^{2}$ cannot be a physical probability density. Therefore we must have%
\begin{equation}
P[g_{ij},\phi,t]\neq\left\vert \Psi\lbrack g_{ij},\phi]\right\vert ^{2}
\label{noneq_QG}%
\end{equation}
at all times. The two quantities $P[g_{ij},\phi,t]$ and $\left\vert
\Psi\lbrack g_{ij},\phi]\right\vert ^{2}$ are different kinds of physical
things. The left-hand side of (\ref{noneq_QG}) can and must be normalised, the
right-hand side can never be. Hence they cannot be equal, neither initially
nor subsequently.

At this point it might be asked what determines the initial $P$. This question
has already been addressed at length in non-gravitational pilot-wave theory
and the same principles apply here. The initial $P$ is not fixed by any law;
it is instead to be determined empirically, or at least constrained as far as
possible by observation, as is generally the case for initial conditions in
physics. In short, $P$ is in principle arbitrary and in practice empirically
constrained \cite{Allori20}.

In non-gravitational pilot-wave theory we have the analogue (\ref{fdot}) of
Liouville's theorem together with the analogue (\ref{Hthm}) of the
coarse-graining $H$-theorem \cite{AV91a}. If we write%
\begin{equation}
P[g_{ij},\phi,t]=\left\vert \Psi\lbrack g_{ij},\phi]\right\vert ^{2}%
f[g_{ij},\phi,t]
\end{equation}
the result (\ref{fdot}) still applies because $P$ and $\left\vert
\Psi\right\vert ^{2}$ obey the same continuity equations ((\ref{cont_P}) and
(\ref{false_cont})). If we define a fine-grained $H$-function%
\begin{equation}
H(t)=\int\int DgD\phi\ P\ln(P/\left\vert \Psi\right\vert ^{2})
\end{equation}
we again have $dH/dt=0$. Furthermore, if we assume that $P$ and $\left\vert
\Psi\right\vert ^{2}$ have no initial fine-grained structure, then even though
$\left\vert \Psi\right\vert ^{2}$ is non-normalisable we can still show that
the coarse-grained $H$-function%
\begin{equation}
\bar{H}(t)=\int\int DgD\phi\ \bar{P}\ln(\bar{P}/\overline{\left\vert
\Psi\right\vert ^{2}})
\end{equation}
obeys the $H$-theorem (\ref{Hthm}) (since none of the steps in ref.
\cite{AV91a} depend on the normalisability of $\left\vert \Psi\right\vert
^{2}$). It might then be thought that quantum relaxation, $\bar{P}%
\rightarrow\overline{\left\vert \Psi\right\vert ^{2}}$, could still take place
on a coarse-grained level. However, when $\Psi$ is non-normalisable, the
$H$-function has no lower bound and any physical distribution $\bar{P}$ will
always be infinitely far away from `equilibrium' (as defined by the minimum
value of $H$).

To see this consider a general system with $H$-function (\ref{Hfn}) for which%
\begin{equation}
\int dq\ \left\vert \psi\right\vert ^{2}=N
\end{equation}
is finite but arbitrarily large (while of course $\int dq\ \rho=1$). Normally
we have $N=1$, in which case $H$ is bounded below by zero. This follows from
the general inequality $x\ln(x/y)\geq x-y$ (with equality if and only if
$x=y$). Putting $x=\rho$ and $y=\left\vert \psi\right\vert ^{2}$ we have
$H\geq\int dq\ (\rho-\left\vert \psi\right\vert ^{2})=0$ (with $H=0$ if and
only if $\rho=\left\vert \psi\right\vert ^{2}$). Thus $H=0$ normally
corresponds to equilibrium. However, for general $N$, the $H$-function is
bounded below by $-\ln N$. To see this write%
\begin{align}
H  &  =\int dq\ \left(  \rho\ln(\rho/\left\vert \psi\right\vert ^{2}%
)-\rho+\left\vert \psi\right\vert ^{2}/N\right) \\
&  =-\ln N+\int dq\ \left(  \rho\ln\left(  \frac{\rho}{(\left\vert
\psi\right\vert ^{2}/N)}\right)  -\rho+\left\vert \psi\right\vert
^{2}/N\right)  \ .
\end{align}
Using $x\ln(x/y)-x+y\geq0$ with $x=\rho$ and $y=\left\vert \psi\right\vert
^{2}/N$ we have%
\begin{equation}
H\geq-\ln N\ ,
\end{equation}
with $H=-\ln N$ if and only if $\rho=\frac{1}{N}\left\vert \psi\right\vert
^{2}$. Clearly, as $N\rightarrow\infty$ there is no lower bound on $H$ and so
there is no physical equilibrium state. The coarse-grained function $\bar
{H}(t)$ could continue to decrease indefinitely, without ever reaching a
minimum and hence without ever reaching equilibrium. Note that if the initial
nonequilibrium distribution $\rho$ is localised in some finite region $R$ of
configuration space, the initial $H$-function $H=\int_{R}dq\ \rho\ln
(\rho/\left\vert \psi\right\vert ^{2})$ will have a finite value. As
$\bar{\rho}$ evolves the coarse-grained function $\bar{H}(t)$ will remain
finite and in this sense the system forever remains infinitely far away from
the `equilibrium' state $\bar{H}=-\infty$.

It would be instructive to explore quantum relaxation numerically for such
systems. This could be done for a minisuperspace model of quantum cosmology,
for which the Wheeler-DeWitt equation has the usual Klein-Gordon-like
structure and the wave function $\psi$ is intrinsically non-normalisable. (For
an example of such a model see Section 7.4.) Given the analogy with a
Klein-Gordon wave function on spacetime, we may expect that $\rho$ will
continue to spread out indefinitely over its unbounded domain. In principle we
might obtain some degree of relaxation in a limited region of configuration
space, but even then the system will necessarily remain infinitely far from
equilibrium overall. We leave such numerical studies for future work
\cite{KV21}. In any case the conclusion is clear: there can be no proper
quantum relaxation for a system with a non-normalisable wave function.

We conclude that, in all circumstances, there is no physical Born rule at the
fundamental level of quantum gravity: $P$ can never be equal to $\left\vert
\Psi\right\vert ^{2}$. We might say that the deep quantum-gravity regime is
necessarily and perpetually in a state of quantum nonequilibrium. It would be
more precise, however, to say that in this regime there is no physical state
of quantum equilibrium.

If we accept this reasoning, it might appear that we now have no hope of ever
recovering the Born rule even in the limit appropriate to ordinary quantum
physics on a classical spacetime background. However, if we follow the
internal logic of pilot-wave theory, we find that in fact the relevant physics
can be recovered. But first we need to consider how probabilities emerge for
systems on a classical spacetime background.

\subsection{Early quantum nonequilibrium as a consequence of quantum gravity}

Let us consider what we can expect to find as the early universe emerges from
the deep quantum-gravity regime and enters the approximate Schr\"{o}dinger
regime. In particular, we would like to know the probability distribution for
a quantum matter field $\phi$ evolving on a classical spacetime background at
the beginning of the Schr\"{o}dinger regime.

We already have the tools required to address this question. At the
fundamental level we have the timeless Wheeler-DeWitt equation (\ref{W-D_ext})
for the wave functional $\Psi\lbrack g_{ij},\phi]$. As we saw in Section
3.4.1, in an appropriate limit we recover the approximate time-dependent
Schr\"{o}dinger equation (\ref{Sch_0}) with an effective wave functional
$\psi^{(0)}[\phi,t]$ for $\phi$ on the classical background. In addition the
fundamental de Broglie equation of motion (\ref{deB2}) for the trajectory
$\phi(x,t)$ reduces to the effective equation (\ref{deB_0}) with the field
velocity $\partial\phi/\partial t$ determined by $\psi^{(0)}$ in the usual
way. The effective Schr\"{o}dinger equation (\ref{Sch_0}) is just the usual
functional Schr\"{o}dinger equation for a free scalar field on a curved
spacetime background. If we allow ourselves to employ the usual regularisation
methods -- for example some form of discretisation, or dimensional
regularisation \cite{Guv89} -- then the emergent wave function $\psi^{(0)}$ is
normalisable as well as time-dependent. In standard quantum mechanics this
might seem sufficient to obtain the Born rule. But in pilot-wave theory so far
all we have done is recover an approximate pilot-wave dynamics for an
individual field system. To speak of probabilities, we must consider ensembles.

Consider, then, a theoretical ensemble of matter fields $\phi$ with the same
effective wave functional $\psi^{(0)}[\phi,t]$. The ensemble will have some
distribution $\rho^{(0)}[\phi,t]$. Before considering how the distribution
evolves in time, let us first focus on its `initial' value $\rho^{(0)}%
[\phi,t_{i}]$ -- defined at some time $t_{i}$ when the system has just emerged
from the deep quantum-gravity regime and entered the approximate
Schr\"{o}dinger regime. What do we expect to find? In Section 3.4.3 we noted
the difficulty of trying to derive $\rho^{(0)}[\phi,t_{i}]$ as a conditional
probability from an underlying non-normalisable distribution $\left\vert
\Psi\lbrack g_{ij},\phi]\right\vert ^{2}$. According to our new approach there
is no such physical distribution. But we can still consider a theoretical
ensemble of systems with the same wave functional $\Psi\lbrack g_{ij},\phi]$.
As we have argued, such an ensemble will have an arbitrary initial
distribution $P[g_{ij},\phi,t_{i}]$ which must be constrained empirically. The
effective distribution $\rho^{(0)}[\phi,t_{i}]$ for the matter field can then
be derived, as a conditional probability, not from $\left\vert \Psi\lbrack
g_{ij},\phi]\right\vert ^{2}$ but from $P[g_{ij},\phi,t_{i}]$ (with the
probability for $\phi$ being conditional on the metric coinciding with the
classical background). Following steps similar to those in Section 3.4.3, but
with $\left\vert \Psi\lbrack g_{ij},\phi]\right\vert ^{2}$ replaced by
$P[g_{ij},\phi,t_{i}]$, we find%
\begin{equation}
\rho^{(0)}[\phi,t_{i}]=\frac{P[g_{ij},\phi,t_{i}]}{\left(  \int P[g_{ij}%
,\phi,t_{i}]D\phi\right)  }\ . \label{condnl_prob_phi}%
\end{equation}
Note that on the right-hand side it is understood that we have inserted the
known value of $g_{ij}$. The resulting conditional distribution $\rho
^{(0)}[\phi,t_{i}]$ is of course normalised by construction.

Now, as we have noted, $P[g_{ij},\phi,t_{i}]$ is arbitrary and can only be
constrained empirically. It follows from (\ref{condnl_prob_phi}) that
$\rho^{(0)}[\phi,t_{i}]$ is also arbitrary and can only be constrained
empirically. Furthermore, because of the fundamental nonequilibrium condition
(\ref{noneq_QG}) of the deep quantum-gravity regime, in general we will have%
\begin{equation}
\rho^{(0)}[\phi,t_{i}]\neq\left\vert \psi^{(0)}[\phi,t_{i}]\right\vert ^{2}
\label{noneq_Sch}%
\end{equation}
(except for the special case where $P[g_{ij},\phi,t_{i}]=\Pi\lbrack
g_{ij}]\left\vert \psi^{(0)}[\phi,t_{i}]\right\vert ^{2}$ for some $\Pi\lbrack
g_{ij}]$). Thus, as we enter the Schr\"{o}dinger regime, the matter-field
system can be expected to be in a state of quantum nonequilibrium.

In effect we have derived the general hypothesis of primordial quantum
nonequilibrium (\ref{noneq_Sch}) -- for matter fields in the early
Schr\"{o}dinger regime -- as a consequence of quantum gravity. This might seem
like a failure, as we have yet to recover the Born rule. However, in
reasonable circumstances, after the initial time $t_{i}$ we can expect the
effective density $\rho^{(0)}[\phi,t]$ to relax towards $\left\vert \psi
^{(0)}[\phi,t]\right\vert ^{2}$ (on a coarse-grained level), simply as a
consequence of the dynamics. Thus we do not derive the Born rule at early
times (at the beginning of the Schr\"{o}dinger regime). On the contrary, we
derive nonequilibrium at early times. The Born rule emerges only later, via
the by-now much-studied process of quantum relaxation.

\subsection{Quantum relaxation in the Schr\"{o}dinger approximation}

To see how the Born rule emerges by quantum relaxation, let us look again at
the effective equations (\ref{Sch_0}) and (\ref{deB_0}) for the pilot-wave
dynamics of the matter field $\phi$ on a classical background spacetime. For
an ensemble of similar systems, with the same effective wave function
$\psi^{(0)}$, at an early `initial' time $t_{i}$ we will have (as argued
above) an arbitrary nonequilibrium distribution (\ref{noneq_Sch}).
Subsequently the distribution $\rho^{(0)}[\phi,t]$ will, by construction,
evolve in time according to the continuity equation%
\begin{equation}
\frac{\partial\rho^{(0)}}{\partial t}+\int d^{3}x\ \frac{\delta}{\delta\phi
}\left(  \rho^{(0)}\frac{\partial\phi}{\partial t}\right)  =0\ ,
\end{equation}
where the velocity $\partial\phi/\partial t$ is given by (\ref{deB_0}). This
is the same continuity equation that is satisfied by $\left\vert \psi
^{(0)}[\phi,t]\right\vert ^{2}$ (as derived from the Schr\"{o}dinger equation
(\ref{Sch_0})). We can then define an $H$-function%
\begin{equation}
H(t)=\int D\phi\ \rho^{(0)}\ln\left(  \rho^{(0)}/\left\vert \psi
^{(0)}\right\vert ^{2}\right)  \ .
\end{equation}
This will have the usual properties. Because $\rho^{(0)}$ and $\left\vert
\psi^{(0)}\right\vert ^{2}$ satisfy the same continuity equation the exact or
fine-grained value of $H(t)$ will be conserved in time. If we assume that
$\rho^{(0)}$ and $\left\vert \psi^{(0)}\right\vert ^{2}$ have no fine-grained
structure at the initial time $t_{i}$, we obtain the coarse-graining
$H$-theorem (\ref{Hthm}). As usual $\bar{H}(t)$ is bounded below by zero and
is equal to zero if and only if $\overline{\rho^{(0)}}=\overline{\left\vert
\psi^{(0)}\right\vert ^{2}}$ everywhere.

For field theory it is more convenient to discuss quantum relaxation in terms
of modes in Fourier space \cite{AV07,AV08a,CV13}. For a scalar field on an
expanding homogeneous background with scale factor $a(t)$, each field mode is
mathematically equivalent to a two-dimensional oscillator with mass $m=a^{3}$
and angular frequency $\omega=k/a$. Extensive numerical simulations have shown
that relaxation occurs efficiently at sub-Hubble wavelengths ($a\lambda
<H^{-1}$) and is slowed or suppressed at super-Hubble wavelengths
($a\lambda>H^{-1}$) \cite{CV13,CV15,CV16}. In this way, at least at the short
wavelengths relevant to laboratory physics, we recover the Born rule as a
consequence of quantum relaxation.

We emphasise that, according to this reasoning, efficient quantum relaxation
emerges only in the Schr\"{o}dinger regime. It is only in this regime that a
physical state of quantum equilibrium exists at all.

It is worth noting that this new approach definitively resolves a controversy
about the origin of the Born rule in pilot-wave theory, which we mentioned at
the end of Section 2.1. The argument for a dynamical origin remains valid, and
is in fact strengthened by the observation that we will generally find
nonequilibrium as we emerge from the deep quantum-gravity regime, with quantum
relaxation to the Born rule taking place only afterwards. In contrast, the
argument from typicality given by the Bohmian mechanics school -- even in its
usual circular form -- can no longer even be formulated: there is no physical
typicality measure $\left\vert \Psi\right\vert ^{2}$ for a universe governed
by quantum gravity.

\subsection{Gravitational instability of the Born rule}

We have argued that, at the fundamental level of quantum gravity, there is no
Born rule and no such thing as a state of quantum equilibrium. Furthermore, as
the very early universe emerges from the deep quantum-gravity regime, it will
be in a state of quantum nonequilibrium as defined in the emergent
Schr\"{o}dinger approximation. And finally, once the Schr\"{o}dinger
approximation is established, quantum relaxation to the Born rule can take
place in the usual way. We are then able to explain why we observe the Born
rule today, even though (according to our interpretation) there is no such
rule in the fundamental quantum-gravitational theory. A natural question then
arises: once quantum equilibrium is reached, can subsequent gravitational
effects drive a system away from quantum equilibrium? In other words, might
quantum gravity render the Born rule unstable?

Consider, for example, a bouncing cosmology \cite{BP17}. On some scenarios
there is a pre-big-bang phase during which the universe is macroscopic,
contracting, and behaving essentially classically. As the universe contracts
it eventually approaches a phase in which quantum effects become important,
resulting in a `bounce' with conventional big-bang cosmology emerging on the
other side as the universe expands. During the bounce quantum-gravitational
effects could be important, in which case the following scenario could arise
\cite{AV14}. During the contracting phase we have an effective Schr\"{o}dinger
approximation with matter fields obeying equations of the form (\ref{Sch_0})
and (\ref{deB_0}). If the contracting phase has lasted a long time, we can
reasonably expect the matter fields to have reached quantum equilibrium. As we
approach the bounce, then, the Born rule should hold. However, once we are
sufficiently close to the bounce the Schr\"{o}dinger approximation may break
down and we could enter the deep quantum-gravity regime. During that phase
there will be no Born rule and no state of quantum equilibrium. We may then
reasonably expect that, as we emerge from the bounce and back into a
Schr\"{o}dinger regime, the matter fields will be in a state of quantum
\textit{non}equilibrium. Thus quantum-gravitational effects during a
cosmological bounce might render the Born rule unstable.

We can envisage similar effects involving black holes. During the contracting
phase of gravitational collapse there is a fairly well-understood regime
described by quantum field theory on a classical spacetime background. In this
regime the Born rule is usually assumed. But according to classical general
relativity, the formation of a black hole entails the formation of a classical
spacetime singularity, where it is expected that quantum-gravitational effects
will be important (possibly preventing the formation of a true singularity).
Close to the singularity we can then expect to find a breakdown of the Born
rule and of the very idea of quantum equilibrium. The semiclassical mechanism
discussed in refs. \cite{AV07,AV04b,KV20} could then come into play. This
mechanism depends on the well-known entanglement between ingoing and outgoing
field modes in the natural vacuum state (for quantised fields propagating on
the background spacetime) \cite{BD82}. In pilot-wave theory entanglement can
provide a channel whereby information can propagate nonlocally \cite{AV91b}.
Such effects are erased in equilibrium but could be relevant if nonequilibrium
already exists for internal degrees of freedom behind the horizon. If the
ingoing field modes interact (locally) with internal nonequilibrium degrees of
freedom, then according to pilot-wave theory the outgoing field modes will
evolve away from quantum equilibrium. In effect, entanglement enables the
nonlocal propagation of quantum nonequilibrium from the interior to the
exterior \cite{AV04b}, as has been confirmed analytically and numerically for
a simple field-theoretical model \cite{KV20}. We then have a mechanism whereby
information can escape from behind the classical event horizon -- provided the
black hole contains nonequilibrium degrees of freedom, presumably because of
quantum-gravitational effects near the singularity.

We may also mention the possible phenomenon of quantum tunnelling from a black
hole to a white hole, which has been argued to be induced by
quantum-gravitational effects \cite{BH2WHth}. As in a bouncing cosmology, we
might expect some degree of quantum nonequilibrium to be generated during the
black-to-white transition or bounce.

Finally, we may consider a more immediately tractable scenario, which will be
developed in detail in the rest of this paper. We have argued that in the deep
quantum-gravity regime there is no state of quantum equilibrium and that such
a state emerges only in the Schr\"{o}dinger approximation. This raises the
question of what might happen in an \textit{intermediate regime} with
quantum-gravitational corrections to the Schr\"{o}dinger approximation. It
seems reasonable to expect that such corrections could generate a small
instability of the Born rule, whereby quantum nonequilibrium can be generated
from a prior equilibrium state. We shall now show that this last expectation
is in fact justified.

\section{Gravitational corrections to the Schr\"{o}dinger approximation}

In this section we discuss quantum-gravitational corrections to the
Schr\"{o}dinger approximation. These take the form of small corrections to the
effective Hamiltonian. The corrections arise from higher-order terms in the
semiclassical expansion (\ref{B-O}) of the Wheeler-DeWitt wave functional.
Such corrections were first derived by Kiefer and Singh \cite{KS91}. Similar
results have been derived for quantum-cosmological models and have been
applied extensively to inflationary cosmology [69--71]. Remarkably, the
quantum-gravitational corrections to the effective Hamiltonian consist of both
Hermitian and non-Hermitian terms. In standard quantum mechanics it is
difficult to interpret the non-Hermitian terms as they appear to violate the
conservation of probability. For this reason the non-Hermitian terms are
usually ignored or dropped by fiat. However, as we shall see in Section 6,
such terms have a clear interpretation in pilot-wave theory: in their presence
probability is still conserved but the Born rule becomes unstable.

\subsection{Corrections to the effective Schr\"{o}dinger equation}

Following ref. \cite{KS91} we continue the analysis of Section 3.4.1 and
consider higher orders in the semiclassical expansion (\ref{B-O}) of the
Wheeler-DeWitt equation for $\Psi\lbrack g_{ij},\phi]$. At order $\mu^{-1}$
Kiefer and Singh obtain an equation for $S_{2}$:%
\begin{equation}
G_{ijkl}\frac{\delta S_{0}}{\delta g_{ij}}\frac{\delta S_{2}}{\delta g_{kl}%
}+\frac{1}{2}G_{ijkl}\frac{\delta S_{1}}{\delta g_{ij}}\frac{\delta S_{1}%
}{\delta g_{kl}}-\frac{i}{2}G_{ijkl}\frac{\delta^{2}S_{1}}{\delta g_{ij}\delta
g_{kl}}+\frac{1}{\sqrt{g}}\frac{\delta S_{1}}{\delta\phi}\frac{\delta S_{2}%
}{\delta\phi}-\frac{i}{2\sqrt{g}}\frac{\delta^{2}S_{2}}{\delta\phi^{2}}=0\ .
\label{OM(-1)}%
\end{equation}
Using (\ref{psi_(0)}) to write $S_{1}$ in terms of $\psi^{(0)}$, and writing%
\begin{equation}
S_{2}=\sigma_{2}[g_{ij}]+\eta\lbrack\phi,g_{ij}] \label{split_S_2}%
\end{equation}
where $\sigma_{2}$ is chosen to satisfy the condition%
\begin{equation}
G_{ijkl}\frac{\delta S_{0}}{\delta g_{ij}}\frac{\delta\sigma_{2}}{\delta
g_{kl}}-\frac{1}{D^{2}}G_{ijkl}\frac{\delta D}{\delta g_{ij}}\frac{\delta
D}{\delta g_{kl}}+\frac{1}{2D}G_{ijkl}\frac{\delta^{2}D}{\delta g_{ij}\delta
g_{kl}}=0\ , \label{sigma_2_condn}%
\end{equation}
the corrected matter wave functional%
\begin{equation}
\psi^{(1)}=\psi^{(0)}\exp(i\eta/\mu) \label{psi_(1)}%
\end{equation}
is found to obey a corrected Schr\"{o}dinger equation%

\begin{equation}
i\frac{\partial\psi^{(1)}}{\partial t}=\int d^{3}x\ \left[  \mathcal{\hat{H}%
}_{\phi}+\frac{1}{8\mu}\frac{1}{\sqrt{g}R}\mathcal{\hat{H}}_{\phi}^{2}%
+i\frac{1}{8\mu}\frac{\delta}{\delta\tau}\left(  \frac{\mathcal{\hat{H}}%
_{\phi}}{\sqrt{g}R}\right)  \right]  \psi^{(1)}\ , \label{Sch_corr_1}%
\end{equation}
where the convenient shorthand%
\begin{equation}
\frac{\delta}{\delta\tau}=2G_{ijkl}\frac{\delta S_{0}}{\delta g_{ij}}%
\frac{\delta}{\delta g_{kl}} \label{mf_time}%
\end{equation}
denotes a many-fingered time derivative. Equation (\ref{Sch_corr_1}) is the
main result of ref. \cite{KS91}.\footnote{As noted by Kiefer and Singh the
correction terms in (\ref{Sch_corr_1}) turn out to be independent of the
factor ordering chosen in the gravitational part of the Wheeler-DeWitt
equation.}

Thus we have%
\begin{equation}
i\frac{\partial\psi^{(1)}}{\partial t}=\int d^{3}x\ \left(  \mathcal{\hat{H}%
}_{\phi}+\mathcal{\hat{H}}_{a}+i\mathcal{\hat{H}}_{b}\right)  \psi^{(1)}\ ,
\label{Sch_correcns}%
\end{equation}
where%
\begin{equation}
\mathcal{\hat{H}}_{a}=\frac{1}{8\mu}\frac{1}{\sqrt{g}R}\mathcal{\hat{H}}%
_{\phi}^{2}%
\end{equation}
and%
\begin{equation}
\mathcal{\hat{H}}_{b}=\frac{1}{8\mu}\frac{\delta}{\delta\tau}\left(
\frac{\mathcal{\hat{H}}_{\phi}}{\sqrt{g}R}\right)
\end{equation}
are both Hermitian operators.

At this order the total effective Hamiltonian takes the form%
\begin{equation}
\hat{H}=\hat{H}_{\phi}+\hat{H}_{a}+i\hat{H}_{b}\ , \label{Ham_eff}%
\end{equation}
where%
\begin{equation}
\hat{H}_{\phi}=\int d^{3}x\ \mathcal{\hat{H}}_{\phi}\ ,\ \ \ \hat{H}_{a}=\int
d^{3}x\ \mathcal{\hat{H}}_{a}\ ,\ \ \ \hat{H}_{b}=\int d^{3}x\ \mathcal{\hat
{H}}_{b}%
\end{equation}
are Hermitian operators. The Hamiltonian has a Hermitian correction $\hat
{H}_{a}$ together with a non-Hermitian correction $i\hat{H}_{b}$.

Similar corrections have been derived and discussed for minisuperspace models
of quantum cosmology [69--71]. As in the above discussion, it is found that
the effective wave function of the matter field $\phi$ obeys a
Schr\"{o}dinger-like equation with a corrected Hamiltonian of the form
(\ref{Ham_eff}), where again $\hat{H}_{\phi}$ is the field (or `matter')
Hamiltonian, the first gravitational correction $\hat{H}_{a}$ is Hermitian,
while the second gravitational correction $i\hat{H}_{b}$ is non-Hermitian.
Several studies have included the term $\hat{H}_{a}$ in models of inflationary
cosmology and have calculated its effect on the primordial power spectrum
[69--71]. It is found that $\hat{H}_{a}$ generates a correction to the
spectrum at large scales, which is however too small to be observable. In
these studies the unitarity-violating term $i\hat{H}_{b}$ is neglected (see
Section 9.3).

The approximate ratio of the non-Hermitian correction $i\hat{H}_{b}$ to the
Hermitian correction $\hat{H}_{a}$ can be easily estimated for an expanding
cosmological background with scale factor $a$. The ratio is, schematically, of
order%
\begin{equation}
\frac{1}{\hat{H}_{\phi}^{2}}\frac{d\hat{H}_{\phi}}{dt}\sim\frac{1}{\hat
{H}_{\phi}^{2}}\frac{d\hat{H}_{\phi}}{da}\dot{a}\sim\frac{H}{E}\ ,
\label{ratio}%
\end{equation}
where $H=\dot{a}/a$ is the Hubble parameter and $E$ is a typical energy scale
for the matter field \cite{KS91}. This ratio is exceedingly small during the
late cosmological era, but it might be large in the very early universe where
$H$ can be large for small $t$ (for example $H=1/2t$ for a radiation-dominated
expansion with $a\propto t^{1/2}$). As we shall see in Section 11, the
non-Hermitian term can be large in the final stages of black-hole evaporation.

\subsection{Origin and status of non-Hermitian corrections}

The origin of the non-Hermitian or unitarity-violating correction $i\hat
{H}_{b}$ may be understood as follows. The full Wheeler-DeWitt equation
implies the conservation of a Klein-Gordon-like current and not of a
Schr\"{o}dinger-like current (as discussed in Section 3.2.1) \cite{DeW67}.
Expanding the Klein-Gordon-like continuity equation implies a
Schr\"{o}dinger-like continuity equation that is modified by small
gravitational corrections in such a way that the usual Schr\"{o}dinger norm is
no longer conserved ($\frac{d}{dt}\int D\phi\ \psi^{\ast}\psi\neq0$)
\cite{K94}. Thus mathematically the appearance of the non-Hermitian correction
$i\hat{H}_{b}$ is readily understandable. Even so, its physical status has
been disputed, because in its presence the standard probabilistic
interpretation of quantum mechanics breaks down. However, as we will see in
Section 6, in pilot-wave theory a non-Hermitian Hamiltonian does not generate
any inconsistency: probability is still fully conserved but the Born rule is
no longer stable.

We have seen that, in pilot-wave quantum gravity, there is no physical
Born-rule state in the deep quantum-gravity regime, and the Born rule emerges
only in a semiclassical Schr\"{o}dinger approximation. It then seems
unsurprising to find that quantum-gravitational corrections to the
Schr\"{o}dinger equation can yield an intermediate regime in which the Born
rule suffers a small instability. In fact it would be reasonable to expect
such a regime on general physical grounds. Thus in pilot-wave theory there
seems to be no reason to be alarmed by the appearance of small non-Hermitian
corrections in the effective Hamiltonian.

From the viewpoint of standard quantum mechanics, however, the non-Hermitian
terms appear unphysical and might be regarded as artifacts. Similar terms are
found if one carries out an analogous approximation scheme in scalar QED: one
may derive an approximate Schr\"{o}dinger equation for a charged scalar field
in a background electromagnetic field, and corrections to that equation
contain similar non-Hermitian terms \cite{KS91}. But the analogy with scalar
QED is misleading. In the latter theory there is a conserved
Schr\"{o}dinger-like current in the full configuration space, and so at the
fundamental level there is a well-defined Born rule from which one may derive
a marginal or conditional distribution for the matter field. In contrast, in
the gravitational case there is no underlying conserved Schr\"{o}dinger
current. On the contrary, the Wheeler-DeWitt equation possesses a conserved
Klein-Gordon-like current whose density is not positive-definite. There
appears to be no well-defined Born-rule distribution at the fundamental level,
and therefore it is no longer obvious that a small violation of unitarity in
the emergent Schr\"{o}dinger regime is necessarily a mathematical artifact.

Some authors working with standard quantum mechanics have proposed to remove
the non-Hermitian terms by an appropriate redefinition of the effective wave
function. To see the general method, consider a Schr\"{o}dinger equation
$i\partial\psi/\partial t=\hat{H}\psi$ with Hamiltonian $\hat{H}=\hat{H}%
_{0}+\Delta\hat{H}$, where the correction term $\Delta\hat{H}$ may or may not
be Hermitian. The term $\Delta\hat{H}$ can be eliminated by redefining the
original wave function $\psi(q,t)$ as%
\begin{equation}
\psi^{\prime}(q,t)=\exp\left(  i\int_{0}^{t}d\tau\ \Delta\hat{H}(\tau)\right)
\psi(q,t) \label{newpsi}%
\end{equation}
(where we assume for simplicity that $\Delta\hat{H}(t)$ commutes with $\hat
{H}_{0}$ and with itself at different times). The new wave function
$\psi^{\prime}$ then satisfies a modified Schr\"{o}dinger equation
$i\partial\psi^{\prime}/\partial t=\hat{H}_{0}\psi^{\prime}$ with Hamiltonian
$\hat{H}=\hat{H}_{0}$ only. Normally this procedure would be regarded as
illegitimate: the new wave function would not be the correct physical wave
function, and in particular $\left\vert \psi^{\prime}\right\vert ^{2}$ would
not be the correct measured probability density for $q$. One cannot simply
remove terms from the Hamiltonian by redefining the wave function.

Even so a similar procedure has been considered by some authors in order to
remove the unwanted non-Hermitian terms. In a quantum-cosmological model Bini
\textit{et al}. \cite{Bini13} show how the term $i\hat{H}_{b}$ can be
eliminated by redefining the wave function in the general form (\ref{newpsi}).
However, as noted by the authors, it is not clear which of the two wave
functions (the original or the new) should be regarded as physically correct
in the sense associated with the Born rule. More recently, a general scheme
for eliminating the non-Hermitian terms has been developed by Kiefer and
Wichmann \cite{KW18}. Their starting point is the observation that the WKB
ansatz $\Psi\lbrack g_{ij},\phi]\approx\Psi_{\mathrm{WKB}}[g_{ij}]\psi
\lbrack\phi,g_{ij}]$ (equation (\ref{WKB})) is invariant under a rescaling%
\begin{equation}
\Psi_{\mathrm{WKB}}\rightarrow e^{A}\Psi_{\mathrm{WKB}}\ ,\ \ \ \psi
\rightarrow e^{-A}\psi\ ,
\end{equation}
where $A[g_{ij}]$ is an arbitrary complex functional analogous to a choice of
`gauge'. Different choices of $A$ will affect the time evolution of $\psi$.
Kiefer and Wichmann obtain a general effective Schr\"{o}dinger equation for
$\psi$ whose Hamiltonian $\hat{H}$ is defined only implicitly. As before
non-Hermitian terms make an appearance. These cannot be eliminated by a simple
redefinition of the form (\ref{newpsi}) as the Hamiltonian is not known
explicitly. However, solving the equations iteratively, an appropriate
redefinition of the form (\ref{newpsi}) can be imposed order by order in such
a way that at each order the redefined wave functions $\psi_{0}^{\prime}$,
$\psi_{1}^{\prime}$, $\psi_{2}^{\prime}$, ... evolve unitarily (with respect
to the usual Schr\"{o}dinger norm). The authors draw an analogy with the
Born-Oppenheimer ansatz for the Schr\"{o}dinger equation as employed in
molecular physics.\footnote{In this context see also refs.
\cite{Chat19,ChatK21}.} But again, as the authors themselves note, the analogy
is limited because the Wheeler-DeWitt equation is fundamentally different from
the Schr\"{o}dinger equation (possessing in particular a quite different form
of conserved current).

At this point we must address a difficult question. Does the method of ref.
\cite{KW18} correctly identify the true physical wave functions $\psi^{\prime
}$ (with purely Hermitian Hamiltonians), or are the wave function
redefinitions an elaborate means of disguising the unwanted but physical
non-Hermitian terms? The correct answer is not known for certain. Our
experience with time-dependent wave functions on a classical spacetime
background tells us that the Hamiltonian should be Hermitian, but that is in
conditions where quantum-gravitational effects can be neglected. Perhaps the
non-Hermitian terms are a natural consequence of quantum gravity (where as we
shall see those terms can be easily understood in pilot-wave theory) and the
modified expansions with wave function redefinitions are really an artificial
means of disguising genuine physical effects. At issue here is exactly how to
identify the familiar Schr\"{o}dinger-regime wave functions as emergent
features of the semiclassical regime. It may be that the question can be
answered only by experiment. Without some empirical input it is difficult to
know exactly how the wave functions we observe in the laboratory are to be
identified from the quantum-gravitational formalism.

Of course in standard quantum mechanics there is good reason to redefine the
effective wave functions so as to eliminate the non-Hermitian terms. But as we
shall see that motivation is lost in pilot-wave theory where such terms are
perfectly consistent with the conservation of probability. Furthermore, again,
given our argument that there is no fundamental Born rule in quantum gravity,
it appears quite plausible on general physical grounds that there will be an
intermediate regime with a small instability of the Born rule. Ultimately,
however, the only way to resolve the dispute will probably be through
experiment. In the rest of this paper we will explore the consequences of the
non-Hermitian terms in pilot-wave theory, with a view to possible experimental tests.

\subsection{Absence of corrections to the de Broglie velocity}

We now continue the analysis of Section 3.4.2 and consider the next order in
the semiclassical expansion (\ref{B-O}) of the de Broglie guidance equation
(\ref{deB2}) for the matter field. Noting that the term $\sigma_{2}$ in
(\ref{split_S_2}) satisfies $\delta\sigma_{2}/\delta\phi=0$, the de Broglie
velocity (\ref{phidot2}) takes the form%
\begin{equation}
\frac{\partial\phi}{\partial t}=\frac{N}{\sqrt{g}}\frac{\delta}{\delta\phi
}\left(  \operatorname{Re}S_{1}+\mu^{-1}\operatorname{Re}\eta+...\right)  \ .
\label{phidot3}%
\end{equation}
Equation (\ref{psi_(1)}) defines a corrected wave function $\psi^{(1)}%
=\psi^{(0)}\exp(i\eta/\mu)$, satisfying a corrected Schr\"{o}dinger equation
(\ref{Sch_correcns}). The total phase of the wave function now includes a
correction equal to $\mu^{-1}\operatorname{Re}\eta$ (in addition to the
uncorrected phase $\operatorname{Re}S_{1}$ of $\psi^{(0)}$). Thus the
corrected wave function $\psi^{(1)}$ has a total phase%
\begin{equation}
\operatorname{Im}\ln\psi^{(1)}=\operatorname{Re}S_{1}+\mu^{-1}%
\operatorname{Re}\eta\ .
\end{equation}
The corrected canonical de Broglie velocity (\ref{phidot3}) can then be
written as%
\begin{equation}
\left(  \frac{\partial\phi}{\partial t}\right)  ^{(1)}=\frac{N}{\sqrt{g}}%
\frac{\delta S^{(1)}}{\delta\phi}\ , \label{deB_no correcn}%
\end{equation}
where $S^{(1)}=\operatorname{Im}\ln\psi^{(1)}$ is the phase of $\psi^{(1)}$.

We arrive at an important conclusion. Despite the presence of the
non-Hermitian term in the Schr\"{o}dinger equation (\ref{Sch_correcns}), the
de Broglie velocity of the matter field is still given by the standard
guidance equation (\ref{deB2}) with the phase $S$ equal to the corrected phase
$S^{(1)}=\operatorname{Im}\ln\psi^{(1)}$ of the effective wave function
$\psi^{(1)}$. This is just the usual guidance equation that we would normally
associate with the Hermitian part of the Hamiltonian. In other words: while
the non-Hermitian term in (\ref{Sch_correcns}) affects the time evolution of
$\psi^{(1)}$ (and hence indirectly affects the trajectories), the presence of
the non-Hermitian term does not change the form of the guidance equation. As
we shall now see, this implies that the Born rule for the matter field is unstable.

Thus, if the semiclassical expansion is to be trusted, the fundamental
equations of pilot-wave quantum gravity imply the instability of the Born rule
-- as described by quantum-gravitational corrections to the Schr\"{o}dinger approximation.

\section{Pilot-wave theory with an unstable Born rule}

By performing a semiclassical expansion of the Wheeler-DeWitt equation
together with the associated de Broglie guidance equation, we have derived an
effective pilot-wave theory of a matter field on a classical background. The
effective Schr\"{o}dinger equation (\ref{Sch_correcns}) contains a
non-Hermitian contribution to the Hamiltonian, while the effective de Broglie
guidance equation (\ref{deB_no correcn}) continues to take the standard form
(associated with only the Hermitian part). Before proceeding it will be
helpful to understand the implications of these results in general terms.

\subsection{Pilot-wave dynamics with a non-Hermitian Hamiltonian}

Let us consider how pilot-wave theory can be applied to a general system with
a Hamiltonian%
\begin{equation}
\hat{H}=\hat{H}_{1}+i\hat{H}_{2}\ , \label{Ham_split}%
\end{equation}
where $\hat{H}_{1}$ and $\hat{H}_{2}$ are both Hermitian.\footnote{Any
operator $\hat{H}$ can of course be decomposed into a sum of a Hermitian part
$\frac{1}{2}(\hat{H}+\hat{H}^{\dag})$ and an anti-Hermitian part $\frac{1}%
{2}(\hat{H}-\hat{H}^{\dag})$.} The wave function $\psi(q,t)$ will evolve
according to the Schr\"{o}dinger equation%
\begin{equation}
i\frac{\partial\psi}{\partial t}=(\hat{H}_{1}+i\hat{H}_{2})\psi\ .
\label{Sch_nonHerm}%
\end{equation}
This implies a continuity equation%
\begin{equation}
\frac{\partial\left\vert \psi\right\vert ^{2}}{\partial t}+\partial_{q}\cdot
j_{1}=s\ , \label{cont_psi2_s}%
\end{equation}
where $j_{1}$ is the standard current associated with $\hat{H}_{1}$
(satisfying $\partial_{q}\cdot j_{1}=2\operatorname{Re}\left(  i\psi^{\ast
}\hat{H}_{1}\psi\right)  $) and%
\begin{equation}
s=2\operatorname{Re}\left(  \psi^{\ast}\hat{H}_{2}\psi\right)  \label{s_def}%
\end{equation}
is an effective `source' term.

Integrating (\ref{cont_psi2_s}) over configuration space we find%
\begin{equation}
\frac{d}{dt}\int dq\ \left\vert \psi\right\vert ^{2}=\int dq\ s=2\left\langle
\hat{H}_{2}\right\rangle \ , \label{dnormdt_result}%
\end{equation}
where the standard quantum expectation value $\left\langle \hat{H}%
_{2}\right\rangle \equiv\int dq\ \left(  \psi^{\ast}\hat{H}_{2}\psi\right)  $
is real for Hermitian $\hat{H}_{2}$ and we assume as usual that $j_{1}$
vanishes at infinity.

We can then define a de Broglie velocity field%
\begin{equation}
v=\frac{j_{1}}{\left\vert \psi\right\vert ^{2}}\ . \label{v_Herm}%
\end{equation}
This amounts to assuming that the velocity is generated by the Hermitian part
of the Hamiltonian only. Note, however, that $j_{1}$ is a functional of $\psi$
whose time evolution is determined via (\ref{Sch_nonHerm}) by the total
Hamiltonian $\hat{H}$ (including the non-Hermitian part).

Equations (\ref{Sch_nonHerm}) and (\ref{v_Herm}) define a pilot-wave dynamics
for an individual system whose Hamiltonian has a non-Hermitian part.
Remarkably, as we have seen, a dynamics of precisely this type emerges
naturally in an appropriate limit of quantum gravity.

\subsection{Instability of the Born rule}

For an ensemble of systems with the same wave function $\psi$, the probability
density $\rho(q,t)$ will satisfy%
\begin{equation}
\frac{\partial\rho}{\partial t}+\partial_{q}\cdot(\rho v)=0
\label{cont_v_Herm}%
\end{equation}
(since each system follows the velocity field $v$). Equations
(\ref{Sch_nonHerm}), (\ref{v_Herm}) and (\ref{cont_v_Herm}) then define a
pilot-wave theory of an ensemble with an unstable Born rule.

To see this, rewrite (\ref{cont_psi2_s}) as%
\begin{equation}
\frac{\partial\left\vert \psi\right\vert ^{2}}{\partial t}+\partial_{q}%
\cdot(\left\vert \psi\right\vert ^{2}v)=s\ . \label{cont_psi2_s_2}%
\end{equation}
From (\ref{cont_v_Herm}) and (\ref{cont_psi2_s_2}) it follows that the ratio%
\begin{equation}
f\equiv\rho/|\psi|^{2}%
\end{equation}
is \textit{not} conserved along trajectories. Instead we find%
\begin{equation}
\frac{df}{dt}=-\frac{s}{|\psi|^{2}}f \label{dfdt}%
\end{equation}
(where again $d/dt=\partial/\partial t+v\cdot\partial_{q}$ is the time
derivative along a trajectory). An initial distribution with $f=1$ everywhere
will generally evolve into a final distribution with $f\neq1$. In other words,
an initial Born-rule distribution $\rho=|\psi|^{2}$ at $t=t_{i}$ generally
evolves into a non-Born-rule distribution $\rho\neq|\psi|^{2}$ at $t>t_{i}$.

Thus (\ref{Sch_nonHerm}), (\ref{v_Herm}) and (\ref{cont_v_Herm}) indeed define
a pilot-wave dynamics of ensembles in which the Born-rule distribution
$\rho=\left\vert \psi\right\vert ^{2}$ is unstable.

\subsection{Timescale for quantum instability. Condition for the growth of
quantum nonequilibrium}

In such a theory quantum nonequilibrium is created on a timescale
$\tau_{\mathrm{noneq}}$ which can be estimated from the rate of change of the
$H$-function (\ref{Hfn}). Taking the time derivative of (\ref{Hfn}) and using
(\ref{cont_v_Herm}) and (\ref{cont_psi2_s_2}) it is easy to show
that\footnote{If $\rho$ and $\left\vert \psi\right\vert ^{2}$ obey the same
continuity equation the exact $H$ is constant in time while the coarse-grained
value tends to decrease (if there is no initial fine-grained structure). But
here the continuity equation (\ref{cont_psi2_s_2}) for $\left\vert
\psi\right\vert ^{2}$ contains a source term $s$ and so the exact $H$ is no
longer constant.}%
\begin{equation}
\frac{dH}{dt}=-\int dq\ \frac{\rho}{\left\vert \psi\right\vert ^{2}}s\ .
\end{equation}
If we are close to equilibrium ($\rho\approx\left\vert \psi\right\vert ^{2}$)
we have%
\begin{equation}
\frac{dH}{dt}\approx-\int dq\ s=-2\left\langle \hat{H}_{2}\right\rangle \ .
\end{equation}
We can then define $\tau_{\mathrm{noneq}}$ as the time required for $H$ to
change by a factor of order unity,%
\begin{equation}
\tau_{\mathrm{noneq}}\left\vert \frac{dH}{dt}\right\vert \approx1\ ,
\label{tau_defn1}%
\end{equation}
yielding the estimate%
\begin{equation}
\tau_{\mathrm{noneq}}\approx\frac{1}{2\left\vert \left\langle \hat{H}%
_{2}\right\rangle \right\vert }\ . \label{tau_est}%
\end{equation}

We might instead estimate $\tau_{\mathrm{noneq}}$ from the rate of change of
the squared-norm $\int dq\ \left\vert \psi\right\vert ^{2}$ by defining%
\begin{equation}
\frac{1}{\tau_{\mathrm{noneq}}}=\frac{1}{\int dq\ \left\vert \psi\right\vert
^{2}}\left\vert \frac{d}{dt}\int dq\ \left\vert \psi\right\vert ^{2}%
\right\vert =\left\vert \frac{d}{dt}\ln\left(  \int dq\ \left\vert
\psi\right\vert ^{2}\right)  \right\vert \ . \label{tau_defn2}%
\end{equation}
Close to equilibrium we have $\int dq\ \left\vert \psi\right\vert ^{2}%
\approx1$ and so from (\ref{dnormdt_result}) we find the same estimate
(\ref{tau_est}).

We saw in Section 4.1 that, for general unnormalised wave functions,%
\[
H\geq-\ln\left(  \int dq\ \left\vert \psi\right\vert ^{2}\right)  \ .
\]
It is then not surprising that the definitions (\ref{tau_defn1}) and
(\ref{tau_defn2}) yield essentially the same timescales.

It must be noted that, for realistic systems, there will usually be two
competing effects: the creation of quantum nonequilibrium on a timescale
$\tau_{\mathrm{noneq}}$ and the relaxation of quantum nonequilibrium on a
timescale $\tau_{\mathrm{relax}}$. The quantum-gravitationally generated
violations of the Born rule can build up over time only if the usual quantum
relaxation process is relatively negligible, that is, only if%
\begin{equation}
\tau_{\mathrm{relax}}>\tau_{\mathrm{noneq}}\ . \label{condn}%
\end{equation}
In other words, quantum nonequilibrium must be generated faster than
relaxation can remove it. As we will see the condition (\ref{condn}) can be
realistically satisfied for a scalar field during inflation (Section 8).
Whether or not it can be satisfied in other circumstances (such as black-hole
evaporation, Section 11) will depend on the results of more detailed modelling
including quantum relaxation.

\section{Quantum cosmology and the Born rule}

Brizuela, Kiefer and Kr\"{a}mer \cite{BKM16} have developed a semiclassical
expansion of the Wheeler-DeWitt equation for a minisuperspace
quantum-cosmological model and have used it to derive quantum-gravitational
corrections to the Schr\"{o}dinger regime for perturbations on a classical
background spacetime. The method is similar to that presented in Sections
3.4.1 and 5.1 for the full superspace, though there are some differences. We
begin by summarising their model and results, adapted to pilot-wave theory,
where again both Hermitian and non-Hermitian corrections appear in the
effective Schr\"{o}dinger equation. We then study the de Broglie velocity and
show that it remains uncorrected at the relevant order. It is then instructive
to consider again why the theory has no fundamental equilibrium state, and to
reconsider the emergence and instability of the Born rule in this
quantum-cosmological context. Finally, we write down a simplified model for
the slow-roll limit, with a view to enabling more tractable calculations.

\subsection{Pilot-wave quantum cosmology}

In the model of ref. \cite{BKM16} both scalar and tensor perturbations
propagate on a background flat Friedmann-Lema\^{\i}tre universe, where the
background is also treated quantum-mechanically. For our purposes we can
ignore the tensor contributions. At the classical level, in terms of conformal
time $\eta$ the background line element is%
\begin{equation}
d\tau^{2}=a^{2}(d\eta^{2}-d\mathbf{x}^{2})\,,
\end{equation}
where $\eta$ is related to the usual cosmic time $t$ by $d\eta/dt=a^{-1}$. We
employ the common convention whereby $\lambda$ denotes a comoving wavelength
at some reference time $\eta_{0}$ when $a_{0}=1$. The physical wavelength of a
field mode at time $\eta$ is then given by $\lambda_{\mathrm{phys}}%
(\eta)=a(\eta)\lambda$.

The background contains a homogeneous scalar field $\phi$ with a potential
$\mathcal{V}(\phi)$ that drives the expansion (here $\phi$ is the homogenous
part of a massive and minimally-coupled inflaton field). Using primes to
denote derivatives with respect to $\eta$, the classical action for the
background is taken to be%
\begin{equation}
S_{\mathrm{b}}=\frac{1}{2}\int d\eta\ \mathfrak{L}^{3}\left[  -\frac{3}{4\pi
G}(a^{\prime})^{2}+a^{2}(\phi^{\prime})^{2}-2a^{4}\mathcal{V}\right]  \ ,
\label{action_0}%
\end{equation}
where $\mathfrak{L}$ is an arbitrary length scale associated with the spatial
integration (to be regarded as an infrared cutoff) \cite{Vent14}. The action
(\ref{action_0}) implies the standard equations of motion%
\begin{align*}
\frac{3}{4\pi G}\frac{a^{\prime\prime}}{a}+(\phi^{\prime})^{2}-4a^{2}%
\mathcal{V}  &  =0\ ,\\
\phi^{\prime\prime}+2\frac{a^{\prime}}{a}\phi^{\prime}+a^{2}\frac
{d\mathcal{V}}{d\phi}  &  =0\ ,
\end{align*}
which in terms of $t$ take the more familiar forms%
\begin{align*}
\frac{\ddot{a}}{a}  &  =-\frac{4\pi G}{3}\left(  \rho+3p\right)  \ ,\\
\ddot{\phi}+3\frac{\dot{a}}{a}\dot{\phi}+\frac{d\mathcal{V}}{d\phi}  &  =0\ ,
\end{align*}
where%
\begin{equation}
\rho=\tfrac{1}{2}\dot{\phi}^{2}+\mathcal{V} \label{en_dens_phi}%
\end{equation}
is the energy density and $p=\rho-2\mathcal{V}$ is the pressure.

Scalar perturbations of the background metric are usually described by scalar
functions $A$, $B$, $\psi$ and $E$:%
\[
d\tau^{2}=a^{2}\left[  (1-2A)d\eta^{2}-2(\partial_{i}B)dx^{i}d\eta-\left(
(1-2\psi)\delta_{ij}+2\partial_{i}\partial_{j}E\right)  dx^{i}dx^{j}\right]
\ .
\]
It is convenient to combine the inflaton perturbation $\varphi\equiv\delta
\phi$ with $A$, $B$, $\psi$, $E$ to form the Mukhanov-Sasaki variable
\cite{MFB92}%
\begin{equation}
\upsilon=a\left(  \tilde{\varphi}+\phi^{\prime}\frac{\Phi_{\mathrm{B}}%
}{\mathcal{H}}\right)  \ , \label{M-S}%
\end{equation}
where%
\begin{align}
\tilde{\varphi}  &  \equiv\varphi+\phi^{\prime}(B-E^{\prime})\ ,\\
\Phi_{\mathrm{B}}  &  \equiv A+\frac{1}{a}\left[  a(B-E^{\prime})\right]
^{\prime}%
\end{align}
and%
\begin{equation}
\mathcal{H}\equiv a^{\prime}/a=Ha
\end{equation}
(where $H=\dot{a}/a$ is the usual Hubble parameter).

The action $S_{\mathrm{p}}$ of the perturbations is found by expanding the
total (including the Einstein-Hilbert)\ action around the background. In terms
of $\upsilon$,%
\begin{equation}
S_{\mathrm{p}}=\frac{1}{2}\int d\eta d^{3}\mathbf{x}\ \left[  (\upsilon
^{\prime})^{2}-\delta^{ij}\partial_{i}\upsilon\partial_{j}\upsilon
+\frac{z^{\prime\prime}}{z}\upsilon^{2}\right]  \ ,
\end{equation}
where%
\begin{equation}
z=a\sqrt{\epsilon}%
\end{equation}
and%
\begin{equation}
\epsilon=-\frac{\dot{H}}{H^{2}}=1-\frac{\mathcal{H}^{\prime}}{\mathcal{H}^{2}}%
\end{equation}
(the first slow-roll parameter). Brizuela \textit{et al}. work with a Fourier
transform%
\begin{equation}
\upsilon(\eta,\mathbf{x})=\int\frac{d^{3}\mathbf{k}}{(2\pi)^{3/2}}%
\upsilon_{\mathbf{k}}(\eta)e^{i\mathbf{k}\cdot\mathbf{x}}\ .
\end{equation}
The wave vector $\mathbf{k}$ is then defined so that the magnitude $k$ is
equal to the inverse of a wavelength (without the usual factor $2\pi$).
Because $\upsilon$ is real, $\upsilon_{\mathbf{k}}^{\ast}=\upsilon
_{-\mathbf{k}}$. In terms of the $\upsilon_{\mathbf{k}}$'s we then have%
\[
S_{\mathrm{p}}=\frac{1}{2}\int d\eta\int d^{3}\mathbf{k}\ \left[
\upsilon_{\mathbf{k}}^{\prime}\upsilon_{\mathbf{k}}^{\ast\prime}%
+\upsilon_{\mathbf{k}}\upsilon_{\mathbf{k}}^{\ast}\left(  \frac{z^{\prime
\prime}}{z}-k^{2}\right)  \right]  \ .
\]
At this point the modes are discretised by the replacement%
\begin{equation}
\int d^{3}\mathbf{k}\rightarrow\frac{1}{\mathfrak{L}^{3}}\sum_{\mathbf{k}}\ .
\end{equation}
In addition, for simplicity, the $\upsilon_{\mathbf{k}}$'s are treated as real
(a proper procedure defines a new set of real variables but gives the same
results). The action of the perturbations is then%
\begin{equation}
S_{\mathrm{p}}=\frac{1}{2}\int d\eta\frac{1}{\mathfrak{L}^{3}}\sum
_{\mathbf{k}}\left[  (\upsilon_{\mathbf{k}}^{\prime})^{2}+\upsilon
_{\mathbf{k}}^{2}\left(  \frac{z^{\prime\prime}}{z}-k^{2}\right)  \right]  \ .
\label{action_p}%
\end{equation}

Following ref. \cite{Vent14}, Brizuela \textit{et al}. eliminate the
appearance of $\mathfrak{L}$ in the equations by appropriate rescalings:%
\begin{equation}
a_{\mathrm{new}}=a_{\mathrm{old}}\mathfrak{L}\ ,\ \eta_{\mathrm{new}}%
=\eta_{\mathrm{old}}/\mathfrak{L}\ ,\ \upsilon_{\mathrm{new}}=\upsilon
_{\mathrm{old}}/\mathfrak{L}^{2}\ ,\ k_{\mathrm{new}}=k_{\mathrm{old}%
}\mathfrak{L\ .} \label{rescale}%
\end{equation}
In terms of these rescaled variables, the background has a Lagrangian%
\[
L_{\mathrm{b}}=\frac{1}{2}\left[  -\frac{3}{4\pi G}(a^{\prime})^{2}+a^{2}%
(\phi^{\prime})^{2}-2a^{4}\mathcal{V}\right]
\]
(where $S_{\mathrm{b}}=\int d\eta\ L_{\mathrm{b}}$) and canonical momenta%
\begin{equation}
\pi_{a}=-\frac{3}{4\pi G}a^{\prime}\ \ \ ,\ \pi_{\phi}=a^{2}\phi^{\prime}\ ,
\label{canmomb}%
\end{equation}
while the perturbations have a Lagrangian%
\begin{equation}
L_{\mathrm{p}}=\frac{1}{2}\sum_{\mathbf{k}}\left[  (\upsilon_{\mathbf{k}%
}^{\prime})^{2}+(\upsilon_{\mathbf{k}})^{2}\left(  \frac{z^{\prime\prime}}%
{z}-k^{2}\right)  \right]
\end{equation}
(with $S_{\mathrm{p}}=\int d\eta\ L_{\mathrm{p}}$) and canonical momenta%
\begin{equation}
\pi_{\mathbf{k}}=\upsilon_{\mathbf{k}}^{\prime}\ . \label{canmomp}%
\end{equation}
The Hamiltonian of the background is then%
\begin{equation}
H_{\mathrm{b}}=-\frac{2\pi G}{3}\pi_{a}^{2}+\frac{1}{2a^{2}}\pi_{\phi}%
^{2}+a^{4}\mathcal{V}\ ,
\end{equation}
(noting the crucial sign difference between the terms in $\pi_{a}^{2}$ and
$\pi_{\phi}^{2}$), while the Hamiltonian of the perturbations is%
\begin{equation}
H_{\mathrm{p}}=\frac{1}{2}\sum_{\mathbf{k}}\left[  \pi_{\mathbf{k}}%
^{2}+\upsilon_{\mathbf{k}}^{2}\left(  k^{2}-\frac{z^{\prime\prime}}{z}\right)
\right]  \ .
\end{equation}

The system is then quantised. We have a Wheeler-DeWitt equation%
\[
\left(  \hat{H}_{\mathrm{b}}+\hat{H}_{\mathrm{p}}\right)  \Psi=0
\]
for a wave functional $\Psi=\Psi(a,\phi,\{\upsilon_{\mathbf{k}}\})$. The sign
difference between the terms in $\pi_{a}^{2}$ and $\pi_{\phi}^{2}$ in the
background Hamiltonian $\hat{H}_{\mathrm{b}}$ signals the Klein-Gordon-like
character of the wave equation.

Taking a product ansatz \cite{KK12}%
\begin{equation}
\Psi=\Psi_{0}(a,\phi)\prod\limits_{\mathbf{k}}\tilde{\Psi}_{\mathbf{k}}%
(a,\phi,\upsilon_{\mathbf{k}})\ ,
\end{equation}
where $\mathcal{\hat{H}}_{\mathrm{b}}\Psi_{0}=0$ defines a wave functional
$\Psi_{0}$ for the background, Brizuela \textit{et al}. define a wave function%
\begin{equation}
\Psi_{\mathbf{k}}=\Psi_{\mathbf{k}}(a,\phi,\upsilon_{\mathbf{k}})=\Psi
_{0}(a,\phi)\tilde{\Psi}_{\mathbf{k}}(a,\phi,\upsilon_{\mathbf{k}})
\end{equation}
for each mode $\mathbf{k}$, where $\Psi_{\mathbf{k}}$ obeys a Wheeler-DeWitt
equation%
\begin{equation}
\frac{1}{2m_{\mathrm{P}}^{2}}\frac{1}{a}\frac{\partial}{\partial a}\left(
a\frac{\partial\Psi_{\mathbf{k}}}{\partial a}\right)  -\frac{1}{2a^{2}}%
\frac{\partial^{2}\Psi_{\mathbf{k}}}{\partial\phi^{2}}+\frac{1}{2}%
m_{\mathrm{P}}^{2}V\Psi_{\mathbf{k}}-\frac{1}{2}\frac{\partial^{2}%
\Psi_{\mathbf{k}}}{\partial\upsilon_{\mathbf{k}}^{2}}+\frac{1}{2}%
\omega_{\mathbf{k}}^{2}\upsilon_{\mathbf{k}}^{2}\Psi_{\mathbf{k}}=0
\label{WD_mode_1}%
\end{equation}
(with appropriate factor ordering), where%
\begin{equation}
m_{\mathrm{P}}^{2}=3/4\pi G
\end{equation}
is a rescaled Planck mass,%
\begin{equation}
V=\frac{2}{m_{\mathrm{P}}^{2}}a^{4}\mathcal{V} \label{auxptl}%
\end{equation}
is an auxiliary potential, and%
\begin{equation}
\omega_{\mathbf{k}}^{2}=k^{2}-\frac{z^{\prime\prime}}{z}%
\end{equation}
is a time-dependent frequency.

It is now straightforward to define a corresponding pilot-wave theory. We can
identify de Broglie velocities for the evolving variables $a(\eta)$,
$\phi(\eta)$ and $\upsilon_{\mathbf{k}}(\eta)$ by setting%
\begin{equation}
\pi_{a}=\frac{\partial S_{\mathbf{k}}}{\partial a}\ \ \ ,\ \pi_{\phi}%
=\frac{\partial S_{\mathbf{k}}}{\partial\phi}\ \ \ ,\ \pi_{\mathbf{k}}%
=\frac{\partial S_{\mathbf{k}}}{\partial\upsilon_{\mathbf{k}}}\ ,
\end{equation}
where $S_{\mathbf{k}}=\operatorname{Im}\ln\Psi_{\mathbf{k}}$ is the phase of
$\Psi_{\mathbf{k}}$. From the classical expressions (\ref{canmomb}) and
(\ref{canmomp}) we then have the de Broglie guidance equations%
\begin{equation}
\frac{da}{d\eta}=-\frac{1}{m_{\mathrm{P}}^{2}}\frac{\partial S_{\mathbf{k}}%
}{\partial a}\ \ \ ,\ \frac{d\phi}{d\eta}=\frac{1}{a^{2}}\frac{\partial
S_{\mathbf{k}}}{\partial\phi} \label{deB_qu_cosm}%
\end{equation}
for the background and%
\begin{equation}
\frac{d\upsilon_{\mathbf{k}}}{d\eta}=\frac{\partial S_{\mathbf{k}}}%
{\partial\upsilon_{\mathbf{k}}} \label{deB_pertns}%
\end{equation}
for the perturbations. Together with the wave equation (\ref{WD_mode_1}) these
define a pilot-wave dynamics for the variables $a$, $\phi$ and $\upsilon
_{\mathbf{k}}$.

It is instructive to check that the expressions (\ref{deB_qu_cosm}) and
(\ref{deB_pertns}) agree with the velocities obtained from the polar
decomposition $\Psi_{\mathbf{k}}=\left\vert \Psi_{\mathbf{k}}\right\vert
e^{iS_{\mathbf{k}}}$ of the wave equation (\ref{WD_mode_1}). This is done by
inserting $\Psi_{\mathbf{k}}=\left\vert \Psi_{\mathbf{k}}\right\vert
e^{iS_{\mathbf{k}}}$ into (\ref{WD_mode_1}) and taking real and imaginary
parts. The real part can be written as a modified Hamilton-Jacobi equation%
\[
-\frac{1}{2m_{\mathrm{P}}^{2}}\left(  \frac{\partial S_{\mathbf{k}}}{\partial
a}\right)  ^{2}+\frac{1}{2a^{2}}\left(  \frac{\partial S_{\mathbf{k}}%
}{\partial\phi}\right)  ^{2}+\frac{1}{2}m_{\mathrm{P}}^{2}V+\frac{1}{2}\left(
\frac{\partial S_{\mathbf{k}}}{\partial\upsilon_{\mathbf{k}}}\right)
^{2}+\frac{1}{2}\omega_{\mathbf{k}}^{2}\upsilon_{\mathbf{k}}^{2}+Q=0
\]
where%
\[
Q=\frac{1}{2m_{\mathrm{P}}^{2}}\frac{1}{\left\vert \Psi_{\mathbf{k}%
}\right\vert }\frac{1}{a}\frac{\partial}{\partial a}\left(  a\frac
{\partial\left\vert \Psi_{\mathbf{k}}\right\vert }{\partial a}\right)
-\frac{1}{2a^{2}}\frac{1}{\left\vert \Psi_{\mathbf{k}}\right\vert }%
\frac{\partial^{2}\left\vert \Psi_{\mathbf{k}}\right\vert }{\partial\phi^{2}%
}-\frac{1}{2}\frac{1}{\left\vert \Psi_{\mathbf{k}}\right\vert }\frac
{\partial^{2}\left\vert \Psi_{\mathbf{k}}\right\vert }{\partial\upsilon
_{\mathbf{k}}^{2}}%
\]
is the quantum potential (terms depending on $\left\vert \Psi_{\mathbf{k}%
}\right\vert $), while the imaginary part can be written in the form of what
we call a pseudo-continuity equation%
\begin{equation}
\frac{\partial}{\partial a}\left(  a\left\vert \Psi_{\mathbf{k}}\right\vert
^{2}a^{\prime}\right)  +\frac{\partial}{\partial\phi}\left(  a\left\vert
\Psi_{\mathbf{k}}\right\vert ^{2}\phi^{\prime}\right)  +\frac{\partial
}{\partial\upsilon_{\mathbf{k}}}\left(  a\left\vert \Psi_{\mathbf{k}%
}\right\vert ^{2}\upsilon_{\mathbf{k}}^{\prime}\right)  =0
\label{p-cont_qucosmo}%
\end{equation}
for a density $a\left\vert \Psi_{\mathbf{k}}\right\vert ^{2}$ with velocities
$a^{\prime}$, $\phi^{\prime}$, $\upsilon_{\mathbf{k}}^{\prime}$ given by the
expressions (\ref{deB_qu_cosm}) and (\ref{deB_pertns}) obtained from the
canonical momenta. The measure%
\begin{equation}
d\mu=\left\vert \Psi_{\mathbf{k}}\right\vert ^{2}adad\phi d\upsilon
_{\mathbf{k}} \label{pseudo-Born_qucosmo}%
\end{equation}
is preserved by the de Broglie velocity field defined by (\ref{deB_qu_cosm})
and (\ref{deB_pertns}).

In pilot-wave theory we can consider the time evolution of a general ensemble,
with probability density $P_{\mathbf{k}}(a,\phi,\upsilon_{\mathbf{k}},\eta)$
(defined with respect to $dad\phi d\upsilon_{\mathbf{k}}$), and with each
system guided by the same wave function $\Psi_{\mathbf{k}}$. By definition
this will satisfy the continuity equation%
\begin{equation}
\frac{\partial P_{\mathbf{k}}}{\partial\eta}+\frac{\partial}{\partial
a}\left(  P_{\mathbf{k}}a^{\prime}\right)  +\frac{\partial}{\partial\phi
}\left(  P_{\mathbf{k}}\phi^{\prime}\right)  +\frac{\partial}{\partial
\upsilon_{\mathbf{k}}}\left(  P_{\mathbf{k}}\upsilon_{\mathbf{k}}^{\prime
}\right)  =0 \label{cont_qucosmo}%
\end{equation}
with velocities $a^{\prime}$, $\phi^{\prime}$, $\upsilon_{\mathbf{k}}^{\prime
}$ again given by (\ref{deB_qu_cosm}) and (\ref{deB_pertns}). This is the same
as equation (\ref{p-cont_qucosmo}) satisfied by the density $a\left\vert
\Psi_{\mathbf{k}}\right\vert ^{2}$ (which has no explicit $\eta$ dependence).
However, as we shall discuss in Section 7.4, there is no equilibrium or
Born-rule state $P_{\mathbf{k}}=a\left\vert \Psi_{\mathbf{k}}\right\vert ^{2}$
since $a\left\vert \Psi_{\mathbf{k}}\right\vert ^{2}$ is non-normalisable.

For completeness we note that the model may of course be expressed in terms of
standard cosmological time $t$ (where $dt=ad\eta$). Since $a^{\prime}=a\dot
{a}$, $\phi^{\prime}=a\dot{\phi}$, and $\upsilon_{\mathbf{k}}^{\prime}%
=a\dot{\upsilon}_{\mathbf{k}}$, (\ref{p-cont_qucosmo}) can be rewritten as a
pseudo-continuity equation%
\begin{equation}
\frac{\partial}{\partial a}\left(  a^{2}\left\vert \Psi_{\mathbf{k}%
}\right\vert ^{2}\dot{a}\right)  +\frac{\partial}{\partial\phi}\left(
a^{2}\left\vert \Psi_{\mathbf{k}}\right\vert ^{2}\dot{\phi}\right)
+\frac{\partial}{\partial\upsilon_{\mathbf{k}}}\left(  a^{2}\left\vert
\Psi_{\mathbf{k}}\right\vert ^{2}\dot{\upsilon}_{\mathbf{k}}\right)  =0
\label{p-cont_qucosmo_t}%
\end{equation}
for a density $a^{2}\left\vert \Psi_{\mathbf{k}}\right\vert ^{2}$, with a
preserved measure%
\begin{equation}
d\mu=\left\vert \Psi_{\mathbf{k}}\right\vert ^{2}a^{2}dad\phi d\upsilon
_{\mathbf{k}}\ , \label{pseudo-Born_qucosmo_t}%
\end{equation}
and with velocities%
\begin{equation}
\dot{a}=-\frac{1}{m_{\mathrm{P}}^{2}}\frac{1}{a}\frac{\partial S_{\mathbf{k}}%
}{\partial a}\ ,\ \ \dot{\phi}=\frac{1}{a^{3}}\frac{\partial S_{\mathbf{k}}%
}{\partial\phi}\ ,\ \ \dot{\upsilon}_{\mathbf{k}}=\frac{1}{a}\frac{\partial
S_{\mathbf{k}}}{\partial\upsilon_{\mathbf{k}}} \label{deB_qu_cosm_t}%
\end{equation}
(where $\dot{a}$ and $\dot{\phi}$ are as originally given by Vink
\cite{Vink92}). A general ensemble with probability density $P_{\mathbf{k}%
}(a,\phi,\upsilon_{\mathbf{k}},t)$ (defined with respect to $dad\phi
d\upsilon_{\mathbf{k}}$) will satisfy the continuity equation%
\[
\frac{\partial P_{\mathbf{k}}}{\partial t}+\frac{\partial}{\partial a}\left(
P_{\mathbf{k}}\dot{a}\right)  +\frac{\partial}{\partial\phi}\left(
P_{\mathbf{k}}\dot{\phi}\right)  +\frac{\partial}{\partial\upsilon
_{\mathbf{k}}}\left(  P_{\mathbf{k}}\dot{\upsilon}_{\mathbf{k}}\right)  =0\ ,
\]
with $\dot{a}$, $\dot{\phi}$, $\dot{\upsilon}_{\mathbf{k}}$ given by
(\ref{deB_qu_cosm_t}). Again, this is the same as equation
(\ref{p-cont_qucosmo_t}) satisfied by $a^{2}\left\vert \Psi_{\mathbf{k}%
}\right\vert ^{2}$, but there is no equilibrium state $P_{\mathbf{k}}%
=a^{2}\left\vert \Psi_{\mathbf{k}}\right\vert ^{2}$ since $a^{2}\left\vert
\Psi_{\mathbf{k}}\right\vert ^{2}$ is non-normalisable.

In standard quantum theory equation (\ref{p-cont_qucosmo}) (or
(\ref{p-cont_qucosmo_t})) would be taken to imply the Born-rule measure
(\ref{pseudo-Born_qucosmo}) (or (\ref{pseudo-Born_qucosmo_t})) as a
probability measure on the minisuperspace with variables $(a,\phi
,\upsilon_{\mathbf{k}})$. We would then encounter the difficulty that the
putative `probability' measure is in general non-normalisable, owing to the
Klein-Gordon-like character of the wave equation (\ref{WD_mode_1}). As we will
discuss in Section 7.4, in pilot-wave theory it is clear that there is in fact
no such physical Born-rule probability measure.

\subsection{Semiclassical expansion}

It is sometimes convenient to write%
\begin{equation}
\alpha=\ln a\ ,
\end{equation}
in terms of which the wave equation (\ref{WD_mode_1}) reads%
\begin{equation}
\frac{1}{2m_{\mathrm{P}}^{2}}e^{-2\alpha}\frac{\partial^{2}\Psi_{\mathbf{k}}%
}{\partial\alpha^{2}}-\frac{1}{2}e^{-2\alpha}\frac{\partial^{2}\Psi
_{\mathbf{k}}}{\partial\phi^{2}}+\frac{1}{2}m_{\mathrm{P}}^{2}V\Psi
_{\mathbf{k}}-\frac{1}{2}\frac{\partial^{2}\Psi_{\mathbf{k}}}{\partial
\upsilon_{\mathbf{k}}^{2}}+\frac{1}{2}\omega_{\mathbf{k}}^{2}\upsilon
_{\mathbf{k}}^{2}\Psi_{\mathbf{k}}=0\ . \label{WD_mode_1prime}%
\end{equation}
Brizuela \textit{et al}. \cite{BKM16} rewrite this in the compact form%
\begin{equation}
-\frac{1}{2m_{\mathrm{P}}^{2}}\mathcal{G}_{AB}\frac{\partial^{2}%
\Psi_{\mathbf{k}}}{\partial q_{A}\partial q_{B}}+\frac{1}{2}m_{\mathrm{P}}%
^{2}V\Psi_{\mathbf{k}}-\frac{1}{2}\frac{\partial^{2}\Psi_{\mathbf{k}}%
}{\partial\upsilon_{\mathbf{k}}^{2}}+\frac{1}{2}\omega_{\mathbf{k}}%
^{2}\upsilon_{\mathbf{k}}^{2}\Psi_{\mathbf{k}}=0\ , \label{WD_mode_2prime}%
\end{equation}
where $\Psi_{\mathbf{k}}=\Psi_{\mathbf{k}}(q_{A},\upsilon_{\mathbf{k}})$ and
the indices $A$, $B$ take values $0$, $1$, with%
\begin{equation}
\mathcal{G}_{AB}=\mathrm{diag}(-e^{-2\alpha},e^{-2\alpha})
\end{equation}
and%
\begin{equation}
(q_{0},q_{1})=(\alpha,m_{\mathrm{P}}^{-1}\phi)\ .
\end{equation}
The Klein-Gordon-like character of the wave equation is now more explicit.

Taking (\ref{WD_mode_2prime}) as a starting point, Brizuela \textit{et al}.
proceed along the lines pioneered by Kiefer and Singh \cite{KS91}. We
summarise their method, since some of the details will be required in what follows.

The method begins by solving (\ref{WD_mode_2prime}) with a WKB-type or
semiclassical expansion%
\begin{equation}
\Psi_{\mathbf{k}}=\exp\left[  i\left(  m_{\mathrm{P}}^{2}S_{0}+m_{\mathrm{P}%
}^{0}S_{1}+m_{\mathrm{P}}^{-2}S_{2}+...\right)  \right]  \label{Psi_k_WKB}%
\end{equation}
in powers of $m_{\mathrm{P}}^{2}$. This is inserted into (\ref{WD_mode_2prime}%
), terms with the same power of $m_{\mathrm{P}}$ are collected and in each
case the sum is set equal to zero.

The highest order, $m_{\mathrm{P}}^{4}$, yields%
\begin{equation}
\partial S_{0}/\partial\upsilon_{\mathbf{k}}=0\ , \label{dS0=0}%
\end{equation}
which tells us that the background part of the wave function does not depend
on the perturbations.

The next order, $m_{\mathrm{P}}^{2}$, yields a Hamilton-Jacobi equation for
the background:%
\begin{equation}
\mathcal{G}_{AB}\frac{\partial S_{0}}{\partial q_{A}}\frac{\partial S_{0}%
}{\partial q_{B}}+V(q_{A})=0\ .
\end{equation}

At order $m_{\mathrm{P}}^{0}$, Brizuela \textit{et al}. find%
\begin{equation}
2\mathcal{G}_{AB}\frac{\partial S_{0}}{\partial q_{A}}\frac{\partial S_{1}%
}{\partial q_{B}}-i\mathcal{G}_{AB}\frac{\partial^{2}S_{0}}{\partial
q_{A}\partial q_{B}}+\left(  \frac{\partial S_{1}}{\partial\upsilon
_{\mathbf{k}}}\right)  ^{2}-i\frac{\partial^{2}S_{1}}{\partial\upsilon
_{\mathbf{k}}^{2}}+\omega_{\mathbf{k}}^{2}\upsilon_{\mathbf{k}}^{2}=0\ ,
\label{eqnforS1}%
\end{equation}
and define an uncorrected wave function%
\begin{equation}
\psi_{\mathbf{k}}^{(0)}=\psi_{\mathbf{k}}^{(0)}(q_{A},\upsilon_{\mathbf{k}%
})=\gamma(q_{A})e^{iS_{1}(q_{A},\upsilon_{\mathbf{k}})}\ , \label{psi_0}%
\end{equation}
where $\gamma$ satisfies%
\begin{equation}
\mathcal{G}_{AB}\frac{\partial}{\partial q_{A}}\left(  \frac{1}{2\gamma^{2}%
}\frac{\partial S_{0}}{\partial q_{B}}\right)  =0\ . \label{eqnforgamma}%
\end{equation}
Introducing a conformal WKB time,%
\begin{equation}
\frac{\partial}{\partial\eta}=\mathcal{G}_{AB}\frac{\partial S_{0}}{\partial
q_{A}}\frac{\partial}{\partial q_{B}}\ , \label{WKBt_qu_cosmo}%
\end{equation}
the equation (\ref{eqnforS1}) for $S_{1}$ can be rewritten as a
Schr\"{o}dinger equation for $\psi_{\mathbf{k}}^{(0)}$:%
\begin{equation}
i\frac{\partial\psi_{\mathbf{k}}^{(0)}}{\partial\eta}=\mathcal{\hat{H}%
}_{\mathbf{k}}\psi_{\mathbf{k}}^{(0)}\ , \label{Sch_0_qucosm}%
\end{equation}
where%
\begin{equation}
\mathcal{\hat{H}}_{\mathbf{k}}=-\frac{1}{2}\frac{\partial^{2}}{\partial
\upsilon_{\mathbf{k}}^{2}}+\frac{1}{2}\omega_{\mathbf{k}}^{2}\upsilon
_{\mathbf{k}}^{2} \label{uncorr_Ham}%
\end{equation}
is the effective uncorrected Hamiltonian for the perturbations.

As before, to this order $\Psi_{\mathbf{k}}$ takes the WKB form%
\begin{equation}
\Psi_{\mathbf{k}}(q_{A},\upsilon_{\mathbf{k}})\approx\Psi_{\mathbf{k}%
}^{\mathrm{WKB}}(q_{A})\psi_{\mathbf{k}}(\upsilon_{\mathbf{k}},q_{A})\ ,
\label{Psi_WKB_qucosm}%
\end{equation}
with%
\begin{equation}
\Psi_{\mathbf{k}}^{\mathrm{WKB}}(q_{A})=\frac{1}{\gamma(q_{A})}\exp
(im_{\mathrm{P}}^{2}S_{0})
\end{equation}
and $\psi_{\mathbf{k}}(\upsilon_{\mathbf{k}},q_{A})=\psi_{\mathbf{k}}%
^{(0)}(\upsilon_{\mathbf{k}},\eta)$.

Note also that, as in Section 3.4.1, the definition (\ref{WKBt_qu_cosmo}) of
WKB time is natural in pilot-wave theory. In terms of the variables $q_{A}$
the de Broglie velocities (\ref{deB_qu_cosm}) for the background can be
written as%
\begin{equation}
\frac{dq_{A}}{d\eta}=\frac{1}{m_{\mathrm{P}}^{2}}\mathcal{G}_{AB}%
\frac{\partial S_{\mathbf{k}}}{\partial q_{B}}\ . \label{deB_qA}%
\end{equation}
In the Hamilton-Jacobi limit we have $S_{\mathbf{k}}\approx m_{\mathrm{P}}%
^{2}S_{0}$. The definition (\ref{WKBt_qu_cosmo}) then reads%
\begin{equation}
\frac{\partial}{\partial\eta}=\frac{dq_{A}}{d\eta}\frac{\partial}{\partial
q_{A}}\ .
\end{equation}
This relation simply reinterprets a dependence on the changing coordinate
$q_{A}(\eta)$ as a dependence on $\eta$, so that in effect we have%
\begin{equation}
\psi_{\mathbf{k}}^{(0)}=\psi_{\mathbf{k}}^{(0)}(\upsilon_{\mathbf{k}},\eta)\ .
\end{equation}

The next order, $m_{\mathrm{P}}^{-2}$, yields the result%
\begin{equation}
\mathcal{G}_{AB}\frac{\partial S_{0}}{\partial q_{A}}\frac{\partial S_{2}%
}{\partial q_{B}}+\frac{1}{2}\mathcal{G}_{AB}\frac{\partial S_{1}}{\partial
q_{A}}\frac{\partial S_{1}}{\partial q_{B}}-\frac{i}{2}\mathcal{G}_{AB}%
\frac{\partial^{2}S_{1}}{\partial q_{A}\partial q_{B}}+\frac{\partial S_{1}%
}{\partial\upsilon_{\mathbf{k}}}\frac{\partial S_{2}}{\partial\upsilon
_{\mathbf{k}}}-\frac{i}{2}\frac{\partial^{2}S_{2}}{\partial\upsilon
_{\mathbf{k}}^{2}}=0\ .
\end{equation}
Writing%
\begin{equation}
S_{2}=S_{2}(q_{A},\upsilon_{\mathbf{k}})=\zeta(q_{A})+\chi(q_{A}%
,\upsilon_{\mathbf{k}})\ , \label{splitS2}%
\end{equation}
$\chi$ is found to satisfy%
\begin{equation}
\frac{\partial\chi}{\partial\eta}=\frac{1}{\psi_{\mathbf{k}}^{(0)}}\left(
-\frac{1}{\gamma}\mathcal{G}_{AB}\frac{\partial\psi_{\mathbf{k}}^{(0)}%
}{\partial q_{A}}\frac{\partial\gamma}{\partial q_{B}}+\frac{1}{2}%
\mathcal{G}_{AB}\frac{\partial^{2}\psi_{\mathbf{k}}^{(0)}}{\partial
q_{A}\partial q_{B}}+i\frac{\partial\psi_{\mathbf{k}}^{(0)}}{\partial
\upsilon_{\mathbf{k}}}\frac{\partial\chi}{\partial\upsilon_{\mathbf{k}}}%
+\frac{i}{2}\psi_{\mathbf{k}}^{(0)}\frac{\partial^{2}\chi}{\partial
\upsilon_{\mathbf{k}}^{2}}\right)  \ . \label{eqnforchi}%
\end{equation}
Brizuela \textit{et al}. then define a corrected wave function%
\begin{equation}
\psi_{\mathbf{k}}^{(1)}=\psi_{\mathbf{k}}^{(1)}(q_{A},\upsilon_{\mathbf{k}%
})=\psi_{\mathbf{k}}^{(0)}(q_{A},\upsilon_{\mathbf{k}})e^{im_{\mathrm{P}}%
^{-2}\chi(q_{A},\upsilon_{\mathbf{k}})}\ , \label{psi1defn}%
\end{equation}
which is shown to obey a quantum-gravitationally corrected Schr\"{o}dinger
equation%
\begin{equation}
i\frac{\partial\psi_{\mathbf{k}}^{(1)}}{\partial\eta}=\mathcal{\hat{H}%
}_{\mathbf{k}}\psi_{\mathbf{k}}^{(1)}-\frac{1}{2m_{\mathrm{P}}^{2}%
\psi_{\mathbf{k}}^{(0)}}\left(  \frac{1}{V}(\mathcal{\hat{H}}_{\mathbf{k}%
})^{2}\psi_{\mathbf{k}}^{(0)}+i\frac{\partial}{\partial\eta}\left(  \frac
{1}{V}\mathcal{\hat{H}}_{\mathbf{k}}\right)  \psi_{\mathbf{k}}^{(0)}\right)
\psi_{\mathbf{k}}^{(1)}\ , \label{Sch_corr}%
\end{equation}
where $V$ is the auxiliary potential (\ref{auxptl}) and in effect
$\psi_{\mathbf{k}}^{(1)}=\psi_{\mathbf{k}}^{(1)}(\upsilon_{\mathbf{k}},\eta)$.

As we saw in the general case (Section 5.1) the corrections have a
non-Hermitian part. In standard quantum mechanics this violates unitarity and
leads to inconsistencies: probability will not be conserved. But as we
described in Section 6, such corrections can arise naturally in pilot-wave
theory while maintaining a fully conserved probability, with the Born rule
rendered unstable.

As noted in Section 5.1 the approximate ratio (\ref{ratio}) of the
non-Hermitian correction to the Hermitian correction is of order $\sim H/E$,
where $H=\dot{a}/a$ is the Hubble parameter and $E$ is a typical energy for
the field. This ratio is extremely small at late times but might be large in
the very early universe.

\subsection{The de Broglie velocity}

As in the general case we can ask what happens to the guidance equation
(\ref{deB_pertns}) for the perturbations $\upsilon_{\mathbf{k}}$ in the
quantum-gravitationally corrected Schr\"{o}dinger approximation. We again find
that the de Broglie velocity reduces to that generated by the Hermitian part
of the effective Hamiltonian.

To show this, let us examine what happens to the expression (\ref{deB_pertns})
under the semiclassical expansion (\ref{Psi_k_WKB}) of the wave function
$\Psi_{\mathbf{k}}$. Noting that $S_{\mathbf{k}}=\operatorname{Im}\ln
\Psi_{\mathbf{k}}$ we have%
\begin{equation}
\frac{d\upsilon_{\mathbf{k}}}{d\eta}=\frac{\partial}{\partial\upsilon
_{\mathbf{k}}}\operatorname{Im}\ln\Psi_{\mathbf{k}}=\frac{\partial}%
{\partial\upsilon_{\mathbf{k}}}\left(  m_{\mathrm{P}}^{2}\operatorname{Re}%
S_{0}+m_{\mathrm{P}}^{0}\operatorname{Re}S_{1}+m_{\mathrm{P}}^{-2}%
\operatorname{Re}S_{2}+...\right)  \ .
\end{equation}
From (\ref{dS0=0}) we then have%
\begin{equation}
\frac{d\upsilon_{\mathbf{k}}}{d\eta}=\frac{\partial}{\partial\upsilon
_{\mathbf{k}}}\left(  \operatorname{Re}S_{1}+m_{\mathrm{P}}^{-2}%
\operatorname{Re}S_{2}+...\right)  \ . \label{deB_expn}%
\end{equation}

By similar reasoning to that given for general case, we can show that the
expression%
\begin{equation}
\operatorname{Re}S_{1}+m_{\mathrm{P}}^{-2}\operatorname{Re}S_{2}+...
\label{phase_psi1}%
\end{equation}
is equal to the overall phase $s_{\mathbf{k}}^{(1)}=\operatorname{Im}\ln
\psi_{\mathbf{k}}^{(1)}$ of the corrected wave function $\psi_{\mathbf{k}%
}^{(1)}$. Note first that, from equation (\ref{eqnforS1}), $S_{1}$ is
generally complex. The definition (\ref{psi_0}) of the uncorrected wave
function $\psi_{\mathbf{k}}^{(0)}=\gamma e^{iS_{1}}$ contains a prefactor
$\gamma$, and from equation (\ref{eqnforgamma}) we see that $\gamma$ can be
chosen to be real. Thus $\operatorname{Re}S_{1}$ is equal to the overall phase
of the uncorrected wave function $\psi_{\mathbf{k}}^{(0)}$. To lowest order,
then, we recover the usual de Broglie velocity. Now in (\ref{splitS2}) we have
$S_{2}=\zeta+\chi$ where from (\ref{eqnforchi}) we see that $\chi$ is complex.
From the definition (\ref{psi1defn}) of the corrected wave function,
$\psi_{\mathbf{k}}^{(1)}=\psi_{\mathbf{k}}^{(0)}e^{im_{\mathrm{P}}^{-2}\chi}$,
we see that the correction to the overall phase is equal to $m_{\mathrm{P}%
}^{-2}\operatorname{Re}\chi$. Thus (\ref{phase_psi1}) is indeed equal to the
overall phase of the corrected wave function $\psi_{\mathbf{k}}^{(1)}$.

To this order, then, the expansion (\ref{deB_expn}) for the de Broglie
velocity can be written as%
\begin{equation}
\frac{d\upsilon_{\mathbf{k}}}{d\eta}=\frac{\partial s_{\mathbf{k}}^{(1)}%
}{\partial\upsilon_{\mathbf{k}}}\ , \label{deB_psi1}%
\end{equation}
where $s_{\mathbf{k}}^{(1)}=\operatorname{Im}\ln\psi_{\mathbf{k}}^{(1)}$ is
the phase of $\psi_{\mathbf{k}}^{(1)}$. The result (\ref{deB_psi1}) is just
the standard guidance equation associated with the Hermitian part of the
Hamiltonian appearing in the corrected Schr\"{o}dinger equation
(\ref{Sch_corr}).

As in the general case (Section 5.3), we find that in the Schr\"{o}dinger
approximation the de Broglie velocity for the perturbations $\upsilon
_{\mathbf{k}}$ is driven by the Hermitian part of the effective Hamiltonian.
Because the quantum-gravitationally corrected Hamiltonian contains a
non-Hermitian part as well, it follows from the fundamental equations of
quantum gravity that the emergent Born rule is unstable.

\subsection{Absence of a fundamental equilibrium state}

In Section 4 we presented a new approach to the Born rule in quantum gravity.
The same reasoning applies to quantum cosmology. The minisuperspace model
considered here provides a convenient testing ground for these proposals.

Let us for a moment ignore the perturbations $\upsilon_{\mathbf{k}}$ and
consider only the variables $a$ and $\phi$ in the deep quantum-gravity regime.
Dropping the terms in $\upsilon_{\mathbf{k}}$ from (\ref{WD_mode_1}) and
recalling the definition (\ref{auxptl}) of $V$, the Wheeler-DeWitt equation
for $\Psi(a,\phi)$ (dropping the subscript $\mathbf{k}$) can be written as%
\begin{equation}
\frac{1}{m_{\mathrm{P}}^{2}}\frac{1}{a}\frac{\partial}{\partial a}\left(
a\frac{\partial\Psi}{\partial a}\right)  -\frac{1}{a^{2}}\frac{\partial
^{2}\Psi}{\partial\phi^{2}}+2a^{4}\mathcal{V}(\phi)\Psi=0\ . \label{WD_mini}%
\end{equation}
The guidance equations for $a$ and $\phi$ are given by (\ref{deB_qu_cosm}).
From (\ref{pseudo-Born_qucosmo}) (without $\upsilon_{\mathbf{k}}$) we know
that the dynamics preserves the measure%
\begin{equation}
d\mu=\left\vert \Psi\right\vert ^{2}adad\phi\ . \label{dmu_mini}%
\end{equation}

This simple model can be used to illustrate our new approach to the Born rule.
Equations (\ref{WD_mini}) and (\ref{deB_qu_cosm}) define a deterministic
pilot-wave dynamics for an individual system. For a theoretical ensemble of
systems guided by the same wave function $\Psi$, the evolving distribution
$P(a,\phi,\eta)$ necessarily satisfies the continuity equation%
\begin{equation}
\frac{\partial P}{\partial\eta}+\frac{\partial}{\partial a}\left(  Pa^{\prime
}\right)  +\frac{\partial}{\partial\phi}\left(  P\phi^{\prime}\right)  =0
\label{Cont_P_mini}%
\end{equation}
(where the probability density $P$ is defined with respect to $dad\phi$). This
is the same as the equation%
\begin{equation}
\frac{\partial}{\partial a}\left(  a\left\vert \Psi\right\vert ^{2}a^{\prime
}\right)  +\frac{\partial}{\partial\phi}\left(  a\left\vert \Psi\right\vert
^{2}\phi^{\prime}\right)  =0
\end{equation}
satisfied by the density $a\left\vert \Psi\right\vert ^{2}$ (cf. equation
(\ref{p-cont_qucosmo})). However, as we have noted, there can be no
equilibrium or Born-rule state $P=a\left\vert \Psi\right\vert ^{2}$ because
$a\left\vert \Psi\right\vert ^{2}$ is non-normalisable.

To see this it is convenient to use the variable $\alpha=\ln a$ in terms of
which (\ref{WD_mini}) reads%
\begin{equation}
\frac{1}{m_{\mathrm{P}}^{2}}\frac{\partial^{2}\Psi}{\partial\alpha^{2}}%
-\frac{\partial^{2}\Psi}{\partial\phi^{2}}+2e^{6\alpha}\mathcal{V}(\phi
)\Psi=0\ . \label{WD_mini_alpha}%
\end{equation}
This is a two-dimensional Klein-Gordon equation with a potential term. The
free part (without the potential) has the general solution%
\begin{equation}
\Psi_{\mathrm{free}}=f(\phi-m_{\mathrm{P}}\alpha)+g(\phi+m_{\mathrm{P}}%
\alpha)\ , \label{Psi_free}%
\end{equation}
where $f$ and $g$ are general packets travelling with the `wave speed'
$c=m_{\mathrm{P}}$ in the minisuperspace $(\alpha,\phi)$. There is then no
question of $\Psi$ being normalisable. In terms of $\alpha$ the measure
(\ref{dmu_mini}) reads%
\begin{equation}
d\mu=\left\vert \Psi\right\vert ^{2}e^{2\alpha}d\alpha d\phi\ .
\end{equation}
For the free part (\ref{Psi_free}) we necessarily have%
\begin{equation}
\int\int e^{2\alpha}d\alpha d\phi\ \left\vert \Psi_{\mathrm{free}}\right\vert
^{2}=\infty\ .
\end{equation}
Indeed even without the factor $e^{2\alpha}$ the integral still diverges. In
terms of $a$ and $\phi$, and for a general solution $\Psi$, we will inevitably
have%
\begin{equation}
\int\int adad\phi\ \left\vert \Psi(a,\phi)\right\vert ^{2}=\infty\ .
\end{equation}

We emphasise once again that the non-normalisability of the wave function
$\Psi$ is not caused by symmetries or unobservable degrees of freedom but is
simply a consequence of the Klein-Gordon-like character of the Wheeler-DeWitt
equation -- which implies a wave-like propagation on the minisuperspace.

By definition $P(a,\phi,\eta)$ is a physical probability density satisfying%
\begin{equation}
\int\int dad\phi\ P(a,\phi,\eta)=1
\end{equation}
at all times $\eta$. Apart from this condition, the initial distribution
$P(a,\phi,\eta_{i})$ (at some initial conformal time $\eta_{i}$) is in
principle arbitrary, and subsequently evolves via (\ref{Cont_P_mini}). In
contrast, $a\left\vert \Psi(a,\phi)\right\vert ^{2}$ is non-integrable and
cannot be a physical density. As in the general quantum-gravitational case, we
must have%
\begin{equation}
P(a,\phi,\eta)\neq a\left\vert \Psi(a,\phi)\right\vert ^{2}
\label{noneq_qucosm}%
\end{equation}
at all times. The quantities $P$ and $a\left\vert \Psi\right\vert ^{2}$ can
never be equal, neither initially nor subsequently.

As in the general quantum-gravitational case, the initial $P(a,\phi,\eta_{i})$
is not fixed by any law but is instead empirical. It must be constrained by
observation, just like any other initial condition in physics.

We can again consider a coarse-grained $H$-function%
\begin{equation}
\bar{H}(t)=\int\int dad\phi\ \bar{P}\ln(\bar{P}/\overline{a\left\vert
\Psi\right\vert ^{2}})\ , \label{Hbar_qucosm}%
\end{equation}
which measures (minus) the relative entropy of $\bar{P}$ with respect to
$\overline{a\left\vert \Psi\right\vert ^{2}}$ and which again obeys the
$H$-theorem (\ref{Hthm}). As we saw in the general case, because $\Psi$ is
non-normalisable (\ref{Hbar_qucosm}) has no lower bound and $\bar{P}$ will
always be infinitely far away from the pseudo-equilibrium state $\overline
{a\left\vert \Psi\right\vert ^{2}}$. In this sense the system will be
perpetually in a state of quantum nonequilibrium.

Further understanding of the time dependence of the $H$-function
(\ref{Hbar_qucosm}) will require numerical simulations for specific solutions
$\Psi$. This will be explored elsewhere \cite{KV21}.

\subsection{Emergence and instability of the Born rule}

In our approach, at the fundamental level, there is no state of quantum
equilibrium and no physical Born rule. Even so, as we discussed in the general
case, the Born rule can be recovered in the effective Schr\"{o}dinger regime
for quantum fields propagating on a classical spacetime background. Let us
summarise how this comes about in terms of the above quantum-cosmological model.

We may reinstate the perturbations $\upsilon_{\mathbf{k}}$ and consider the
variables $a$ and $\phi$ in a semiclassical regime where $a$ and $\phi$ define
an approximately classical background. We have seen that the system can then
be described by a Schr\"{o}dinger approximation with a time-dependent wave
function $\psi_{\mathbf{k}}(\upsilon_{\mathbf{k}},\eta)$. To zeroth-order
$\psi_{\mathbf{k}}$ satisfies the time-dependent Schr\"{o}dinger equation
(\ref{Sch_0_qucosm}) so that $\psi_{\mathbf{k}}=\psi_{\mathbf{k}}^{(0)}$. The
de Broglie velocity is then given by the usual formula%
\begin{equation}
\frac{d\upsilon_{\mathbf{k}}}{d\eta}=\frac{\partial s_{\mathbf{k}}^{(0)}%
}{\partial\upsilon_{\mathbf{k}}}\ , \label{deB_psi0}%
\end{equation}
where now $s_{\mathbf{k}}^{(0)}=\operatorname{Im}\ln\psi_{\mathbf{k}}^{(0)}$
is the phase of $\psi_{\mathbf{k}}^{(0)}$. For a theoretical ensemble with the
same wave function $\psi_{\mathbf{k}}^{(0)}$ we can consider an arbitrary
probability distribution $\rho_{\mathbf{k}}^{(0)}(\upsilon_{\mathbf{k}},\eta)$
for $\upsilon_{\mathbf{k}}$ on a given background $(a,\phi)$, which
necessarily evolves according to the continuity equation%
\begin{equation}
\frac{\partial\rho_{\mathbf{k}}^{(0)}}{\partial\eta}+\frac{\partial}%
{\partial\upsilon_{\mathbf{k}}}\left(  \rho_{\mathbf{k}}^{(0)}\upsilon
_{\mathbf{k}}^{\prime}\right)  =0\ . \label{rhok_cont}%
\end{equation}

As in our general discussion we first need to consider $\rho_{\mathbf{k}%
}^{(0)}(\upsilon_{\mathbf{k}},\eta_{i})$ at some initial time $\eta_{i}$ soon
after the system enters the Schr\"{o}dinger regime. This arises as a
conditional probability from a theoretical ensemble of universes with
Wheeler-DeWitt wave function $\Psi_{\mathbf{k}}(a,\phi,\upsilon_{\mathbf{k}})$
and arbitrary distribution $P_{\mathbf{k}}(a,\phi,\upsilon_{\mathbf{k}}%
,\eta_{i})$. Conditioning on the background $(a,\phi)$ we then have%
\begin{equation}
\rho_{\mathbf{k}}^{(0)}(\upsilon_{\mathbf{k}},\eta_{i})=\frac{P_{\mathbf{k}%
}(a,\phi,\upsilon_{\mathbf{k}},\eta_{i})}{\left(  \int P_{\mathbf{k}}%
(a,\phi,\upsilon_{\mathbf{k}},\eta_{i})d\upsilon_{\mathbf{k}}\right)  }\ .
\end{equation}
Like $P_{\mathbf{k}}(a,\phi,\upsilon_{\mathbf{k}},\eta_{i})$ itself,
$\rho_{\mathbf{k}}^{(0)}(\upsilon_{\mathbf{k}},\eta_{i})$ is in principle
arbitrary and can only be constrained empirically. Because of the fundamental
nonequilibrium condition%
\begin{equation}
P_{\mathbf{k}}(a,\phi,\upsilon_{\mathbf{k}},\eta)\neq a\left\vert
\Psi_{\mathbf{k}}(a,\phi,\upsilon_{\mathbf{k}})\right\vert ^{2}%
\end{equation}
(for all $\eta$), in general we have%
\[
\rho_{\mathbf{k}}^{(0)}(\upsilon_{\mathbf{k}},\eta_{i})\neq\left\vert
\psi_{\mathbf{k}}^{(0)}(\upsilon_{\mathbf{k}},\eta_{i})\right\vert ^{2}%
\]
(except in the special case $P_{\mathbf{k}}(a,\phi,\upsilon_{\mathbf{k}}%
,\eta_{i})=\Pi(a,\phi)\left\vert \psi_{\mathbf{k}}^{(0)}(\upsilon_{\mathbf{k}%
},\eta_{i})\right\vert ^{2}$ for some $\Pi(a,\phi)$). Thus we can expect the
perturbations $\upsilon_{\mathbf{k}}$ to be in a state of quantum
nonequilibrium at the beginning of the Schr\"{o}dinger regime.

At later times the Born rule%
\begin{equation}
\rho_{\mathbf{k}}^{(0)}(\upsilon_{\mathbf{k}},\eta)=\left\vert \psi
_{\mathbf{k}}^{(0)}(\upsilon_{\mathbf{k}},\eta)\right\vert ^{2}%
\end{equation}
can emerge by quantum relaxation in the usual way (on a coarse-grained level).
The zeroth-order Schr\"{o}dinger equation (\ref{Sch_0_qucosm}) implies that
$\left\vert \psi_{\mathbf{k}}^{(0)}\right\vert ^{2}$ obeys the same continuity
equation%
\begin{equation}
\frac{\partial\left\vert \psi_{\mathbf{k}}^{(0)}\right\vert ^{2}}{\partial
\eta}+\frac{\partial}{\partial\upsilon_{\mathbf{k}}}\left(  \left\vert
\psi_{\mathbf{k}}^{(0)}\right\vert ^{2}\upsilon_{\mathbf{k}}^{\prime}\right)
=0
\end{equation}
as is satisfied by $\rho_{\mathbf{k}}^{(0)}$ (equation (\ref{rhok_cont})). We
can then define a coarse-grained $H$-function%
\begin{equation}
\bar{H}(t)=\int d\upsilon_{\mathbf{k}}\ \overline{\rho_{\mathbf{k}}^{(0)}}%
\ln\left(  \overline{\rho_{\mathbf{k}}^{(0)}}/\overline{\left\vert
\psi_{\mathbf{k}}^{(0)}\right\vert ^{2}}\right)  \ ,
\end{equation}
which will satisfy the $H$-theorem (\ref{Hthm}). Quantum relaxation will in
fact be rather limited for a one-dimensional system with a single degree of
freedom $\upsilon_{\mathbf{k}}$ \cite{AV92,AV01}. But in a realistic cosmology
different modes will be entangled at early times. Extensive numerical
simulations show that even just two dimensions suffice for efficient
relaxation to occur [38, 40--42]. More precisely, for quantum fields on
expanding space, efficient quantum relaxation is found for field modes with
sub-Hubble wavelengths, while relaxation is suppressed at super-Hubble
wavelengths \cite{CV13,CV15,CV16}. Thus the Born rule is certainly recovered
at the wavelengths observed in the laboratory.

Even so, if we include first-order gravitational corrections to the
Schr\"{o}dinger equation, we find small non-Hermitian terms in the Hamiltonian
that can in principle drive the system away from equilibrium. The corrected
Schr\"{o}dinger equation (\ref{Sch_corr}) contains such a term, implying that
the corrected wave function $\psi_{\mathbf{k}}^{(1)}$ obeys a continuity
equation%
\begin{equation}
\frac{\partial\left\vert \psi_{\mathbf{k}}^{(1)}\right\vert ^{2}}{\partial
\eta}+\frac{\partial}{\partial\upsilon_{\mathbf{k}}}\left(  \left\vert
\psi_{\mathbf{k}}^{(1)}\right\vert ^{2}\upsilon_{\mathbf{k}}^{\prime}\right)
=s \label{psi1_cont}%
\end{equation}
with a source term $s$ (as discussed in Section 6). However, the de Broglie
velocity still takes the standard form (\ref{deB_psi1}), and so the actual
distribution $\rho_{\mathbf{k}}^{(1)}(\upsilon_{\mathbf{k}},\eta)$ satisfies
the usual continuity equation%
\begin{equation}
\frac{\partial\rho_{\mathbf{k}}^{(1)}}{\partial\eta}+\frac{\partial}%
{\partial\upsilon_{\mathbf{k}}}\left(  \rho_{\mathbf{k}}^{(1)}\upsilon
_{\mathbf{k}}^{\prime}\right)  =0 \label{rho1_cont}%
\end{equation}
(where in both (\ref{psi1_cont}) and (\ref{rho1_cont}) we have $\upsilon
_{\mathbf{k}}^{\prime}=\partial s_{\mathbf{k}}^{(1)}/\partial\upsilon
_{\mathbf{k}}$). As we saw in Section 6.2, the mismatch between the continuity
equations (\ref{psi1_cont}) and (\ref{rho1_cont}) implies that an initial
equilibrium distribution $\rho_{\mathbf{k}}^{(1)}=\left\vert \psi_{\mathbf{k}%
}^{(1)}\right\vert ^{2}$ can evolve away from equilibrium ($\rho_{\mathbf{k}%
}^{(1)}\neq\left\vert \psi_{\mathbf{k}}^{(1)}\right\vert ^{2}$ at later
times). As we have noted, in practice this can occur only if nonequilibrium is
generated on a timescale $\tau_{\mathrm{noneq}}$ that is shorter than the
usual timescale $\tau_{\mathrm{relax}}$ for quantum relaxation.

\subsection{Simplified model for the slow-roll limit}

We have seen how a semiclassical expansion of the Wheeler-DeWitt equation
together with the de Broglie guidance equation leads to a pilot-wave model of
quantum cosmology in which the Born rule can be unstable. We have a
quantum-gravitationally corrected Schr\"{o}dinger equation (\ref{Sch_corr})
for the wave function $\psi_{\mathbf{k}}^{(1)}(\upsilon_{\mathbf{k}},\eta)$
together with the standard de Broglie guidance equation (\ref{deB_psi1}) for
the perturbations $\upsilon_{\mathbf{k}}$. We will now write these equations
in a far slow-roll limit. The resulting simplified model can then be used to
perform tractable calculations of the quantum-gravitational production of
quantum nonequilibrium in cosmology.

Note first that the result (\ref{Sch_corr}) is written in terms of rescaled
variables $a_{\mathrm{new}}$, $\eta_{\mathrm{new}}$, $\upsilon_{\mathbf{k}%
\mathrm{new}}$, $k_{\mathrm{new}}$ (defined by (\ref{rescale})), where the
subscript `new' has been dropped. We will find it convenient to undo these
rescalings and write the equations in terms of the original variables
$a_{\mathrm{old}}$, $\eta_{\mathrm{old}}$, $\upsilon_{\mathbf{k}\mathrm{old}}%
$, $k_{\mathrm{old}}$ (now dropping the subscript `old'). The lengthscale
$\mathfrak{L}$ then reappears. Making this change, from now on it is
understood that we use the symbols $a$, $\eta$, $\upsilon_{\mathbf{k}}$, $k$
to denote the original (unrescaled) variables. We also revert to standard
cosmic time $t$ via the relation $d\eta=a^{-1}dt$. In (\ref{Sch_corr}) we then
need to make the replacements%
\begin{equation}
\frac{\partial}{\partial\eta}\rightarrow\mathfrak{L}\frac{\partial}%
{\partial\eta}=\mathfrak{L}a\frac{\partial}{\partial t}\,,\,\,V\rightarrow
\mathfrak{L}^{4}V=\frac{2}{m_{\mathrm{P}}^{2}}\mathfrak{L}^{4}a^{4}%
\mathcal{V}\ . \label{restore}%
\end{equation}

Our simplified limit is taken as follows. First, we ignore the factor
$z^{\prime\prime}/z$ and write%
\begin{equation}
\omega_{\mathbf{k}}^{2}=k^{2}-\frac{z^{\prime\prime}}{z}\approx k^{2}\ .
\end{equation}
Second, in the expression (\ref{M-S}) for the Mukhanov-Sasaki variable
$\upsilon$ we neglect terms containing $\phi^{\prime}$, so that $\upsilon
\approx a\varphi$ and%
\begin{equation}
\upsilon_{\mathbf{k}}\approx a\varphi_{\mathbf{k}}\ . \label{MS_approx}%
\end{equation}

It is also convenient to rescale the variable $\varphi_{\mathbf{k}}$ and to
write%
\begin{equation}
\varphi_{\mathbf{k}}=\mathfrak{L}^{3/2}q_{\mathbf{k}}\ , \label{phi_rescale}%
\end{equation}
remembering that here both $\varphi_{\mathbf{k}}$ and $q_{\mathbf{k}}$ are
treated as real.

With these changes and simplifications, the uncorrected Hamiltonian
$\mathcal{\hat{H}}_{\mathbf{k}}$ (given by (\ref{uncorr_Ham})) appearing in
(\ref{Sch_corr}) becomes%
\begin{equation}
\mathcal{\hat{H}}_{\mathbf{k}}=\mathfrak{L}a\hat{H}_{\mathbf{k}}\ ,
\label{restore_Ham}%
\end{equation}
where%
\begin{equation}
\hat{H}_{\mathbf{k}}=-\frac{1}{2a^{3}}\frac{\partial^{2}}{\partial
q_{\mathbf{k}}^{2}}+\frac{1}{2}ak^{2}q_{\mathbf{k}}^{2} \label{H_k}%
\end{equation}
is the Hamiltonian for the (real) degree of freedom $q_{\mathbf{k}}$.

Taking into account (\ref{restore}) and (\ref{restore_Ham}), and defining%
\begin{equation}
\bar{k}=\frac{1}{\mathfrak{L}}\,,
\end{equation}
the quantum-gravitationally corrected Schr\"{o}dinger equation (\ref{Sch_corr}%
) becomes%
\begin{equation}
i\frac{\partial\psi_{\mathbf{k}}^{(1)}}{\partial t}=\hat{H}_{\mathbf{k}}%
\psi_{\mathbf{k}}^{(1)}-\frac{\bar{k}^{3}}{2m_{\mathrm{P}}^{2}}\frac{1}%
{\psi_{\mathbf{k}}^{(0)}}\left[  \frac{1}{a^{3}}\frac{(\hat{H}_{\mathbf{k}%
})^{2}}{(2\mathcal{V}/m_{\mathrm{P}}^{2})}\psi_{\mathbf{k}}^{(0)}%
+i\frac{\partial}{\partial t}\left(  \frac{1}{a^{3}}\frac{\hat{H}_{\mathbf{k}%
}}{(2\mathcal{V}/m_{\mathrm{P}}^{2})}\right)  \psi_{\mathbf{k}}^{(0)}\right]
\psi_{\mathbf{k}}^{(1)}\ . \label{corr_Sch_qc_restored}%
\end{equation}

To complete our simplified model, we also need to rewrite the de Broglie
guidance equation (\ref{deB_psi1}). We undo the rescalings (\ref{rescale}),
restore $t$, and use the approximation (\ref{MS_approx}) with the rescaling
(\ref{phi_rescale}). We find%
\begin{equation}
\frac{dq_{\mathbf{k}}}{dt}=\frac{1}{a^{3}}\frac{\partial s_{\mathbf{k}}^{(1)}%
}{\partial q_{\mathbf{k}}}\ . \label{deB_qc_restored}%
\end{equation}
For a general theoretical ensemble with wave function $\psi_{\mathbf{k}}%
^{(1)}(q_{\mathbf{k}},t)$, the probability density $\rho_{\mathbf{k}}%
^{(1)}(q_{\mathbf{k}},t)$ will evolve by the continuity equation%
\begin{equation}
\frac{\partial\rho_{\mathbf{k}}^{(1)}}{\partial t}+\frac{\partial}{\partial
q_{\mathbf{k}}}\left(  \rho_{\mathbf{k}}^{(1)}\dot{q}_{\mathbf{k}}\right)
=0\ , \label{cont_qc}%
\end{equation}
with velocity field $\dot{q}_{\mathbf{k}}=dq_{\mathbf{k}}/dt$ given by
(\ref{deB_qc_restored}).

Equations (\ref{corr_Sch_qc_restored}), (\ref{deB_qc_restored}) and
(\ref{cont_qc}) define a simplified pilot-wave quantum-cosmological model in
which the Born rule can be unstable. This model will be used as a starting
point for some of the calculations that follow.

\section{Quantum instability for a scalar field on de Sitter space}

In the slow-roll limit the energy density (\ref{en_dens_phi}) of the
background scalar field may be written as $\rho\approx\mathcal{V}$ and is
approximately constant in time. The Friedmann--Lema\^{\i}tre equation
$(\dot{a}/a)^{2}=(8\pi G/3)\rho$ then implies an approximate de Sitter
expansion, $a\propto e^{Ht}$, with an approximately constant Hubble parameter%
\begin{equation}
H=\sqrt{(8\pi G/3)\mathcal{V}}\ .
\end{equation}
Using $m_{\mathrm{P}}^{2}=3/4\pi G$ we can then write%
\begin{equation}
2\mathcal{V}/m_{\mathrm{P}}^{2}=H^{2}\ .
\end{equation}
Inserting this into the corrected Schr\"{o}dinger equation
(\ref{corr_Sch_qc_restored}) we then have%
\begin{equation}
i\frac{\partial\psi_{\mathbf{k}}^{(1)}}{\partial t}=\hat{H}_{\mathbf{k}}%
\psi_{\mathbf{k}}^{(1)}-\frac{\bar{k}^{3}}{2m_{\mathrm{P}}^{2}H^{2}}\frac
{1}{\psi_{\mathbf{k}}^{(0)}}\left[  \frac{1}{a^{3}}(\hat{H}_{\mathbf{k}}%
)^{2}\psi_{\mathbf{k}}^{(0)}+i\frac{\partial}{\partial t}\left(  \frac
{1}{a^{3}}\hat{H}_{\mathbf{k}}\right)  \psi_{\mathbf{k}}^{(0)}\right]
\psi_{\mathbf{k}}^{(1)}\ . \label{Sch_corr_deS}%
\end{equation}
The de Broglie guidance equation remains, of course, as given by
(\ref{deB_qc_restored}).

We will now apply this model -- defined by (\ref{Sch_corr_deS}),
(\ref{deB_qc_restored}) and (\ref{cont_qc}) -- to the gravitational production
of quantum nonequilibrium on de Sitter space. Our results in effect provide a
mechanism for the gravitational production of quantum nonequilibrium during
inflation that was suggested in refs. \cite{AV10,AV14}.

\subsection{Approximation for the Bunch-Davies vacuum}

First we estimate the timescale (\ref{tau_est}) for quantum instability on de
Sitter space, where the term $\hat{H}_{2}$ comes from quantum-gravitational
corrections to the Schr\"{o}dinger equation. To a first approximation we may
take%
\begin{equation}
\left\langle \hat{H}_{2}\right\rangle \approx\left\langle \hat{H}%
_{2}\right\rangle _{\mathrm{B-D}}\ ,
\end{equation}
where $\left\langle ...\right\rangle _{\mathrm{BD}}$ denotes a quantum
expectation value calculated with the zeroth-order Bunch-Davies vacuum wave
function $\psi_{\mathbf{k}}^{(0)}$ (without gravitational corrections). The
gravitationally-corrected wave function $\psi_{\mathbf{k}}^{(1)}$ obeys the
equation (\ref{Sch_corr_deS}), where $H$ is the quasi-static Hubble parameter,
$\hat{H}_{\mathbf{k}}$ is the single-mode Hamiltonian (\ref{H_k}) for a
massless scalar field, and $\bar{k}=1/\mathfrak{L}$ is an arbitrary scale.
Note that the corrections in (\ref{Sch_corr_deS}) are multiplicative, one real
and one imaginary.

In our notation, with $\hat{H}=\hat{H}_{1}+i\hat{H}_{2}$, we have%
\begin{equation}
\hat{H}_{1}=\hat{H}_{\mathbf{k}}-\frac{\bar{k}^{3}}{2m_{\mathrm{P}}^{2}%
H^{2}\psi_{\mathbf{k}}^{(0)}}\frac{1}{a^{3}}(\hat{H}_{\mathbf{k}})^{2}%
\psi_{\mathbf{k}}^{(0)}%
\end{equation}
(where to lowest order $\hat{H}_{1}\simeq\hat{H}_{\mathbf{k}}$) and%
\begin{equation}
\hat{H}_{2}=-\frac{\bar{k}^{3}}{2m_{\mathrm{P}}^{2}H^{2}}\frac{1}%
{\psi_{\mathbf{k}}^{(0)}}\frac{\partial}{\partial t}\left(  \frac{1}{a^{3}%
}H_{\mathbf{k}}\right)  \psi_{\mathbf{k}}^{(0)}\ , \label{H_2}%
\end{equation}
where the Bunch-Davies wave function $\psi_{\mathbf{k}}^{(0)}$ satisfies%
\begin{equation}
i\frac{\partial\psi_{\mathbf{k}}^{(0)}}{\partial t}=\hat{H}_{\mathbf{k}}%
\psi_{\mathbf{k}}^{(0)} \label{Sch_BD}%
\end{equation}
(with the Minkowski boundary condition for the limit $H\longrightarrow0$).

Using $\dot{a}=Ha$ we may write%
\begin{equation}
\hat{H}_{2}=\frac{\bar{k}^{3}}{2m_{\mathrm{P}}^{2}H}\frac{1}{a^{3}}\frac
{1}{\psi_{\mathbf{k}}^{(0)}}\hat{H}_{\mathbf{k}}^{\prime}\psi_{\mathbf{k}%
}^{(0)}\ , \label{H2}%
\end{equation}
where we have defined a `distorted' single-mode Hamiltonian%
\begin{equation}
\hat{H}_{\mathbf{k}}^{\prime}=-\frac{3}{a^{3}}\frac{\partial^{2}}{\partial
q_{\mathbf{k}}^{2}}+ak^{2}q_{\mathbf{k}}^{2} \label{Hdist}%
\end{equation}
(of the same form as (\ref{H_k}) but with different numerical coefficients).

Since (\ref{H2}) is purely multiplicative we have%
\begin{equation}
\left\langle \hat{H}_{2}\right\rangle _{\mathrm{B-D}}=\frac{\bar{k}^{3}%
}{2m_{\mathrm{P}}^{2}H}\frac{1}{a^{3}}\left\langle \hat{H}_{\mathbf{k}%
}^{\prime}\right\rangle _{\mathrm{B-D}}\ , \label{H_2_BD}%
\end{equation}
where%
\begin{equation}
\left\langle \hat{H}_{\mathbf{k}}^{\prime}\right\rangle _{\mathrm{B-D}}%
=\frac{3}{a^{3}}\left\langle -\partial^{2}/\partial q_{\mathbf{k}}%
^{2}\right\rangle _{\mathrm{B-D}}+ak^{2}\left\langle q_{\mathbf{k}}%
^{2}\right\rangle _{\mathrm{B-D}}\ . \label{H_dist_av}%
\end{equation}
This can be calculated from the Bunch-Davies wave function $\psi_{\mathbf{k}%
}^{(0)}=\left\vert \psi_{\mathbf{k}}^{(0)}\right\vert e^{is_{\mathbf{k}}%
^{(0)}}$, which has amplitude and phase \cite{AV10}%
\begin{align}
\left\vert \psi_{\mathbf{k}}^{(0)}\right\vert  &  =\frac{1}{(2\pi\Delta
_{k}^{2})^{1/4}}e^{-q_{\mathbf{k}}^{2}/4\Delta_{k}^{2}}\ ,\label{amp}\\
s_{\mathbf{k}}^{(0)}  &  =-\frac{ak^{2}q_{\mathbf{k}}^{2}}{2H(1+k^{2}%
/H^{2}a^{2})}+\frac{1}{2}\frac{k}{Ha}-\frac{1}{2}\tan^{-1}\left(  \frac{k}%
{Ha}\right)  \ , \label{phase}%
\end{align}
and width%
\begin{equation}
\left\langle q_{\mathbf{k}}^{2}\right\rangle _{\mathrm{B-D}}=\Delta_{k}%
^{2}=\frac{H^{2}}{2k^{3}}\left(  1+\frac{k^{2}}{H^{2}a^{2}}\right)  \ .
\label{BD_width}%
\end{equation}
We obtain%
\begin{equation}
\left\langle \hat{H}_{\mathbf{k}}^{\prime}\right\rangle _{\mathrm{B-D}}%
=\frac{1}{2}\frac{k}{a}\left(  4+\frac{H^{2}a^{2}}{k^{2}}\right)  \ .
\label{H_dist_av_result}%
\end{equation}
From this we may find our estimate (\ref{tau_est}) for $\tau_{\mathrm{noneq}}$.

\subsection{Timescale for quantum instability}

From (\ref{tau_est}), (\ref{H_2_BD}) and (\ref{H_dist_av_result}) we have%
\begin{equation}
\tau_{\mathrm{noneq}}\sim\frac{2m_{\mathrm{P}}^{2}}{H^{3}}\left(  \frac
{k}{\bar{k}}\right)  ^{3}\frac{x^{4}}{4+x^{2}}\ , \label{tau_noneq}%
\end{equation}
where it is convenient to define%
\begin{equation}
x=\frac{Ha}{k}=\frac{1}{2\pi}\frac{\lambda_{\mathrm{phys}}}{H^{-1}}\ .
\end{equation}
The result (\ref{tau_noneq}) gives an estimated timescale for the
gravitational production of quantum nonequilibrium during an approximate de
Sitter phase.

We may consider a given mode with fixed $k$. According to (\ref{tau_noneq})
the effect is significant -- that is, $\tau_{\mathrm{noneq}}$ is small -- only
when $x$ is small (when the mode is inside the Hubble radius). As the mode
evolves in time, both $x$ and $\tau_{\mathrm{noneq}}$ grow unboundedly large
and the gravitational effect is frozen. Thus there is no gravitational
production of quantum nonequilibrium for $\lambda_{\mathrm{phys}}>>H^{-1}$.

To ensure the validity and relevance of our approximate calculations, we can
restrict ourselves to a range of physical wavelengths%
\begin{equation}
10^{3}l_{\mathrm{P}}\lesssim\lambda_{\mathrm{phys}}\lesssim10H^{-1}%
\ .\label{lambda_range}%
\end{equation}
The lower bound is justified by the need to avoid the deep quantum-gravity
regime where our calculations are invalid, while the upper bound may be
justified by the standard inflationary result that the mode becomes
effectively classical soon after exiting the Hubble radius (this rough
approach is needed because we are using our simplified model for the slow-roll
limit). In terms of $x$ the range (\ref{lambda_range}) corresponds
(approximately) to $10^{-1}\lesssim x\lesssim1$.

To get a sense of orders of magnitude, let us consider the result
(\ref{tau_noneq}) for $k\sim\bar{k}$ and $x\sim1$, so that roughly%
\begin{equation}
\tau_{\mathrm{noneq}}\sim m_{\mathrm{P}}^{2}/H^{3}\ . \label{tau_noneq_rough}%
\end{equation}
As an illustrative example, if we take the inflationary phase to have an
energy scale of order $H\sim10^{16}\ \mathrm{GeV}\sim10^{-3}m_{\mathrm{P}}$
(which is likely to be an upper bound), we find $\tau_{\mathrm{noneq}}%
\sim10^{9}t_{\mathrm{P}}$ (where $t_{\mathrm{P}}$ is the Planck time). By
comparison the time $H^{-1}\sim10^{3}t_{\mathrm{P}}$ the mode spends in the
regime $x\sim1$ is only a fraction $\sim10^{-6}$ of $\tau_{\mathrm{noneq}}$.
Thus the timescale over which our estimate applies is only about $10^{-6}$ of
the timescale $\tau_{\mathrm{noneq}}$ for the creation of nonequilibrium.
These crude estimates suggest that the effect will indeed be small (as expected).

As noted in Section 6.3 the created quantum nonequilibrium can build up over
time only if the condition (\ref{condn}) is satisfied. In the ideal limit of a
Bunch-Davies vacuum there is in fact no quantum relaxation at all (essentially
because the trajectories for the field modes are too simple to generate
relaxation) \cite{AV10}. In effect $\tau_{\mathrm{relax}}=\infty$ and the
condition (\ref{condn}) is indeed satisfied. Any nonequilibrium created during
a de Sitter expansion will not be erased by relaxation. Small perturbations to
the Bunch-Davies vacuum are unlikely to change this conclusion, since
numerical studies for the oscillator (analogous to a field mode) indicate that
small perturbations do not cause relaxation \cite{KV19}. Thus de Sitter space
provides a particularly simple and effective example of the gravitational
creation of quantum nonequilibrium building up over time.

\section{Corrections to the cosmic microwave background}

The temperature anisotropy $\Delta T(\mathbf{\hat{n}})\equiv T(\mathbf{\hat
{n}})-\bar{T}$ of the cosmic microwave background (CMB) can be expressed in
spherical harmonics%
\begin{equation}
\frac{\Delta T(\mathbf{\hat{n}})}{\bar{T}}=\sum_{l=2}^{\infty}\sum_{m=-l}%
^{+l}a_{lm}Y_{lm}(\mathbf{\hat{n}})\ , \label{sph_har}%
\end{equation}
where $\mathbf{\hat{n}}$ labels points on the sky and $\bar{T}$ is the mean
temperature. The coefficients $a_{lm}$ are generated by the primordial
curvature perturbation $\mathcal{R}_{\mathbf{k}}$ \cite{LR99},%
\begin{equation}
a_{lm}=\frac{i^{l}}{2\pi^{2}}\int d^{3}\mathbf{k}\ \mathcal{T}(k,l)\mathcal{R}%
_{\mathbf{k}}Y_{lm}(\mathbf{\hat{k}})\ , \label{alm}%
\end{equation}
where $\mathcal{T}(k,l)$ is the transfer function. It is usually assumed that
$\Delta T(\mathbf{\hat{n}})$ is drawn from a theoretical ensemble with an
isotropic probability distribution. This implies that the angular power
spectrum $C_{l}\equiv\left\langle \left\vert a_{lm}\right\vert ^{2}%
\right\rangle $ is independent of $m$ (where $\left\langle ...\right\rangle $
denotes an ensemble average). Thus statistical isotropy implies that for given
$l$ the $2l+1$ quantities $\left\vert a_{lm}\right\vert ^{2}$ all have the
same ensemble mean $C_{l}$. This allows us to probe the underlying theoretical
ensemble from measurements made on our one observed sky. The measured mean
statistic%
\begin{equation}
C_{l}^{\mathrm{sky}}\equiv\frac{1}{2l+1}\sum_{m=-l}^{+l}\left\vert
a_{lm}\right\vert ^{2}%
\end{equation}
satisfies $\left\langle C_{l}^{\mathrm{sky}}\right\rangle =C_{l}$ and
therefore provides an unbiased estimate of $C_{l}$ for the ensemble. For a
Gaussian distribution $C_{l}^{\mathrm{sky}}$ has a cosmic variance%
\begin{equation}
\frac{\Delta C_{l}^{\mathrm{sky}}}{C_{l}}=\sqrt{\frac{2}{2l+1}}\ . \label{CV}%
\end{equation}
We can probe $C_{l}$ accurately at large $l$, while for small $l$ the accuracy
is limited. It is also usually assumed that the theoretical ensemble for
$\mathcal{R}$ is statistically homogeneous, which implies $\left\langle
\mathcal{R}_{\mathbf{k%
\acute{}%
}}^{\ast}\mathcal{R}_{\mathbf{k}}\right\rangle =\delta_{\mathbf{kk}%
\acute{}%
}\left\langle \left\vert \mathcal{R}_{\mathbf{k}}\right\vert ^{2}\right\rangle
$. From (\ref{alm}) we then have the formula%
\begin{equation}
C_{l}=\frac{1}{2\pi^{2}}\int_{0}^{\infty}\frac{dk}{k}\ \mathcal{T}%
^{2}(k,l)\mathcal{P}_{\mathcal{R}}(k)\ , \label{Cl}%
\end{equation}
where%
\begin{equation}
\mathcal{P}_{\mathcal{R}}(k)\equiv\frac{4\pi k^{3}}{V}\left\langle \left\vert
\mathcal{R}_{\mathbf{k}}\right\vert ^{2}\right\rangle \label{PPS_R}%
\end{equation}
is the primordial power spectrum for $\mathcal{R}_{\mathbf{k}}$. It is
sometimes claimed that probabilities are meaningless for a single
universe.\footnote{This is a central tenet of the Bohmian mechanics school of
de Broglie-Bohm theory \cite{DGZ92,DT09}.} But in fact, by assuming
statistical isotropy and statistical homogeneity, measurements made on a
single sky can constrain the primordial spectrum $\mathcal{P}_{\mathcal{R}%
}(k)$ -- even though $\mathcal{P}_{\mathcal{R}}(k)$ is a property of a
theoretical ensemble.\footnote{For a careful discussion see ref.
\cite{Allori20}.}

According to inflationary cosmology the primordial spectrum (\ref{PPS_R})
originates from quantum vacuum fluctuations in a scalar inflaton field
[109--111]. An inflaton perturbation $\varphi_{\mathbf{k}}$ generates a
curvature perturbation $\mathcal{R}_{\mathbf{k}}\propto\varphi_{\mathbf{k}}$
(once the physical wavelength exits the Hubble radius). The measured power
spectrum $\mathcal{P}_{\mathcal{R}}(k)$ is then proportional to the variance
$\left\langle \left\vert \varphi_{\mathbf{k}}\right\vert ^{2}\right\rangle $
of $\varphi_{\mathbf{k}}$ -- which is usually calculated in quantum field
theory by applying the Born rule. But in pilot-wave theory we can consider
nonequilibrium probabilities for $\varphi_{\mathbf{k}}$ and hence a general
variance \cite{AV10,AV07}%
\begin{equation}
\left\langle \left\vert \phi_{\mathbf{k}}\right\vert ^{2}\right\rangle
=\left\langle \left\vert \phi_{\mathbf{k}}\right\vert ^{2}\right\rangle
_{\mathrm{QT}}\xi(k)\ ,\label{var_noneq}%
\end{equation}
where $\left\langle |\phi_{\mathbf{k}}|^{2}\right\rangle _{\mathrm{QT}}$ is
the standard quantum-theoretical variance and the factor $\xi(k)$ quantifies
the departure from equilibrium at wavenumber $k$. For a quantum nonequilibrium
ensemble of inflaton perturbations we have a primordial power spectrum%
\begin{equation}
\mathcal{P}_{\mathcal{R}}(k)=\mathcal{P}_{\mathcal{R}}^{\mathrm{QT}}%
(k)\xi(k)\ ,\label{PS_noneq}%
\end{equation}
where $\mathcal{P}_{\mathcal{R}}^{\mathrm{QT}}(k)$ is the quantum-theoretical
spectrum. From (\ref{Cl}) we then obtain a corrected angular power spectrum
$C_{l}$.

Thus measurements of the CMB can probe primordial corrections to the Born rule
as quantified by $\xi(k)$ \cite{AV10,VPV19}. We will now show how to calculate
(approximately) a new contribution to $\xi(k)$ from the quantum-gravitational
creation of quantum nonequilibrium during inflation.

\subsection{Approximate calculation of the nonequilibrium function $f$}

In inflationary cosmology observations of the CMB can probe the width (or
variance) of the distribution $\rho_{\mathbf{k}}$ of inflaton perturbations
$\varphi_{\mathbf{k}}\propto q_{\mathbf{k}}$ as a function of wave number $k$.
In quantum nonequilibrium we have $\rho_{\mathbf{k}}=\left\vert \psi
_{\mathbf{k}}\right\vert ^{2}f_{\mathbf{k}}$, with $f_{\mathbf{k}}\neq1$, so
that in general $\rho_{\mathbf{k}}$ is different from $\left\vert
\psi_{\mathbf{k}}\right\vert ^{2}$. To compare with observation we need to
know how the gravitational corrections affect $\rho_{\mathbf{k}}$. As well as
knowing how the corrections affect $\psi_{\mathbf{k}}$ we also need to know
how they affect the nonequilibrium function $f_{\mathbf{k}}$. An approximate
result for $f_{\mathbf{k}}$ will now be derived.

We use the following method. If we assume that the corrected wave function
$\psi_{\mathbf{k}}$ is already known, we may calculate $\rho_{\mathbf{k}}$ by
calculating $f_{\mathbf{k}}=\rho_{\mathbf{k}}/\left\vert \psi_{\mathbf{k}%
}\right\vert ^{2}$. As we saw in Section 6.2, in the presence of non-Hermitian
gravitational corrections, $f_{\mathbf{k}}$ is no longer conserved along
trajectories but instead changes in accordance with the differential equation
(\ref{dfdt}). This has the solution%
\begin{equation}
f_{\mathbf{k}}(q_{\mathbf{k}}(t_{f}),t_{f})=f_{\mathbf{k}}(q_{\mathbf{k}%
}(t_{i}),t_{i})\exp\left(  -\int_{t_{i}}^{t_{f}}dt\ u_{\mathbf{k}%
}(q_{\mathbf{k}}(t),t)\right)  \ , \label{f_k}%
\end{equation}
where%
\begin{equation}
u_{\mathbf{k}}=s_{\mathbf{k}}/|\psi_{\mathbf{k}}|^{2}%
\end{equation}
and%
\begin{equation}
s_{\mathbf{k}}=2\operatorname{Re}\left(  \psi_{\mathbf{k}}^{\ast}\hat{H}%
_{2}\psi_{\mathbf{k}}\right)  \ ,
\end{equation}
with $\hat{H}_{2}$ given by (\ref{H2}) and (\ref{Hdist}). The trajectories are
of course determined by the exact $\psi_{\mathbf{k}}$. However, as a first
approximation, in the expressions for $u_{\mathbf{k}}$ and $s_{\mathbf{k}}$ we
may insert the lowest-order Bunch-Davies wave function $\psi_{\mathbf{k}%
}^{(0)}$ and perform the integration in (\ref{f_k}) along the uncorrected
trajectories generated by $\psi_{\mathbf{k}}^{(0)}$.

We first find an expression for $u_{\mathbf{k}}$ (with $\psi_{\mathbf{k}}%
=\psi_{\mathbf{k}}^{(0)}$). Since $\hat{H}_{2}$ is a purely multiplicative
operator we have%
\begin{equation}
u_{\mathbf{k}}=2\operatorname{Re}\left(  \hat{H}_{2}\right)  \ .
\end{equation}
Employing the Schr\"{o}dinger equation (\ref{Sch_BD}) and writing
$\psi_{\mathbf{k}}^{(0)}=\left\vert \psi_{\mathbf{k}}^{(0)}\right\vert
e^{is_{\mathbf{k}}^{(0)}}$ we have%
\begin{equation}
u_{\mathbf{k}}=-\frac{\bar{k}^{3}}{m_{\mathrm{P}}^{2}H}\frac{1}{a^{3}}\left(
6\frac{\partial s_{\mathbf{k}}^{(0)}}{\partial t}+2ak^{2}q_{\mathbf{k}}%
^{2}\right)  \ ,
\end{equation}
where from (\ref{phase})%
\begin{equation}
6\frac{\partial s_{\mathbf{k}}^{(0)}}{\partial t}+2ak^{2}q_{\mathbf{k}}%
^{2}=ak^{2}q_{\mathbf{k}}^{2}\left(  \frac{2-5x^{2}-x^{4}}{(1+x^{2})^{2}%
}\right)  -\frac{3k}{a}\frac{1}{1+x^{2}}%
\end{equation}
(again defining $x=Ha/k$). Thus, with $a=kx/H$, we may write%
\begin{equation}
u_{\mathbf{k}}=-\frac{\bar{k}^{3}}{m_{\mathrm{P}}^{2}}\frac{H^{2}}{k^{3}x^{3}%
}\left(  \frac{1}{H}k^{3}\frac{(2-5x^{2}-x^{4})x}{(1+x^{2})^{2}}q_{\mathbf{k}%
}^{2}-3H\frac{1}{x}\frac{1}{1+x^{2}}\right)  \ .
\end{equation}

We will be integrating $u_{\mathbf{k}}$ along trajectories $q_{\mathbf{k}}%
(t)$, from arbitrary initial points $q_{\mathbf{k}}(t_{i})$ to arbitrary final
points $q_{\mathbf{k}}(t_{f})$. It will be convenient to write $q_{\mathbf{k}%
}^{2}$ in terms of $x$ and $q_{\mathbf{k}}(x_{f})$. We may then integrate with
respect to $x$ from $x_{i}=(H/k)a_{i}$ to $x_{f}=(H/k)a_{f}$. The trajectories
are generated by the lowest-order Bunch-Davies wave function $\psi
_{\mathbf{k}}^{(0)}$ whose phase $s_{\mathbf{k}}^{(0)}$ is given by
(\ref{phase}). The relevant de Broglie equation of motion%
\begin{equation}
\frac{dq_{\mathbf{k}}}{dt}=\frac{1}{a^{3}}\frac{\partial s_{\mathbf{k}}^{(0)}%
}{\partial q_{\mathbf{k}}}%
\end{equation}
((\ref{deB_qc_restored}) with $s_{\mathbf{k}}^{(1)}$ replaced by
$s_{\mathbf{k}}^{(0)}$) can be solved exactly to yield the trajectories
$q_{\mathbf{k}}(t)$ \cite{AV10}. Reverting for a moment to conformal time
$\eta$ (where on de Sitter space $\eta=-1/Ha$), the solution $q_{\mathbf{k}%
}(\eta)=q_{\mathbf{k}}(0)\sqrt{1+k^{2}\eta^{2}}$ can be written as%
\begin{equation}
q_{\mathbf{k}}(\eta)=q_{\mathbf{k}}(\eta_{f})\sqrt{\frac{1+k^{2}\eta^{2}%
}{1+k^{2}\eta_{f}^{2}}}\ ,
\end{equation}
where $\eta_{f}$ is a final time. Writing in terms of $x=Ha/k=-1/k\eta$ we
have%
\begin{equation}
q_{\mathbf{k}}^{2}(x)=q_{\mathbf{k}}^{2}(x_{f})\frac{(x_{f})^{2}}{x^{2}}%
\frac{1+x^{2}}{1+(x_{f})^{2}}\ .
\end{equation}

Our expression for $u_{\mathbf{k}}$ as a function of $x$ then reads
\begin{equation}
u_{\mathbf{k}}=-\frac{\bar{k}^{3}}{m_{\mathrm{P}}^{2}}\frac{H^{2}}{k^{3}x^{3}%
}\left(  \frac{1}{H}k^{3}\frac{(2-5x^{2}-x^{4})x}{(1+x^{2})^{2}}q_{\mathbf{k}%
}^{2}(x_{f})\frac{(x_{f})^{2}}{x^{2}}\frac{1+x^{2}}{1+(x_{f})^{2}}-3H\frac
{1}{x}\frac{1}{1+x^{2}}\right)  \ , \label{u_k}%
\end{equation}
where we need to evaluate the integral%
\begin{equation}
I_{\mathbf{k}}=\int_{t_{i}}^{t_{f}}dt\ u_{\mathbf{k}}(q_{\mathbf{k}}(t),t)
\end{equation}
along trajectories, in order to find the function%
\begin{equation}
f_{\mathbf{k}}(q_{\mathbf{k}}(t_{f}),t_{f})=f_{\mathbf{k}}(q_{\mathbf{k}%
}(t_{i}),t_{i})\exp\left(  -I_{\mathbf{k}}\right)  \label{f_k2}%
\end{equation}
(at arbitrary final points $q_{\mathbf{k}}(t_{f})$). Since $dt=(1/H)(1/x)dx$
we have%
\begin{equation}
I_{\mathbf{k}}=\int_{t_{i}}^{t_{f}}u_{\mathbf{k}}(q_{\mathbf{k}}%
(t),t)dt=\frac{1}{H}\int_{x_{i}}^{x_{f}}\frac{1}{x}u_{\mathbf{k}%
}(q_{\mathbf{k}}(x),x)dx\ .
\end{equation}
Inserting (\ref{u_k}) we then have%
\begin{equation}
I_{\mathbf{k}}=-q_{\mathbf{k}}^{2}(x_{f})\frac{(x_{f})^{2}}{1+(x_{f})^{2}%
}\frac{\bar{k}^{3}}{m_{\mathrm{P}}^{2}}I_{1}+3\frac{H^{2}}{k^{3}}\frac{\bar
{k}^{3}}{m_{\mathrm{P}}^{2}}I_{2}%
\end{equation}
where%
\begin{align}
I_{1}  &  =\int_{x_{i}}^{x_{f}}\frac{(2-5x^{2}-x^{4})}{(1+x^{2})x^{5}%
}dx\ ,\label{I_1}\\
I_{2}  &  =\int_{x_{i}}^{x_{f}}\frac{1}{(1+x^{2})x^{5}}dx\ . \label{I_2}%
\end{align}
For given $x_{i}$ and $x_{f}$ these integrals can easily be evaluated numerically.

To specify $x_{i}$ and $x_{f}$ let us take a range (\ref{lambda_range}) of
physical wavelengths, which corresponds approximately to%
\begin{equation}
x_{i}=10^{-1}\ ,\ \ x_{f}=1\ .
\end{equation}
For these values the integrals (\ref{I_1}) and (\ref{I_2}) are found to be%
\begin{equation}
I_{1}\simeq4.7\times10^{3}\ ,\ \ \ I_{2}\simeq\allowbreak2.5\times10^{3}\ .
\end{equation}
We then have%
\begin{equation}
I_{\mathbf{k}}\simeq-2.4\times10^{3}\frac{\bar{k}^{3}}{m_{\mathrm{P}}^{2}%
}\allowbreak q_{\mathbf{k}}^{2}(x_{f})+0.8\times10^{4}\frac{\bar{k}^{3}%
}{m_{\mathrm{P}}^{2}}\allowbreak\frac{H^{2}}{k^{3}}\ . \label{I_calc}%
\end{equation}
Note that for $k\sim\bar{k}$ the magnitude of the second term is $\sim
10^{4}H^{2}/m_{\mathrm{P}}^{2}$, which for an illustrative (and probable upper
bound) value $H\sim10^{-3}m_{\mathrm{P}}$ is of order $\sim10^{-2}$, and so
$I_{\mathbf{k}}$ is indeed small.

Having evaluated $I_{\mathbf{k}}$ we can write down the nonequilibrium
function (\ref{f_k2}). Recalling that $f_{\mathbf{k}}=\rho_{\mathbf{k}}%
/|\psi_{\mathbf{k}}|^{2}$, and reverting to standard time $t$, we can then
find the corrected density $\rho_{\mathbf{k}}(q_{\mathbf{k}}(t_{f}),t_{f})$
from the formula%
\begin{equation}
\frac{\rho_{\mathbf{k}}(q_{\mathbf{k}}(t_{f}),t_{f})}{|\psi_{\mathbf{k}%
}(q_{\mathbf{k}}(t_{f}),t_{f})|^{2}}=\frac{\rho_{\mathbf{k}}(q_{\mathbf{k}%
}(t_{i}),t_{i})}{|\psi_{\mathbf{k}}(q_{\mathbf{k}}(t_{i}),t_{i})|^{2}}%
\exp\left(  -I_{\mathbf{k}}\right)  \ . \label{rho_formula}%
\end{equation}

\subsection{Nonequilibrium correction to the primordial power spectrum}

We wish to calculate the correction to the primordial power spectrum as a
function of wave number $k$. Focussing on a single mode with wave vector
$\mathbf{k}$, let us assume for simplicity that our mode is initially in
quantum equilibrium:%
\begin{equation}
\rho_{\mathbf{k}}(q_{\mathbf{k}}(t_{i}),t_{i})=|\psi_{\mathbf{k}%
}(q_{\mathbf{k}}(t_{i}),t_{i})|^{2}\ .
\end{equation}
From (\ref{rho_formula}) we then have a final nonequilibrium density%
\begin{equation}
\rho_{\mathbf{k}}(q_{\mathbf{k}}(t_{f}),t_{f})=|\psi_{\mathbf{k}%
}(q_{\mathbf{k}}(t_{f}),t_{f})|^{2}\exp\left(  -I_{\mathbf{k}}\right)  \ .
\label{rho_f}%
\end{equation}

In (\ref{I_calc}) we have an expression for $I_{\mathbf{k}}$ as a function of
arbitrary final points $q_{\mathbf{k}}(t_{f})$ (or $q_{\mathbf{k}}(x_{f})$).
In (\ref{rho_f}) we have an expression for the density $\rho_{\mathbf{k}}$ at
the final time $t_{f}$ and at arbitrary final points $q_{\mathbf{k}}(t_{f})$.
We can then simply write $\rho_{\mathbf{k}}$ at time $t_{f}$ as a function of
the general variable $q_{\mathbf{k}}$:%
\begin{equation}
\rho_{\mathbf{k}}(q_{\mathbf{k}},t_{f})=|\psi_{\mathbf{k}}(q_{\mathbf{k}%
},t_{f})|^{2}\exp\left(  -I_{\mathbf{k}}(q_{\mathbf{k}})\right)  \ ,
\label{rho_f_new}%
\end{equation}
where%
\begin{equation}
I_{\mathbf{k}}(q_{\mathbf{k}})\simeq-2.4\times10^{3}\frac{\bar{k}^{3}%
}{m_{\mathrm{P}}^{2}}\allowbreak q_{\mathbf{k}}^{2}+0.8\times10^{4}\frac
{\bar{k}^{3}}{m_{\mathrm{P}}^{2}}\allowbreak\frac{H^{2}}{k^{3}}\ .
\label{I_calc_new}%
\end{equation}

Note that $\psi_{\mathbf{k}}(q_{\mathbf{k}},t_{f})$ will be gravitationally
corrected. A complete calculation of $\rho_{\mathbf{k}}(q_{\mathbf{k}},t_{f})$
then requires an expression for $|\psi_{\mathbf{k}}(q_{\mathbf{k}},t_{f}%
)|^{2}$ including the gravitational corrections. But the primordial power
spectrum depends only on the width of $\rho_{\mathbf{k}}(q_{\mathbf{k}}%
,t_{f})$, so it suffices to consider only how $|\psi_{\mathbf{k}%
}(q_{\mathbf{k}},t_{f})|^{2}$ contributes to the width. Writing the corrected
wave function as%
\begin{equation}
\psi_{\mathbf{k}}\simeq\psi_{\mathbf{k}}^{(1)}=\psi_{\mathbf{k}}^{(0)}%
+\delta\psi_{\mathbf{k}}^{(1)} \label{psi_corr}%
\end{equation}
(for simplicity omitting the arguments $q_{\mathbf{k}}$, $t_{f}$), and using
the fact that $I_{\mathbf{k}}$ is small, we may write%
\[
\rho_{\mathbf{k}}=|\psi_{\mathbf{k}}|^{2}\exp\left(  -I_{\mathbf{k}}\right)
\simeq|\psi_{\mathbf{k}}^{(0)}+\delta\psi_{\mathbf{k}}^{(1)}|^{2}\left(
1-I_{\mathbf{k}}\right)
\]
where $\delta\psi_{\mathbf{k}}^{(1)}\sim O(1/m_{\mathrm{P}}^{2})$ and
$I_{\mathbf{k}}\sim O(1/m_{\mathrm{P}}^{2})$. This may be written as%
\begin{equation}
\rho_{\mathbf{k}}=|\psi_{\mathbf{k}}^{(0)}|^{2}+\psi_{\mathbf{k}}^{(0)\ast
}\delta\psi_{\mathbf{k}}^{(1)}+\psi_{\mathbf{k}}^{(0)}\delta\psi_{\mathbf{k}%
}^{(1)\ast}-I_{\mathbf{k}}|\psi_{\mathbf{k}}^{(0)}|^{2}+O(1/m_{\mathrm{P}}%
^{4})
\end{equation}
(again omitting the arguments $q_{\mathbf{k}}$, $t_{f}$). The second and third
terms are an $O(1/m_{\mathrm{P}}^{2})$ correction coming from the
gravitational correction to $\psi_{\mathbf{k}}$, while the fourth term is an
$O(1/m_{\mathrm{P}}^{2})$ correction coming from the gravitational production
of quantum nonequilibrium.

We may now consider how the separate effects change the width of
$\rho_{\mathbf{k}}$. The mean-square $\left\langle q_{\mathbf{k}}%
^{2}\right\rangle _{f}=\int dq_{\mathbf{k}}\ q_{\mathbf{k}}^{2}\rho
(q_{\mathbf{k}},t_{f})$ (evaluated at the final time $t_{f}$) takes the form%
\begin{equation}
\left\langle q_{\mathbf{k}}^{2}\right\rangle _{f}=\left\langle q_{\mathbf{k}%
}^{2}\right\rangle _{f}^{(0)}+\left(  \delta\left\langle q_{\mathbf{k}}%
^{2}\right\rangle _{f}\right)  ^{(1)}+\left(  \delta\left\langle
q_{\mathbf{k}}^{2}\right\rangle _{f}\right)  ^{\mathrm{noneq}}%
+O(1/m_{\mathrm{P}}^{4})\ , \label{q2total}%
\end{equation}
where%
\begin{equation}
\left\langle q_{\mathbf{k}}^{2}\right\rangle _{f}^{(0)}=\int dq_{\mathbf{k}%
}\ q_{\mathbf{k}}^{2}|\psi_{\mathbf{k}}^{(0)}(q_{\mathbf{k}},t_{f})|^{2}
\label{A}%
\end{equation}
is the uncorrected Bunch-Davies width (at time $t_{f}$),%
\begin{equation}
\left(  \delta\left\langle q_{\mathbf{k}}^{2}\right\rangle _{f}\right)
^{(1)}=\int dq_{\mathbf{k}}\ q_{\mathbf{k}}^{2}\left(  \psi_{\mathbf{k}%
}^{(0)\ast}(q_{\mathbf{k}},t_{f})\delta\psi_{\mathbf{k}}^{(1)}(q_{\mathbf{k}%
},t_{f})+\psi_{\mathbf{k}}^{(0)}(q_{\mathbf{k}},t_{f})\delta\psi_{\mathbf{k}%
}^{(1)\ast}(q_{\mathbf{k}},t_{f})\right)  \label{B}%
\end{equation}
is the correction to the width from the gravitational correction to
$\psi_{\mathbf{k}}$ itself, and%
\begin{equation}
\left(  \delta\left\langle q_{\mathbf{k}}^{2}\right\rangle _{f}\right)
^{\mathrm{noneq}}=-\int dq_{\mathbf{k}}\ q_{\mathbf{k}}^{2}I_{\mathbf{k}%
}(q_{\mathbf{k}})|\psi_{\mathbf{k}}^{(0)}(q_{\mathbf{k}},t_{f})|^{2} \label{C}%
\end{equation}
is the correction to the width from the gravitational production of quantum
nonequilibrium. We now calculate the term (\ref{C}) and obtain an estimate of
its effect on the primordial power spectrum.

We are really interested in the primordial deficit function $\xi(k)$, defined
by%
\begin{equation}
\xi(k)=\frac{\left\langle q_{\mathbf{k}}^{2}\right\rangle _{f}}{\left\langle
q_{\mathbf{k}}^{2}\right\rangle _{f}^{(0)}}\ , \label{ksi total}%
\end{equation}
where $\left\langle q_{\mathbf{k}}^{2}\right\rangle _{f}$ is given by
(\ref{q2total}). We may write%
\begin{equation}
\xi(k)=1+\delta\xi^{(1)}(k)+\delta\xi^{\mathrm{noneq}}(k)+...\ ,
\label{ksi_expn}%
\end{equation}
where%
\begin{equation}
\delta\xi^{(1)}(k)=\frac{\left(  \delta\left\langle q_{\mathbf{k}}%
^{2}\right\rangle _{f}\right)  ^{(1)}}{\left\langle q_{\mathbf{k}}%
^{2}\right\rangle _{f}^{(0)}}\ ,\ \ \ \delta\xi^{\mathrm{noneq}}%
(k)=\frac{\left(  \delta\left\langle q_{\mathbf{k}}^{2}\right\rangle
_{f}\right)  ^{\mathrm{noneq}}}{\left\langle q_{\mathbf{k}}^{2}\right\rangle
_{f}^{(0)}}\ .
\end{equation}
Note that, as defined by (\ref{ksi total}), the deficit function $\xi(k)$
contains contributions from both effects (gravitational corrections to
$\psi_{\mathbf{k}}$ and the gravitational production of nonequilibrium).

We are particularly interested in the term $\delta\xi^{\mathrm{noneq}}$, which
from (\ref{C}) and (\ref{I_calc_new}) takes the form%
\begin{equation}
\delta\xi^{\mathrm{noneq}}(k)\simeq10^{4}\frac{\bar{k}^{3}}{m_{\mathrm{P}}%
^{2}}\left(  \frac{1}{4}\allowbreak\frac{\left\langle q_{\mathbf{k}}%
^{4}\right\rangle _{f}^{(0)}}{\left\langle q_{\mathbf{k}}^{2}\right\rangle
_{f}^{(0)}}-\frac{4}{5}\allowbreak\frac{H^{2}}{k^{3}}\right)  \ ,
\end{equation}
where $\left\langle q_{\mathbf{k}}^{2}\right\rangle _{f}^{(0)}$ and
$\left\langle q_{\mathbf{k}}^{4}\right\rangle _{f}^{(0)}$ are the respective
(quantum-theoretical) means of $q_{\mathbf{k}}^{2}$ and $q_{\mathbf{k}}^{4}$
in the Bunch-Davies vacuum at the final time $t_{f}$. From (\ref{BD_width})
the mean-square width $\left\langle q_{\mathbf{k}}^{2}\right\rangle _{f}%
^{(0)}=\Delta_{k}^{2}(t_{f})$ can be written as%
\begin{equation}
\left\langle q_{\mathbf{k}}^{2}\right\rangle _{f}^{(0)}=\frac{H^{2}}{2k^{3}%
}\left(  1+\frac{1}{x_{f}^{2}}\right)  \ .
\end{equation}
Taking $x_{f}=1$ we have $\left\langle q_{\mathbf{k}}^{2}\right\rangle
_{f}^{(0)}=H^{2}/k^{3}$. For the Gaussian wave-function amplitude (\ref{amp})
we also have $\left\langle q_{\mathbf{k}}^{4}\right\rangle _{f}^{(0)}=3\left(
\left\langle q_{\mathbf{k}}^{2}\right\rangle _{f}^{(0)}\right)  ^{2}$ yielding
a ratio%
\begin{equation}
\frac{\left\langle q_{\mathbf{k}}^{4}\right\rangle _{f}^{(0)}}{\left\langle
q_{\mathbf{k}}^{2}\right\rangle _{f}^{(0)}}=\frac{3H^{2}}{k^{3}}\ .
\end{equation}
Thus we find, finally,%
\begin{equation}
\delta\xi^{\mathrm{noneq}}(k)\simeq-5\times10^{2}\frac{H^{2}}{m_{\mathrm{P}%
}^{2}}\left(  \frac{\bar{k}}{k}\right)  ^{3}\ . \label{dksi_noneq_algeb}%
\end{equation}
According to this approximate calculation, the nonequilibrium correction to
$\xi$ amounts to a power deficit scaling as $\sim1/k^{3}$.

To get a sense of orders of magnitude, again taking our illustrative (and
probable upper bound) value $H\sim10^{16}\ \mathrm{GeV}\sim10^{-3}%
m_{\mathrm{P}}$, in the region $k\sim\bar{k}$ the result
(\ref{dksi_noneq_algeb}) yields%
\begin{equation}
\delta\xi^{\mathrm{noneq}}\sim-5\times10^{-4}\ . \label{dksi_noneq}%
\end{equation}
According to our estimate, the nonequilibrium correction to $\xi$ is small and negative.

\subsection{Comparison with gravitational corrections to the wave function}

Our quantum-gravitational correction to the Born rule yields a large-scale
power deficit (\ref{dksi_noneq_algeb}). In contrast, the quantum-gravitational
correction to the wave function $\psi_{\mathbf{k}}$, as derived by Brizuela
\textit{et al}. \cite{BKM16}, yields a large-scale power excess%
\begin{equation}
\mathcal{P}^{(1)}(k)=\mathcal{P}^{(0)}(k)\left(  1+0.988\frac{H^{2}%
}{m_{\mathrm{P}}^{2}}\left(  \frac{\bar{k}}{k}\right)  ^{3}+O\left(
\frac{H^{4}}{m_{\mathrm{P}}^{4}}\right)  \right)  \label{BKM_PS}%
\end{equation}
or%
\begin{equation}
\delta\xi^{(1)}(k)\simeq\frac{H^{2}}{m_{\mathrm{P}}^{2}}\left(  \frac{\bar{k}%
}{k}\right)  ^{3}\ . \label{dksi_Kiefer_algeb}%
\end{equation}
The opposing signs, $\delta\xi^{\mathrm{noneq}}<0$ and $\delta\xi^{(1)}>0$,
mean that the overall sign of the total correction $\delta\xi^{\mathrm{noneq}%
}+\delta\xi^{(1)}$ cannot be known without more accurate calculations. In
particular, in our derivation of $\delta\xi^{\mathrm{noneq}}$ the magnitude of
the coefficient is not determined accurately. Thus at present we cannot say if
the overall effect will be a power deficit or a power excess.

The magnitudes of both effects depend on the arbitrary scale $\bar{k}$ which
appears in the correction (\ref{H_2}) to the Hamiltonian. However the ratio%
\begin{equation}
\left\vert \frac{\delta\xi^{\mathrm{noneq}}(k)}{\delta\xi^{(1)}(k)}\right\vert
\simeq5\times10^{2}%
\end{equation}
is independent of $\bar{k}$ (and indeed independent of $k$). The difference in
magnitude, by about two orders, is however not very significant given our
crude estimate for $\delta\xi^{\mathrm{noneq}}$.

It is noteworthy that such different physical effects -- gravitational
corrections to the wave function on the one hand, and gravitational
corrections to the Born rule on the other -- both yield essentially the same
results except for an overall sign difference. In both cases we find a scaling
$\propto\left(  \bar{k}/k\right)  ^{3}$.

Finally we make a few comments on the derivation of (\ref{BKM_PS}) by Brizuela
\textit{et al}. \cite{BKM16}. The calculation keeps only the Hermitian
correction appearing in the Schr\"{o}dinger equation (\ref{Sch_corr}) and
drops the non-Hermitian term. As we have seen, the ratio (\ref{ratio}) of the
latter correction to the former is generally of order $\sim H/E$ where $E$ is
a typical energy for the field, so it is not entirely clear if the
non-Hermitian term can be neglected. However, Brizuela \textit{et al}.
motivate dropping the non-Hermitian term for other reasons. To solve
(\ref{Sch_corr}) for the corrected wave function $\psi_{\mathbf{k}}^{(1)}$ a
Gaussian ansatz is assumed. If the non-Hermitian term is included, two
perceived problems arise. First, the normalisation of $\psi_{\mathbf{k}}%
^{(1)}$ is not conserved in time and the standard probability interpretation
becomes problematic: it is not possible to unambiguously take expectation
values and compute the power spectrum. Second, solving the Gaussian ansatz
numerically leads to large oscillations in the amplitude of $\psi_{\mathbf{k}%
}^{(1)}$ in the distant past ($\eta\rightarrow-\infty$), which does not happen
if the non-Hermitian term is dropped. From the point of view of standard
quantum mechanics these issues motivate dropping the non-Hermitian term (even
if the semiclassical expansion of the Wheeler-DeWitt equation has naturally
produced it). But in pilot-wave theory the two perceived problems are only
apparent. First, as we discussed in Section 6 for a general system with a
non-Hermitian Hamiltonian, even though the normalisation of the wave function
$\psi$ is not conserved, the actual probability density $\rho$ remains
normalised (by construction) at all times, and the effect of the non-Hermitian
term is to generate a deviation of $\rho$ from $\left\vert \psi\right\vert
^{2}$. We can then use $\rho$ to unambiguously take expectation values and
compute the power spectrum. Second, because observed quantities such as power
spectra are now calculated from the actual density $\rho$ (which in general
will not equal $\left\vert \psi\right\vert ^{2}$), the fact that the wave
amplitude $\left\vert \psi\right\vert $ becomes large in the remote past need
not by itself signal a difficulty.

\subsection{Effect on the angular power spectrum}

We may ask if there is a realistic prospect of measuring the correction
$\delta\xi^{\mathrm{noneq}}$ to the primordial power spectrum. In addition to
having to consider the other contribution $\delta\xi^{(1)}$, an immediate
question is whether or not the effect on the $C_{l}$'s will be swamped by the
cosmic variance (\ref{CV}), in which case it will be unobservable in
principle. This will depend on the region of $k$-space that is affected --
that is, on the value of $\bar{k}$.

From (\ref{Cl}), (\ref{PS_noneq}) and (\ref{ksi_expn}) we can write%
\begin{equation}
C_{l}=C_{l}^{\mathrm{QT}}+\delta C_{l}^{(1)}+\delta C_{l}^{\mathrm{noneq}}+...
\label{Cl_expn}%
\end{equation}
where%
\begin{align}
C_{l}^{\mathrm{QT}}  &  =\frac{1}{2\pi^{2}}\int_{0}^{\infty}\frac{dk}%
{k}\ \mathcal{T}^{2}(k,l)\mathcal{P}_{\mathcal{R}}^{\mathrm{QT}}%
(k)\ ,\nonumber\\
\delta C_{l}^{(1)}  &  =\frac{1}{2\pi^{2}}\int_{0}^{\infty}\frac{dk}%
{k}\ \mathcal{T}^{2}(k,l)\mathcal{P}_{\mathcal{R}}^{\mathrm{QT}}(k)\delta
\xi^{(1)}(k)\ ,\\
\delta C_{l}^{\mathrm{noneq}}  &  =\frac{1}{2\pi^{2}}\int_{0}^{\infty}%
\frac{dk}{k}\ \mathcal{T}^{2}(k,l)\mathcal{P}_{\mathcal{R}}^{\mathrm{QT}%
}(k)\delta\xi^{\mathrm{noneq}}(k)\ .\nonumber
\end{align}
We then have a fractional correction%
\begin{equation}
\frac{\delta C_{l}^{\mathrm{noneq}}}{C_{l}^{\mathrm{QT}}}=\frac{\int
_{0}^{\infty}\frac{dk}{k}\ \mathcal{T}^{2}(k,l)\mathcal{P}_{\mathcal{R}%
}^{\mathrm{QT}}(k)\delta\xi^{\mathrm{noneq}}(k)}{\int_{0}^{\infty}\frac{dk}%
{k}\ \mathcal{T}^{2}(k,l)\mathcal{P}_{\mathcal{R}}^{\mathrm{QT}}(k)}\ .
\label{dCl_frac}%
\end{equation}

Let us first consider (\ref{dCl_frac}) at low $l$ (say $l\lesssim20$), where
the angular power spectrum is dominated by the Sachs-Wolfe effect and the
(square of the) transfer function takes the approximate analytical form
\cite{LL00}%
\begin{equation}
\mathcal{T}^{2}(k,l)=\pi H_{0}^{4}j_{l}^{2}(2k/H_{0})\ ,
\end{equation}
where $j_{l}$ is the $l$th order spherical Bessel function and $H_{0}$ is the
Hubble parameter today. Taking (roughly) $\mathcal{P}_{\mathcal{R}%
}^{\mathrm{QT}}(k)\simeq\mathrm{const}.$, (\ref{dCl_frac}) becomes%
\begin{equation}
\frac{\delta C_{l}^{\mathrm{noneq}}}{C_{l}^{\mathrm{QT}}}\simeq2l(l+1)\int
_{0}^{\infty}\frac{dk}{k}\ j_{l}^{2}(2k/H_{0})\delta\xi^{\mathrm{noneq}}(k)\ ,
\label{rat_Cl_noneq}%
\end{equation}
where we have used $\int_{0}^{\infty}\frac{dx}{x}\ j_{l}^{2}(x)=\frac
{1}{2l(l+1)}$. The integral (\ref{rat_Cl_noneq}) is dominated by the scale
$k\approx lH_{0}/2$, so a significant effect requires $\delta\xi
^{\mathrm{noneq}}$ to be significant for $k$ in this region -- that is, for
wavelengths $\lambda\approx(4\pi/l)H_{0}^{-1}$, which for low $l$ is
comparable to the Hubble radius today ($H_{0}^{-1}\simeq3000\ \mathrm{Mpc}$).
If the scale $\bar{k}$ is far from this region, the effect will be negligible.
And even if $\bar{k}\approx lH_{0}/2$, the effect will still be negligible if
the coefficient $H^{2}/m_{\mathrm{P}}^{2}$ in (\ref{dksi_noneq_algeb}) is very
small. For example, even for $H$ as large as $H\sim10^{16}\ \mathrm{GeV}%
\sim10^{-3}m_{\mathrm{P}}$, we have $H^{2}/m_{\mathrm{P}}^{2}\sim10^{-6}$.

If the correction $\delta\xi^{\mathrm{noneq}}$ is indeed present in the region
$k\approx lH_{0}/2$ then from (\ref{rat_Cl_noneq}) we may roughly expect to
find a fractional deficit%
\begin{equation}
\left\vert \delta C_{l}^{\mathrm{noneq}}/C_{l}^{\mathrm{QT}}\right\vert
\sim\left\vert \delta\xi^{\mathrm{noneq}}\right\vert \label{frac_def}%
\end{equation}
in the angular power spectrum at the $l$th multipole. Similarly, at large
values of $l$, a multipole of order $l$ probes a comoving wavenumber $k\simeq
H_{0}l/2$ \cite{LL00}, so that if the correction $\delta\xi^{\mathrm{noneq}}$
is present in the region $k\simeq lH_{0}/2$ we again roughly expect to find a
fractional deficit of order (\ref{frac_def}). However such a deficit can be
observed in principle only if it is larger than the cosmic variance (\ref{CV}).

Brizuela \textit{et al}. \cite{BKM16} suggest that $\bar{k}$ can be
interpreted as an infrared cutoff. For the purpose of estimating the magnitude
of $\delta\xi^{(1)}$, however, $\bar{k}$ is taken to coincide with the pivot
scale $k_{\ast}=0.05\ \mathrm{Mpc}^{-1}$ chosen by the \textit{Planck} team.
Taking an upper bound $H\lesssim10^{-5}m_{\mathrm{P}}$ (motivated by limits on
the tensor-to-scalar ratio) yields an estimated upper bound $\left\vert
\delta\xi^{(1)}\right\vert \lesssim2\times10^{-10}$ at $k\simeq k_{\ast}$,
which is far too small to be observable \cite{BKM16}. Taking the same bound
$H\lesssim10^{-5}m_{\mathrm{P}}$ our result (\ref{dksi_noneq_algeb}) implies
an upper bound%
\begin{equation}
\left\vert \delta\xi^{\mathrm{noneq}}\right\vert \lesssim5\times10^{-8}
\label{ub}%
\end{equation}
at $k\simeq k_{\ast}$, which is again far too small to be observable.

\section{Quantum instability in a radiation-dominated universe}

We have seen that quantum-gravitational effects can create quantum
nonequilibrium for a scalar field on exponentially expanding de Sitter space.
We expect there will be similar effects for fields on an expanding background
generally. For example, we might consider the radiation-dominated expansion
that took place during the early phase of the hot big bang. Various quantum
fields were propagating on this background: the electromagnetic field,
fermionic fields, and others. Quantum-gravitational corrections to the
Schr\"{o}dinger evolution for these fields will create quantum nonequilibrium,
in particular at very early times when the expansion was extremely rapid. The
particle-like excitations of those fields later correspond to relic
cosmological particles, which include the photons of the CMB, the neutrinos of
the expected cosmic neutrino background, as well as other more exotic
particles such as gravitinos whose existence is predicted by some theories of
particle physics (and which might be a significant component of dark matter).
It is then conceivable that such relic particles could violate the Born rule
even today. This possibility has already been studied in some detail in a
scenario where it is assumed that the universe begins in a state of quantum
nonequilibrium: quantum relaxation will be important in the early universe but
can be suppressed for some special systems that decouple very early
\cite{AV01,AV07,AV08a,UV15}. We are now concerned with a novel possibility:
that even particles initially in equilibrium could develop nonequilibrium over
time as a result of quantum-gravitational effects. In a realistic scenario, we
would have to compare the timescale $\tau_{\mathrm{noneq}}$ over which
nonequilibrium is created with the timescale $\tau_{\mathrm{relax}}$ for
quantum relaxation. As noted in Section 6.3, deviations from the Born rule can
build up over time only if $\tau_{\mathrm{relax}}>\tau_{\mathrm{noneq}}$.
Estimates for $\tau_{\mathrm{relax}}$ for relic cosmological particles have
been discussed elsewhere \cite{AV01,AV07,AV08a,UV15}. Here we focus on
developing an estimate for $\tau_{\mathrm{noneq}}$ for such particles. As we
will see, the result is found to be so large as to make this phenomenon
seemingly of theoretical interest only.

In Section 7 we studied the instability of the Born rule for a simple model of
quantum cosmology with a background expansion driven by a scalar field $\phi$
with a potential $\mathcal{V}(\phi)$. As shown in ref. \cite{BKM16}, scalar
perturbations on the background satisfy the gravitationally-corrected
Schr\"{o}dinger equation (\ref{Sch_corr}), which we have written in the
simplified form (\ref{corr_Sch_qc_restored}) by taking a slow-roll limit. The
gravitational corrections depend on $\mathcal{V}(\phi)$. A glance at the
uncorrected part of (\ref{corr_Sch_qc_restored}) shows that it is just the
usual Schr\"{o}dinger equation for a scalar-field Fourier mode $q_{\mathbf{k}%
}$ propagating on a background expanding space with scale factor $a=a(t)$. If
we were to consider a scalar field propagating on a radiation-dominated
background (with $a\propto t^{1/2}$), then the uncorrected Schr\"{o}dinger
equation would take precisely this form. The question is: what will the
gravitational corrections look like for this case?

To answer this question rigorously and from first principles, we would need to
develop a quantum-cosmological model with a background radiation-dominated
expansion. This might be done along the lines of the above model for an
appropriate choice of $\mathcal{V}(\phi)$. It is well known that, for a
potential $\mathcal{V}(\phi)=\frac{1}{4}\lambda\phi^{4}$, averaging over
oscillations in $\phi$ yields an effective equation of state $p=\rho/3$
corresponding to a radiation-dominated universe \cite{MSTGott}. However,
instead, here we will use a shortcut to obtain a simple estimate of the
correction terms.

We assume that some appropriate potential $\mathcal{V}(\phi)$ can be used to
model the radiation-dominated expansion of the background, and that the
semiclassical expansion of the Wheeler-DeWitt equation leads to an effective
Schr\"{o}dinger equation of the form (\ref{corr_Sch_qc_restored}) for the
Fourier mode $q_{\mathbf{k}}$ of some field propagating on the background,
where the relevant potential $\mathcal{V}(\phi)$ appears in the correction
terms. We treat the propagating field as a scalar, but we expect similar
equations to apply for example to modes of the electromagnetic field (only
with more components).

The correction terms in the Schr\"{o}dinger equation
(\ref{corr_Sch_qc_restored}) depend on $\mathcal{V}$ which depends on the
time-evolving background field $\phi$. We can crudely estimate the magnitude
of $\mathcal{V}$ as follows. Classically we know that the energy density
(\ref{en_dens_phi}) is a sum of kinetic and potential terms. The
Friedmann--Lema\^{\i}tre equation $(\dot{a}/a)^{2}=(8\pi G/3)\rho$ implies
that $H^{2}=(8\pi G/3)\rho$, where of course now $H$ is time dependent. As a
rough order-of-magnitude estimate we might ignore the kinetic term in $\rho$
and write $\rho\approx\mathcal{V}(\phi)$. Using $m_{\mathrm{P}}^{2}=3/4\pi G$
we then have, roughly,%
\begin{equation}
2\mathcal{V}(\phi)/m_{\mathrm{P}}^{2}\approx H^{2}\ .
\end{equation}
Inserting this into (\ref{corr_Sch_qc_restored}) we have an approximate
Schr\"{o}dinger equation%
\[
i\frac{\partial\psi_{\mathbf{k}}^{(1)}}{\partial t}\approx\hat{H}_{\mathbf{k}%
}\psi_{\mathbf{k}}^{(1)}-\frac{\bar{k}^{3}}{2m_{\mathrm{P}}^{2}}\frac{1}%
{\psi_{\mathbf{k}}^{(0)}}\left[  \frac{1}{a^{3}}\frac{(\hat{H}_{\mathbf{k}%
})^{2}}{H^{2}}\psi_{\mathbf{k}}^{(0)}+i\frac{\partial}{\partial t}\left(
\frac{1}{a^{3}}\frac{\hat{H}_{\mathbf{k}}}{H^{2}}\right)  \psi_{\mathbf{k}%
}^{(0)}\right]  \psi_{\mathbf{k}}^{(1)}%
\]
for a field mode on an expanding background with time-dependent Hubble
parameter $H=H(t)$. This should suffice for the purpose of order-of-magnitude
estimates. In our notation (\ref{Ham_split}) the correction $\hat{H}_{2}$ is
then estimated to be%
\begin{equation}
\hat{H}_{2}\approx-\frac{\bar{k}^{3}}{2m_{\mathrm{P}}^{2}}\frac{1}%
{\psi_{\mathbf{k}}^{(0)}}\frac{\partial}{\partial t}\left(  \frac{1}{a^{3}%
}\frac{\hat{H}_{\mathbf{k}}}{H^{2}}\right)  \psi_{\mathbf{k}}^{(0)}\ .
\end{equation}
This is the term that causes the instability of the Born rule over time.

For a radiation-dominated expansion $a\propto t^{1/2}$ and $H=\dot{a}%
/a=1/2t$.\textbf{ }Following common convention we take the scale factor today,
at time $t_{0}$, to be $a_{0}=1$ (so that comoving wavelengths $\lambda$
correspond to physical wavelengths today). Writing $a=(t/t_{0})^{1/2}$ we have
$a^{3}H^{2}=H_{0}^{2}/a$, where $H_{0}=1/2t_{0}$ is the Hubble parameter
today. Using $\dot{a}=Ha$ and the expression (\ref{H_k}) for $\hat
{H}_{\mathbf{k}}$ we can then write%
\begin{equation}
\hat{H}_{2}\approx\frac{\bar{k}^{3}}{m_{\mathrm{P}}^{2}H_{0}}\frac{1}{a}%
\frac{1}{\psi_{\mathbf{k}}^{(0)}}\hat{H}_{\mathbf{k}}^{\prime}\psi
_{\mathbf{k}}^{(0)}\ , \label{H2_raddom}%
\end{equation}
where now we have defined a distorted Hamiltonian%
\begin{equation}
\hat{H}_{\mathbf{k}}^{\prime}=-\frac{1}{2a^{3}}\frac{\partial^{2}}{\partial
q_{\mathbf{k}}^{2}}-\frac{1}{2}ak^{2}q_{\mathbf{k}}^{2}\ .
\end{equation}

To estimate the timescale $\tau_{\mathrm{noneq}}$ for the gravitational
production of quantum nonequilibrium, as given by (\ref{tau_est}), we need to
estimate the magnitude of the quantum expectation value $\left\langle \hat
{H}_{2}\right\rangle $. Following the same reasoning as in Section 8.1 for de
Sitter space, to a first approximation we take%
\begin{equation}
\left\langle \hat{H}_{2}\right\rangle \approx\left\langle \hat{H}%
_{2}\right\rangle ^{(0)}\ ,
\end{equation}
where $\left\langle ...\right\rangle ^{(0)}$ denotes a quantum expectation
value calculated with the uncorrected wave function $\psi_{\mathbf{k}}^{(0)}$.
As before (\ref{H2_raddom}) is purely multiplicative and so%
\begin{equation}
\left\langle \hat{H}_{2}\right\rangle ^{(0)}\approx\frac{\bar{k}^{3}%
}{m_{\mathrm{P}}^{2}H_{0}}\frac{1}{a}\left\langle \hat{H}_{\mathbf{k}}%
^{\prime}\right\rangle ^{(0)}\ ,
\end{equation}
where our sought-for timescale is given approximately by%
\begin{equation}
\tau_{\mathrm{noneq}}\approx\frac{1}{2\left\vert \left\langle \hat{H}%
_{2}\right\rangle ^{(0)}\right\vert }\ . \label{tau_rad_dom}%
\end{equation}

Let us write, crudely, $\left\langle \hat{H}_{\mathbf{k}}^{\prime
}\right\rangle ^{(0)}\sim E$ where $E=E_{0}/a$ is the energy of a single
particle propagating on the expanding background (where $E_{0}$ is the energy
today). We also reintroduce the lengthscale $\mathfrak{L}$ via the definition
$\bar{k}=1/\mathfrak{L}$. We then have%
\begin{equation}
\left\langle \hat{H}_{2}\right\rangle ^{(0)}\sim\frac{H_{0}^{-1}}%
{\mathfrak{L}^{3}m_{\mathrm{P}}^{2}}\frac{E}{a}\ .
\end{equation}
Using $m_{\mathrm{P}}\sim1/l_{\mathrm{P}}$ and inserting $c$, the timescale
(\ref{tau_rad_dom}) then takes the suggestive form%
\begin{equation}
\tau_{\mathrm{noneq}}\sim\frac{\mathfrak{L}}{c}\left(  \frac{a\mathfrak{L}%
}{l_{\mathrm{P}}}\right)  \left(  \frac{\mathfrak{L}}{H_{0}^{-1}}\right)
\left(  \frac{m_{\mathrm{P}}}{E}\right)  \ , \label{tau_rad_dom_1}%
\end{equation}
where $a\mathfrak{L}$ is the physical lengthscale at time $t$ corresponding to
the comoving lengthscale $\mathfrak{L}$.

Writing $a=(t/t_{0})^{1/2}$ and employing the standard temperature clock
$t\sim(1\ \mathrm{s})\left(  1\ \mathrm{MeV}/k_{\mathrm{B}}T\right)  ^{2}$ for
a radiation-dominated phase with ambient temperature $T$, the result
(\ref{tau_rad_dom_1}) takes the form%
\begin{equation}
\tau_{\mathrm{noneq}}\sim\frac{\mathfrak{L}}{c}\sqrt{\frac{(1\ \mathrm{s}%
)}{t_{0}}}\left(  \frac{1\ \mathrm{MeV}}{k_{\mathrm{B}}T}\right)  \left(
\frac{\mathfrak{L}}{l_{\mathrm{P}}}\right)  \left(  \frac{\mathfrak{L}}%
{H_{0}^{-1}}\right)  \left(  \frac{m_{\mathrm{P}}}{E}\right)  \ .
\label{tau_rad_dom_2}%
\end{equation}
Writing $E\sim k_{\mathrm{B}}T$ and $m_{\mathrm{P}}\sim k_{\mathrm{B}%
}T_{\mathrm{P}}$, we then have our estimated timescale%
\begin{equation}
\tau_{\mathrm{noneq}}\sim\frac{\mathfrak{L}}{c}\sqrt{\frac{(1\ \mathrm{s}%
)}{t_{0}}}\left(  \frac{\mathfrak{L}}{l_{\mathrm{P}}}\right)  \left(
\frac{\mathfrak{L}}{H_{0}^{-1}}\right)  \left(  \frac{1\ \mathrm{MeV}%
}{k_{\mathrm{B}}T}\right)  \left(  \frac{k_{\mathrm{B}}T_{\mathrm{P}}%
}{k_{\mathrm{B}}T}\right)  \ . \label{tau_rad_dom_3}%
\end{equation}

To get an idea of the magnitude of $\tau_{\mathrm{noneq}}$, let us apply our
result to the deep Planck era where $k_{\mathrm{B}}T\sim k_{\mathrm{B}%
}T_{\mathrm{P}}\sim10^{19}\ \mathrm{GeV}$. From (\ref{tau_rad_dom_3}) we find%
\begin{equation}
\tau_{\mathrm{noneq}}\sim10^{-22}\frac{\mathfrak{L}}{c}\sqrt{\frac
{(1\ \mathrm{s})}{t_{0}}}\left(  \frac{\mathfrak{L}}{l_{\mathrm{P}}}\right)
\left(  \frac{\mathfrak{L}}{H_{0}^{-1}}\right)  \ . \label{tau_rad_dom_4}%
\end{equation}
At this point we must choose a value for the lengthscale $\mathfrak{L}$. It
seems reasonable to set%
\begin{equation}
\mathfrak{L}\sim H_{0}^{-1}\simeq10^{28}\ \mathrm{cm}\ .
\end{equation}
With the current age $t_{0}\sim10^{17}\ \mathrm{s}$ of the universe we then
obtain the huge result%
\begin{equation}
\tau_{\mathrm{noneq}}\sim10^{48}\ \mathrm{s}\ .
\end{equation}
This is overwhelmingly large even compared to $t_{0}$. At lower temperatures
$\tau_{\mathrm{noneq}}$ is even larger.

Our result for $\tau_{\mathrm{noneq}}$ is so huge because of the factor
$\left(  \mathfrak{L}/l_{\mathrm{P}}\right)  $. Without some changed
understanding of the infrared cutoff lengthscale $\mathfrak{L}$ there seems to
be no way around this. We therefore tentatively conclude that there is no hope
of a significant effect building up over realistic cosmological timescales --
even in a situation where the usual quantum relaxation can be neglected. It
is, however, still possible that relic cosmological particles today could show
residual violations of the Born rule as a result of the universe beginning in
a state of primordial quantum nonequilibrium \cite{AV01,AV07,AV08a,UV15}. As
we saw in Sections 4.2 and 7.5, such primordial nonequilibrium is to be
expected as the universe emerges from the deep quantum-gravity regime.

\section{Quantum instability in the spacetime of an evaporating black hole}

Kiefer, M\"{u}ller and Singh \cite{KMS94} considered how the procedure of ref.
\cite{KS91} -- which we summarised in Sections 3.4.1 and 5.1 -- can be
generalised to the asymptotically-flat spacetime of an evaporating black hole.
This requires the inclusion of a boundary term $Mc^{2}$ in the
(integrated)\ Wheeler-DeWitt equation, where the mass $M$ of the black hole is
asymptotically defined by the usual ADM energy. Kiefer \textit{et al}. argue
that the resulting quantum-gravitational corrections to the Schr\"{o}dinger
equation, for a matter field propagating in the spacetime of the black hole,
will be of the same form as in equation (\ref{Sch_corr_1}) but with the
replacement%
\begin{equation}
\sqrt{g}R\rightarrow-16\pi GM/c^{2}\ , \label{rep_BH}%
\end{equation}
where the Schwarzchild radius $r_{\mathrm{S}}=2GM/c^{2}$ provides a natural
lengthscale. It is argued that, for an evaporating black hole, the Hermitian
correction to the Hamiltonian will be negligible compared to the dominant term
$\hat{H}_{\phi}=\int d^{3}x\,\mathcal{\hat{H}}_{\phi}$ (where $\mathcal{\hat
{H}}_{\phi}$ is the matter Hamiltonian density), since the ratio will be of
order the energy of the field divided by $Mc^{2}$. The general non-Hermitian
correction in (\ref{Sch_corr_1}) takes the suggestive form%
\begin{equation}
\Delta\hat{H}_{\phi}=i\frac{4\pi\hslash G}{c^{4}}\int d^{3}x\ \frac{\delta
}{\delta\tau}\left(  \frac{\mathcal{\hat{H}}_{\phi}}{\sqrt{g}R}\right)  \ ,
\label{non-H_dt}%
\end{equation}
where $\delta/\delta\tau$ is the many-fingered time derivative (\ref{mf_time})
and we have inserted the expansion parameter $\mu=c^{2}/32\pi G$ as well as
$\hbar$ and $c$. With the replacement (\ref{rep_BH}), for an evaporating black
hole, (\ref{non-H_dt}) then takes the approximate form%
\begin{equation}
\Delta\hat{H}_{\phi}\simeq-i\frac{4\pi\hslash G}{c^{4}}\frac{d}{dt}\left(
\frac{c^{2}}{16\pi GM}\right)  \int d^{3}x\,\mathcal{\hat{H}}_{\phi}%
=i\frac{\hslash}{4c^{2}}\frac{1}{M^{2}}\frac{dM}{dt}\hat{H}_{\phi}
\label{KS_BH}%
\end{equation}
(neglecting the rate of change of $\mathcal{\hat{H}}_{\phi}$ compared with the
rate of change of the background geometry). Kiefer \textit{et al}. suggest
that this term might play a role in alleviating the problem of black-hole
information loss.

Assuming that the time-dependent mass $M(t)$ takes the phenomenological form
\cite{Wald84,DeW75}%
\begin{equation}
M(t)\simeq M_{0}\left(  1-\kappa\frac{m_{\mathrm{P}}^{3}}{M_{0}^{3}}\left(
\frac{t}{t_{\mathrm{P}}}\right)  \right)  ^{1/3}\ , \label{M(t)}%
\end{equation}
where $M_{0}$ is the initial mass, $\kappa$ is a numerical factor, and here
$m_{\mathrm{P}}=\sqrt{\hbar c/G}$ denotes the standard Planck mass, the
non-Hermitian correction (\ref{KS_BH}) becomes important (compared to the
dominant term $\hat{H}_{\phi}$) if $M$ approaches the Planck mass
$m_{\mathrm{P}}$, which happens after a time%
\begin{equation}
t_{\ast}\simeq(M_{0}/m_{\mathrm{P}})^{3}t_{\mathrm{P}}\ .
\end{equation}
In the same regime we can then expect the gravitational production of quantum
nonequilibrium to become important.

We can now estimate the rate of production of quantum nonequilibrium during
the evaporation of a black hole. Assuming the time-dependent mass (\ref{M(t)})
we have%
\begin{equation}
\frac{dM}{dt}\simeq-\frac{1}{3}\kappa\frac{m_{\mathrm{P}}}{t_{\mathrm{P}}%
}\left(  \frac{m_{\mathrm{P}}}{M}\right)  ^{2}\ , \label{Mdot}%
\end{equation}
and so from (\ref{KS_BH}) we have%
\begin{equation}
\Delta\hat{H}_{\phi}\simeq-\frac{1}{12}i\kappa\left(  \frac{m_{\mathrm{P}}}%
{M}\right)  ^{4}\hat{H}_{\phi}\ , \label{non-H_BH}%
\end{equation}
where $M=M(t)$ is the mass of the black hole and $\hat{H}_{\phi}$ is the
(uncorrected)\ Hamiltonian of a matter field $\phi$ propagating on the
background spacetime. In particular, $\hat{H}_{\phi}$ could be the Hamiltonian
of a field in the exterior region, in the vicinity of the event horizon.
According to our framework, the effective Schr\"{o}dinger equation for the
field will contain a small non-Hermitian term of the form (\ref{non-H_BH}),
which will cause the field to evolve away from quantum equilibrium.

Let us focus on a single field mode with outgoing wave vector $\mathbf{k}$
(defined in the asymptotically flat region). In our notation (\ref{Ham_split})
the correction $\hat{H}_{2}$ is then%
\begin{equation}
\hat{H}_{2}\simeq-\frac{1}{12}\kappa\left(  \frac{m_{\mathrm{P}}}{M}\right)
^{4}\hat{H}_{\mathbf{k}}\ , \label{H2_BH}%
\end{equation}
where $\hat{H}_{\mathbf{k}}$ is the (uncorrected) Hamiltonian for the field mode.

The timescale $\tau_{\mathrm{noneq}}$ for the gravitational production of
quantum nonequilibrium, caused by the dynamical spacetime of the evaporating
black hole, is given by (\ref{tau_est}). From (\ref{H2_BH}) we see that
$\tau_{\mathrm{noneq}}$ will depend inversely on the equilibrium mean energy
$E_{\mathbf{k}}=\left\langle \hat{H}_{\mathbf{k}}\right\rangle $ of the field
mode. The evaporating black hole has a Hawking temperature $T=1/8\pi M$ (in
natural units)\ or%
\begin{equation}
k_{\mathrm{B}}T=\frac{\hbar c^{3}}{G}\frac{1}{8\pi M}=\frac{1}{8\pi
}m_{\mathrm{P}}c^{2}\frac{m_{\mathrm{P}}}{M}\ .
\end{equation}
For the purposes of our estimate we take $E_{\mathbf{k}}\sim k_{\mathrm{B}}T$.
From (\ref{tau_est}) and (\ref{H2_BH}) we then have%
\begin{equation}
\tau_{\mathrm{noneq}}\simeq\frac{6}{\kappa}\left(  \frac{M}{m_{\mathrm{P}}%
}\right)  ^{4}\frac{1}{E_{\mathbf{k}}}\sim\frac{48\pi}{\kappa}t_{\mathrm{P}%
}\left(  \frac{M}{m_{\mathrm{P}}}\right)  ^{5}\ . \label{tau_BH}%
\end{equation}

The creation of quantum nonequilibrium will be significant if the timescale
$\tau_{\mathrm{noneq}}$ is not too large compared to the timescale
$t_{\mathrm{evap}}$ over which the black hole evaporates, where%
\begin{equation}
\frac{1}{t_{\mathrm{evap}}}=\frac{1}{M}\left\vert \frac{dM}{dt}\right\vert \ .
\end{equation}
From (\ref{Mdot}) we have%
\begin{equation}
t_{\mathrm{evap}}\simeq\frac{3}{\kappa}t_{\mathrm{P}}\left(  \frac
{M}{m_{\mathrm{P}}}\right)  ^{3}\ . \label{t_evap}%
\end{equation}
The ratio of the timescales is then%
\begin{equation}
\frac{\tau_{\mathrm{noneq}}}{t_{\mathrm{evap}}}\simeq\frac{48\pi}{3}\left(
\frac{M}{m_{\mathrm{P}}}\right)  ^{2}%
\end{equation}
(where conveniently the numerical factor $\kappa$ cancels). Thus for
$M>>m_{\mathrm{P}}$ we find $\tau_{\mathrm{noneq}}>>t_{\mathrm{evap}}$ and it
seems safe to conclude that in that regime the creation of quantum
nonequilibrium will be negligible. On the other hand, in the final stages of
evaporation, as $M$ approaches $m_{\mathrm{P}}$, according to our estimate
quantum nonequilibrium will be created on a timescale that exceeds the
evaporation timescale by only one or two orders of magnitude, suggesting that
a significant degree of nonequilibrium will be created. In other words, in the
very final stage of black-hole evaporation, outgoing field modes could show
significant departures from the Born rule.

The question remains as to whether or not such departures from the Born rule
can survive quantum relaxation. In the usual calculation of Hawking radiation
it is assumed that the field is in a relatively simple quantum state
corresponding to an appropriately defined vacuum \cite{BD82}. Back reaction on
the background spacetime is neglected. This suffices to obtain the rate of
radiation. But in the final stages of evaporation the background spacetime
changes rapidly and the quantum state is likely to be more complicated than is
usually assumed. We may then expect the field to be subject to quantum
relaxation on some timescale $\tau_{\mathrm{relax}}$. As noted in Section 6.3
quantum nonequilibrium can grow or build up over time only if the condition
(\ref{condn}), or $\tau_{\mathrm{relax}}>\tau_{\mathrm{noneq}}$, is satisfied.
Whether or not nonequilibrium survives in the outgoing radiation then depends
on which effect (instability or relaxation) dominates as $M$ approaches
$m_{\mathrm{P}}$. To answer this we need to know how $\tau_{\mathrm{relax}}$
scales with $M$ in the final stages (where the above estimate has
$\tau_{\mathrm{noneq}}\propto\left(  M/m_{\mathrm{P}}\right)  ^{5}$). That
will require a more detailed model. If it turns out that $\tau_{\mathrm{relax}%
}<\tau_{\mathrm{noneq}}$ as $M$ approaches $m_{\mathrm{P}}$, then any
nonequilibrium that is created will be dissipated before it can build up: the
outgoing nonequilibrium field modes will rapidly relax to equilibrium and so
the resulting Hawking radiation (observed in the asymptotically flat region)
will still obey the Born rule. If instead $\tau_{\mathrm{relax}}%
>\tau_{\mathrm{noneq}}$ as $M$ approaches $m_{\mathrm{P}}$, then we can expect
the outgoing field modes to maintain some degree of nonequilibrium, and the
resulting Hawking radiation could show deviations from the Born rule. It is
perhaps optimistic to expect to be able to observe effects from the final
Planckian phase of Hawking evaporation. But in principle the effects are
predicted to exist. More precise predictions require a more realistic model,
including quantum relaxation, with a reliable estimate of $\tau
_{\mathrm{relax}}$. This is a matter for future work. It seems likely from the
above estimate that significant nonequilibrium will be created in the very
final stage of black-hole evaporation, but whether or not it quickly relaxes
again remains to be seen.

Realistically, Hawking radiation might be observed from evaporating primordial
black holes. It is expected that black holes with a range of masses will have
formed in the early universe \cite{Hawk71}. The evaporation timescale
(\ref{t_evap}) is of order the current age $t_{0}\sim10^{17}\ \mathrm{s}$ of
the universe for microscopic black holes with $M\sim10^{15}\ \mathrm{g}$. Thus
today it might be possible to detect radiation from primordial black holes
with masses $\lesssim10^{15}\ \mathrm{g}$ (if they are sufficiently abundant,
where on some scenarios they may form a significant component of dark matter
\cite{primBHs}). The calculated radiated power $\sim(-t)^{-2/3}$ formally
diverges at $t=0$. It is widely believed that the hole will disappear in an
explosion, whose products will depend on the high-temperature behaviour of
matter. For $M\sim10^{15}\ \mathrm{g}$ it is estimated that a significant
fraction of the luminosity will be in the $\gamma$-ray region (peaked at
$\sim100\ \mathrm{MeV}$) \cite{BHP}. It is therefore plausible that we might
be able to detect $\gamma$-rays from the evaporation of primordial black
holes. If so we could probe their quantum properties, and in particular search
for signs of deviations from the Born rule. Such deviations could manifest as
a blurring of the usual single-photon two-slit interference pattern or as
anomalous polarisation probabilities (deviations from the standard $\cos
^{2}\theta$ modulation with angle $\theta$ \cite{AV04a}).

Finally, as noted in Section 4.4, it has been argued that primordial black
holes can decay via quantum-gravitationally-induced tunnelling (from black
holes to white holes) \cite{BH2WHth}, with potentially observable signatures
in radio and gamma-ray astronomy \cite{BH2WHobs}. In future work it would be
of interest to consider how quantum nonequilibrium might be created during
such a process.

\section{Quantum instability for atomic systems}

It is of theoretical interest to ask if, at least in principle, the Born rule
could be unstable for a laboratory atomic system in the gravitational field of
the earth.

Consider again the general non-Hermitian correction (\ref{non-H_dt}). We saw
that in the spacetime of a Schwarzchild black hole the quantity $\sqrt{g}R$,
which has the dimensions of length, is replaced by $-8\pi r_{\mathrm{S}}$
(equation (\ref{rep_BH})) where the Schwarzchild radius $r_{\mathrm{S}%
}=2GM/c^{2}$ is the natural lengthscale of the background spacetime. For a
system in the gravitational field of the earth we may then expect to implement
a replacement of the form%
\begin{equation}
\sqrt{g}R\rightarrow-8\pi r_{\mathrm{c}}\ , \label{lab}%
\end{equation}
where $r_{\mathrm{c}}$ is the local radius of curvature ($r_{\mathrm{c}}%
\simeq10^{13}\ \mathrm{cm}$ at the surface of the earth).

A related suggestion was made on dimensional grounds by Kiefer and Singh
\cite{KS91}, who considered the effect of the dominant Hermitian term
$\sim\hat{H}_{a}^{2}$ on atomic energy levels (where $\hat{H}_{a}$ is the
uncorrected atomic Hamiltonian), with $r_{\mathrm{c}}$ taken to be the
curvature lengthscale associated with the gravitational field of the atomic
nucleus. A correction term of the form $\sim\left(  G\hslash^{4}/c^{4}%
m^{2}r_{\mathrm{c}}\right)  \nabla^{2}\nabla^{2}$ yields tiny shifts in the
energy levels, which are of course far too small to be observable.

Our interest is in the non-Hermitian term (\ref{non-H_dt}) for a laboratory
atomic system in the gravitational field of the earth. Implementing the
replacement (\ref{lab}), and writing $\mathcal{\hat{H}}_{\phi}$ as
$\mathcal{\hat{H}}_{a}$, the atomic Hamiltonian $\hat{H}_{a}=\int
d^{3}x\,\mathcal{\hat{H}}_{a}$ suffers a non-Hermitian correction%
\begin{equation}
\Delta\hat{H}_{a}\sim-i\frac{\hslash G}{c^{4}}\frac{1}{r_{\mathrm{c}}}\int
d^{3}x\,\frac{\delta}{\delta\tau}\left(  \mathcal{\hat{H}}_{a}\right)
=-i\frac{l_{\mathrm{P}}}{r_{\mathrm{c}}}\frac{\partial\hat{H}_{a}}{\partial
t}t_{\mathrm{P}}\ , \label{Ham_corr_atomic}%
\end{equation}
which is non-zero only if the atom has a time-dependent (uncorrected)
Hamiltonian $\hat{H}_{a}$. The correction (\ref{Ham_corr_atomic}) is roughly
equal to the change in $\hat{H}_{a}$ over a Planck time and is suppressed by
the tiny ratio $l_{\mathrm{P}}/r_{\mathrm{c}}$ (and with a factor of $-i$).

We can now estimate the timescale $\tau_{\mathrm{noneq}}$ for the
gravitational production of quantum nonequilibrium for an atomic system,
caused by a dynamical Hamiltonian in a background curved space, where
$\tau_{\mathrm{noneq}}$ is given by (\ref{tau_est}). In our notation
(\ref{Ham_split}) the correction $\hat{H}_{2}$ is%
\begin{equation}
\hat{H}_{2}\sim-\frac{l_{\mathrm{P}}}{r_{\mathrm{c}}}\frac{\partial\hat{H}%
_{a}}{\partial t}t_{\mathrm{P}}\ .
\end{equation}
If the atomic Hamiltonian $\hat{H}_{a}$ changes rapidly over a small timescale
$t_{a}$, we can apply the sudden approximation where the atomic wave function
$\psi_{a}$ hardly changes over a time $t_{a}$. We can then write $\left\langle
\partial\hat{H}_{a}/\partial t\right\rangle \approx d\left\langle \hat{H}%
_{a}\right\rangle /dt$ and define%
\begin{equation}
\frac{1}{t_{a}}=\left\vert \frac{1}{\left\langle \hat{H}_{a}\right\rangle
}\frac{d\left\langle \hat{H}_{a}\right\rangle }{dt}\right\vert \ .
\end{equation}
We then have%
\begin{equation}
\left\vert \left\langle \hat{H}_{2}\right\rangle \right\vert \sim
\frac{l_{\mathrm{P}}}{r_{\mathrm{c}}}\frac{t_{\mathrm{P}}}{t_{a}}\left\langle
\hat{H}_{a}\right\rangle
\end{equation}
and a timescale (inserting $\hbar$)%
\begin{equation}
\tau_{\mathrm{noneq}}\sim\frac{r_{\mathrm{c}}}{l_{\mathrm{P}}}\frac{t_{a}%
}{t_{\mathrm{P}}}\frac{\hbar}{\left\langle \hat{H}_{a}\right\rangle }\ .
\label{tau_atomic}%
\end{equation}
In realistic conditions we will of course have $r_{\mathrm{c}}/l_{\mathrm{P}%
}>>>1$ and $t_{a}/t_{\mathrm{P}}>>>1$, and so $\tau_{\mathrm{noneq}}$ will be
huge compared with the natural quantum timescale $\hbar/\left\langle \hat
{H}_{a}\right\rangle $ for the atomic system. We can also compare
$\tau_{\mathrm{noneq}}$ with $t_{a}$. Writing $E_{\mathrm{P}}=\hbar
/t_{\mathrm{P}}$ we have%
\begin{equation}
\frac{\tau_{\mathrm{noneq}}}{t_{a}}\sim\frac{r_{\mathrm{c}}}{l_{\mathrm{P}}%
}\frac{E_{\mathrm{P}}}{\left\langle \hat{H}_{a}\right\rangle }\ .
\end{equation}
For $r_{\mathrm{c}}/l_{\mathrm{P}}>>>1$ and $E_{\mathrm{P}}/\left\langle
\hat{H}_{a}\right\rangle >>>1$ the ratio $\tau_{\mathrm{noneq}}/t_{a}$ will
again be prohibitively large.

Note that, because of the rapidly-changing Hamiltonian, the atomic wave
function will necessarily be a superposition of energy eigenstates. This will
ensure quantum relaxation over timescales $\tau_{\mathrm{relax}}%
<<\tau_{\mathrm{noneq}}$. Thus, even if we were able to probe an atomic
ensemble over times of order $\tau_{\mathrm{noneq}}$ (which realistically will
far exceed the age of the universe $t_{0}\sim10^{17}\ \mathrm{s}$), any
nonequilibrium that was gravitationally-generated would have quickly
dissipated. It would appear that, as a point of principle, the gravitational
creation of quantum nonequilibrium at the atomic scale is of theoretical
interest only.

\section{Conclusion}

We have argued that there is no well-defined Born-rule state at the
fundamental level of quantum gravity. A theoretical ensemble with a
(non-normalisable) Wheeler-DeWitt wave functional $\Psi$ is necessarily in a
state of quantum nonequilibrium $P\neq\left\vert \Psi\right\vert ^{2}$
(initially and always). An equilibrium Born-rule state can exist only after we
enter the semiclassical regime of quantum systems on a classical spacetime
background, with time-dependent and normalisable wave functions $\psi$
satisfying a Schr\"{o}dinger equation. At the beginning of the semiclassical
regime, however, `initial' quantum nonequilibrium $\rho\neq\left\vert
\psi\right\vert ^{2}$ is inherited from the deep quantum-gravity regime as a
conditional probability. Thus quantum gravity naturally creates an early
nonequilibrium universe. Quantum relaxation $\rho\rightarrow\left\vert
\psi\right\vert ^{2}$ takes place only afterwards, finally yielding the Born
rule as an emergent equilibrium state. We have also shown how small
quantum-gravitational corrections to the Schr\"{o}dinger equation yield an
intermediate regime in which the Born rule is unstable: an initial ensemble
$\rho=\left\vert \psi\right\vert ^{2}$ can evolve to $\rho\neq\left\vert
\psi\right\vert ^{2}$. We have seen that the latter effects are generally very
small, though perhaps significant in the final stages of black-hole
evaporation. These results have been obtained by applying the de Broglie-Bohm
pilot-wave formulation of quantum mechanics to canonical quantum gravity. The
results emerge naturally by following the internal logic of this particular
approach to quantum physics.

A key starting point for our argument has been the Klein-Gordon-like structure
of the Wheeler-DeWitt equation. We have argued that this is not a peculiarity
to work around but a sign that the timeless Wheeler-DeWitt wave functional
$\Psi\lbrack g_{ij},\phi]$ is not a wave functional as we usually understand
the term. The quantity $\left\vert \Psi\lbrack g_{ij},\phi]\right\vert ^{2}$
is non-normalisable not for some technical reason to be fixed but because,
while it may superficially resemble the familiar probability densities
$\left\vert \psi(q,t)\right\vert ^{2}$ encountered in non-gravitational
physics, in reality $\left\vert \Psi\lbrack g_{ij},\phi]\right\vert ^{2}$ is a
quite different kind of physical thing. We are able to make sense of this in
de Broglie-Bohm pilot-wave theory, because in this formulation of quantum
mechanics there is no law-like relationship between the probability density
$P$ and the wave function $\Psi$. Instead, in pilot-wave theory, the Born rule
emerges as an equilibrium state by dynamical relaxation. Because $\left\vert
\Psi\lbrack g_{ij},\phi]\right\vert ^{2}$ is non-normalisable, however, there
is no equilibrium state in the deep quantum-gravity regime, where inevitably
$P[g_{ij},\phi,t]\neq\left\vert \Psi\lbrack g_{ij},\phi]\right\vert ^{2}$
always. An equilibrium Born-rule state can emerge only in the semiclassical
regime of time-dependent and normalisable wave functions $\psi(q,t)$, by means
of (coarse-grained) relaxation $\rho(q,t)\rightarrow\left\vert \psi
(q,t)\right\vert ^{2}$. The small non-Hermitian gravitational corrections to
the Schr\"{o}dinger equation for $\psi$, first derived by Kiefer and Singh,
then make it possible for initial equilibrium $\rho=\left\vert \psi\right\vert
^{2}$ to evolve to nonequilibrium $\rho\neq\left\vert \psi\right\vert ^{2}$,
though the latter effects are generally very small.

It should be emphasised that our intermediate regime, with a small instability
of the Born rule, has been derived from the fundamental equations of
pilot-wave quantum gravity. Performing a semiclassical expansion of the
Wheeler-DeWitt wave functional results in small non-Hermitian terms in the
effective Hamiltonian of the Schr\"{o}dinger regime, while the de Broglie
velocities remain those associated with only the Hermitian part of the
Hamiltonian. As a consequence, there is a mismatch between the continuity
equations satisfied by the actual density $\rho$ and by the Born-rule density
$\left\vert \psi\right\vert ^{2}$, resulting in a small
gravitationally-induced instability of the Born rule.

The derivation of non-Hermitian corrections to the Schr\"{o}dinger equation
depends, however, on a semiclassical expansion whose validity might be
questioned. Previous authors have generally ignored the non-Hermitian terms
because there is no consistent way to interpret them in standard quantum
mechanics. It is then tempting to view such terms as an artifact to be
eliminated from the effective Hamiltonian by appropriately redefining the
effective wave function (as discussed in Section 5.2). But in pilot-wave
theory the non-Hermitian terms are fully consistent, thereby removing the
ground for viewing them as artificial or for seeking to eliminate them. It may
however be that the question of whether or not the non-Hermitian terms really
exist can only be settled by experiment, in the sense that some empirical
input may be needed in order to know exactly how our familiar time-dependent
wave functions are to be identified as they emerge from the underlying
quantum-gravitational formalism.

It is also worth remarking that, even if the non-Hermitian terms do turn out
to be a mathematical artifact, this will only invalidate the intermediate
regime studied in this paper. It will still be the case that there is no
well-defined Born rule in the deep quantum-gravity regime, which remains in a
perpetual state of nonequilibrium. It will also still be the case that as the
universe enters the semiclassical regime it will be in a state of
nonequilibrium, with relaxation to the Born rule taking place only afterwards.
The Born rule will still be unstable, in the sense that when an equilibrium
system enters the deep quantum-gravity regime it can be expected to emerge in
a state of nonequilibrium.

Pilot-wave theory solves the notorious quantum measurement problem
\cite{Bell87}, but for as long as we are confined to quantum equilibrium it
remains indistinguishable from other formulations or interpretations of
quantum physics. In this paper we have shown that pilot-wave theory offers a
natural understanding of certain peculiarities of canonical quantum gravity.
The non-normalisability of the Wheeler-DeWitt wave functional and the failure
of the naive Schr\"{o}dinger interpretation are explained by the absence of a
physical Born-rule equilibrium state. The emergence of the Born rule in the
semiclassical regime is explained by quantum relaxation. The small
non-Hermitian terms in the gravitationally-corrected Schr\"{o}dinger equation
no longer present an inconsistency but instead generate a small instability of
the Born rule. Finally, the presence of actual de Broglie-Bohm trajectories
for all wave functionals arguably gives a clearer account of WKB time and of
the emergence of time-dependent wave functions in the semiclassical regime.
Thus in several respects the de Broglie-Bohm formulation of quantum mechanics
appears advantageous in understanding gravitation.

We have seen that a system emerging from the deep quantum-gravity regime can
be expected to violate the Born rule. In future work it would therefore be of
interest to study the potential creation of quantum nonequilibrium in bouncing
cosmologies and in black-hole to white-hole transitions. We may also expect
quantum-gravitational effects to generate nonequilibrium close to classical
spacetime singularities, thereby potentially enabling the resolution of
black-hole information loss by the mechanism studied in refs.
\cite{AV04b,KV20} (where ingoing Hawking field modes interact with
nonequilibrium degrees of freedom behind the horizon and thereby transmit
nonequilibrium to the exterior region via entanglement with the outgoing
modes). Finally, it would also be of interest to reconsider the effects
discussed in this paper in terms of loop quantum gravity.

The ultimate test of any physical theory is of course by experiment. We have
provided approximate calculations of the effects of quantum instability for
various systems. Experimentally the most promising candidates for a test
appear to be exploding primordial black holes, which may form a significant
component of decaying dark matter. As we have seen, radiation from the final
stages of black-hole evaporation can potentially show deviations from the Born
rule, for example in the form of anomalous photon polarisation probabilities.
Should such effects ever be observed, a new window would be opened on the
relationship between quantum theory and gravitation.

\end{document}